Super-hot (T > 30 MK) Thermal Plasma in Solar Flares

by

Amir Caspi

A dissertation submitted in partial satisfaction of the

requirements for the degree of

Doctor of Philosophy

in

Physics

in the

Graduate Division

of the

University of California, Berkeley

Committee in charge:

Professor Robert P. Lin, Chair
Professor Steven E. Boggs
Professor Gibor Basri

Spring 2010

Super-hot (T > 30 MK) Thermal Plasma in Solar Flares

by

Amir Caspi

Abstract

Super-hot (T > 30 MK) Thermal Plasma in Solar Flares

by

Amir Caspi

Doctor of Philosophy in Physics

University of California, Berkeley

Professor Robert P. Lin, Chair

The Sun offers a convenient nearby laboratory to study the physical processes of particle acceleration and impulsive energy release in magnetized plasmas that occur throughout the universe, from planetary magnetospheres to black hole accretion disks. Solar flares are the most powerful explosions in the solar system, releasing up to $10^{32}$-$10^{33}$ ergs over only 100-1,000 seconds. These events can accelerate electrons up to hundreds of MeV and can heat plasma to tens of MK, exceeding ~40 MK in the most intense flares. The accelerated electrons and the hot plasma each contain tens of percent of the total flare energy, indicating an intimate link between particle acceleration, plasma heating, and flare energy release.

X-ray emission is the most direct signature of these processes; accelerated electrons emit hard X-ray bremsstrahlung as they collide with the ambient atmosphere, while hot plasma emits soft X-rays from both bremsstrahlung and excitation lines of highly-ionized atoms. The *Reuven Ramaty High Energy Solar Spectroscopic Imager* (RHESSI) observes this emission from ~3 keV to ~17 MeV with unprecedented spectral, spatial, and temporal resolution, providing the most precise measurements of the X-ray flare spectrum and enabling the most accurate characterization of the X-ray-emitting hot and accelerated electron populations.

RHESSI observations show that "super-hot" temperatures exceeding ~30 MK are common in large flares but are achieved almost exclusively by X-class events and appear to be strictly associated with coronal magnetic field strengths exceeding ~170 Gauss; these results suggest a direct link between the magnetic field and heating of super-hot plasma, and that super-hot flares may require a minimum threshold of field strength and overall flare intensity.

Imaging and spectroscopic observations of the 2002 July 23 X4.8 event show that the super-hot plasma is both spectrally and spatially distinct from the usual ~10-20 MK plasma observed in nearly all flares, and is located above rather than at the top of the loop containing the cooler plasma. It exists with high density even during the pre-impulsive phase, which is dominated by coronal non-thermal emission with negligible footpoints, suggesting that particle acceleration and plasma heating are intrinsically related but that, rather than the traditional picture of chromospheric evaporation, the origins of super-hot plasma may be the compression and subsequent thermalization of ambient material accelerated in the reconnection region above the flare loop, a physically-plausible process not detectable with current instruments but potentially observable with future telescopes. Explaining the origins of super-hot plasma would thus ultimately help to understand the mechanisms of particle acceleration and impulsive energy release in solar flares.

To my parents, Rachel & Ehud,
and to my wife, Heather,
whose constant love
and encouragement
has kept me afloat

And to Fiver and Pitzi,
my constant companions,
whom I dearly miss

# Table of Contents

# List of Figures

ix

# Acknowledgements

First and foremost, I would like to thank my advisor, Bob Lin, whose encyclopedic knowledge, scientific rigor, and infectious curiosity and enthusiasm are a constant inspiration; I greatly appreciate his patience and dedication in mentoring me these past years. I sincerely thank my committee, Steve Boggs and Gibor Basri, for their time and willingness to review this dissertation. I am also deeply grateful to many colleagues at the Space Sciences Lab, at NASA Goddard Space Flight Center, and at other institutions for their invaluable insights and helpful conversations; I hope they forgive not being named individually, although I am particularly indebted to Brian Dennis, who has been a sort of mentor-from-afar and whose continued interest kept me encouraged throughout this work. Of course, I also thank NASA contract NAS5-98033 and NASA grant NNX08AJ18G for supporting me financially during my graduate career.

More personally, I would like to thank my friends who have provided support and words of encouragement throughout my journey, including Greg and James, with whom I wish I spoke more often; Mike, who understands that science is cool; and Doug and David, who always had faith in me. Niv and Sharona deserve special thanks; they have become like family, and comprehend first-hand what a graduate career entails. I am also grateful for Josh and Penny, good friends and neighbors who consistently spurred me on and provided much-needed company, not to mention countless gourmet dinners.

Finally, I would like to thank my family, especially my parents, Rachel and Ehud, without whom none of this would have been possible. An accomplished scientist and an outstanding engineer, they fostered my skeptical nature and nurtured my scientific curiosity without reservation from as far back as I can remember. I am forever thankful for my loving wife, Heather, who followed me across the country and who has unquestioningly supported me in every way possible. To my parents and my wife, I owe more than words can express.

# Chapter 1: The Sun and Solar Flares

*Introduction*

The Sun is the most prominent of the heavenly bodies, dominating the entire daytime sky. It is no coincidence that the Sun figures prominently in nearly every recorded mythology. It is the source of life-giving light and heat, and is the origin (whether directly or indirectly) of nearly all the energy consumed on our planet. The Sun also provides a wealth of knowledge, however, as its proximity provides the perfect opportunity to study the physics behind a host of astrophysical phenomena up close and, practically, without leaving home.

Solar flares are the most explosive phenomenon in the solar system, releasing $10^{32}$-$10^{33}$ ergs over timescales of only $10^2$-$10^3$ seconds; few phenomena in the universe are as explosive, in terms of energy released per time, with the rare exceptions including flares on other stars, accretion disk flares, and accretion-related gamma-ray bursts from compact objects (e.g. magnetars). Solar flares thus provide us with a convenient backyard laboratory to study these enormously energetic processes. Tens of percent of the energy released in a solar flare can go into accelerated particles, making flares efficient particle accelerators, although the exact mechanisms behind this process remain poorly understood. Solar flares can also heat plasma to temperatures exceeding ~40 MK, possibly higher. Such hot plasmas are difficult to create and to contain; studying their evolution and understanding their origins can therefore hopefully shed insight into how energy is released and transported in solar flares and similar phenomena.

X-rays are the spectral signature of the energetic processes within a flare (Chapter 2) and are therefore probes of the X-ray-emitting accelerated and/or heated electron populations; it is thus by observing X-rays that we can determine the characteristics (e.g. the energy spectra and locations) of these energetic electrons and the ambient solar atmosphere in which they interact. X-ray observations of flare emission started in the late 1950s with the dawn of the space age, and are continued today by various instruments including the *Reuven Ramaty High Energy Solar Spectroscopic Imager* (RHESSI). RHESSI (Chapter 3) combines high-resolution spectroscopy and imaging of X-rays and gamma rays from ~3 keV up to ~17 MeV to provide an unprecedented look at the physical processes within solar flares. RHESSI is especially suited to observations of the hot flare plasmas, including the "super-hot" plasmas with temperatures above ~30 MK often found in large flares (Chapter 4); its rich data set has provided valuable insight into the evolution of such super-hot temperatures and has shown that the super-hot component is observed from the very beginning of the flare. RHESSI's sensitivity and spectral resolution also allowed the discovery of an unusual "pre-impulsive" phase found in certain flares (Chapter 5), which precedes the primary flare emission and which appears to be dominated by strong nonthermal emission from the corona; this intriguing time period may hold the clues to what triggers the explosive release of energy during the impulsive phase. These observations have, of course, brought new questions to light, such as how super-hot plasmas can reach such high temperatures and densities (1-2 orders of magnitude higher than the ambient values) so early in the flare; this is especially interesting at times such as the pre-impulsive phase when no footpoint emission is evident, suggesting that the super-hot plasma cannot originate from the traditional picture of accelerated particle-driven chromospheric evaporation often thought to be the source of the cooler, ~10-20 MK plasma observed in nearly all flares. If flares are driven by magnetic reconnection, this suggests that the super-hot plasma may be heated directly by reconnection and by subsequent compression during relaxation of flare loops (Chapter 6); although we have no current ob-servations of such a process, rough calculations yield temperatures and densities that are consistent (to order of magnitude) with observed quantities, suggesting that the scenario is physically plausible and that next-generation instruments with higher sensitivity may be able to answer this question, which would reveal not only the origins of super-hot plasma but also the ultimate mechanisms of flare energy transport and release.

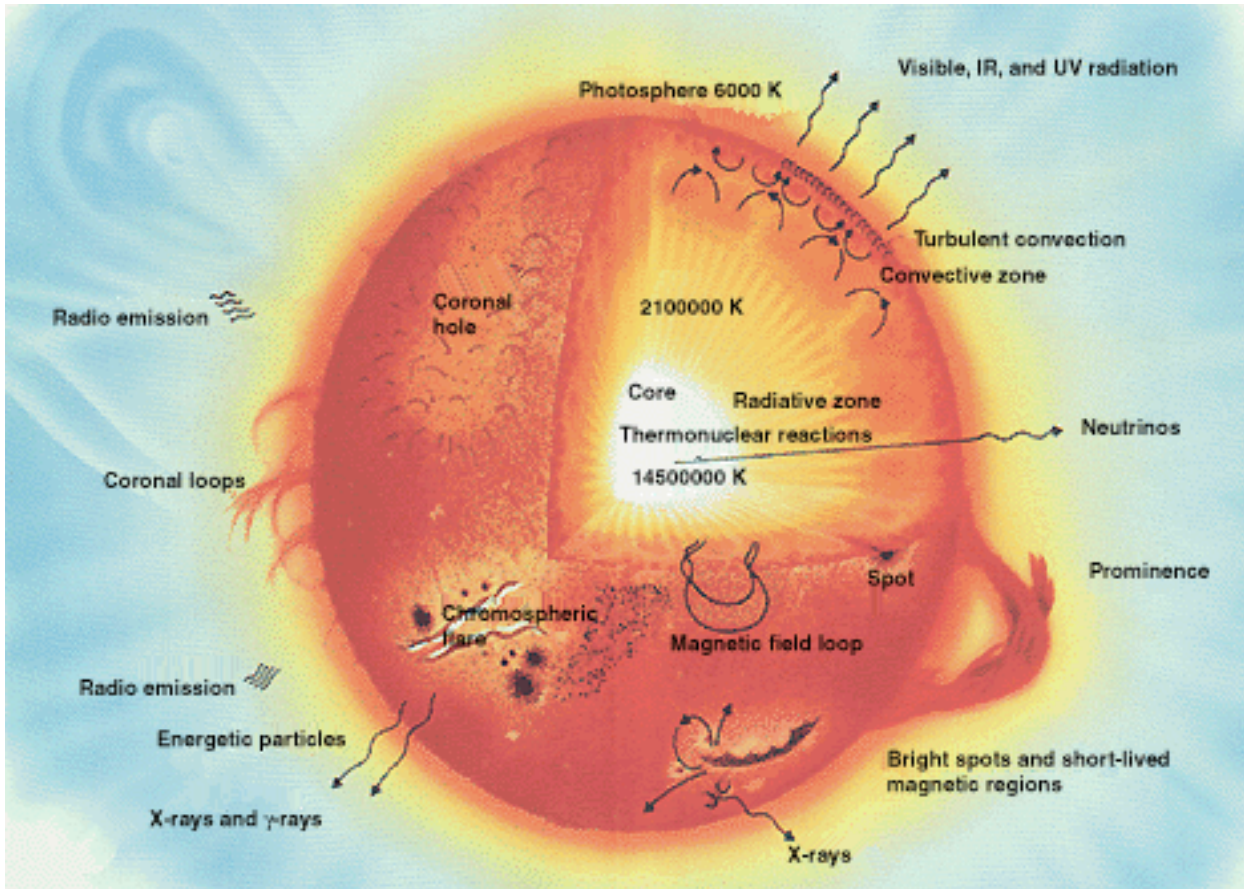

**Figure 1.1** – Cutaway view of the Sun showing the interior layers and average temperature values, and schematic representations of surface features and emission phenomena. (Image credit: SolarViews.org)

### 1.1 Solar structure

The Sun, as any other star, is a giant ball of gas and plasma; it is composed (by mass) of ~73% hydrogen, ~25% helium, and ~2% heavier elements such as iron, nickel, carbon, oxygen, and others. The gravitational compression of its enormous mass (~$2 \times 10^{30}$ kg, about $3.3 \times 10^5$ times more massive than the Earth) generates pressures and temperatures sufficiently high to maintain stable thermonuclear fusion at the core. The combination of gravitational compression, thermal/radiation pressure, and atomic physics lead to a layered structure of the Sun (Figure 1.1 and Figure 1.2, left), with each layer having distinctive characteristics. Detailed descriptions can be found in Gibson (1973) or, for a more modern treatment, Phillips (1992).

The Sun's core contains about half of its mass but extends to only about 25% of its radius ($R_\odot \approx 7.0 \times 10^8$ m). Here, at temperatures of up to ~15 MK, densities of up to ~160 g/cm$^3$, and pressures of up to ~2.5×10^11 atm, hydrogen nuclei are fused into helium through the three-step *p-p* reaction; each helium nucleus is less massive than the four initial hydrogen nuclei, with the remaining ~0.7% of the mass converted to energy and released as a gamma-ray photon (and a neutrino). These photons carry ~99% of the Sun's total generated energy, but this energy is not transported to the surface directly; rather, because of the high densities within the solar interior, the photons will scatter many times – their mean-free-path is only ~1 cm, and an individual photon may take ~10^5-10^6 years to escape the solar interior. Outwards from the core up to ~0.7 R_⊙ lies the *radiative zone*; here, the density and temperature drop from ~20 g/cm^3 and ~7 MK to ~0.2 g/cm^3 and ~2 MK, respectively. These values are sufficiently high that the atmosphere remains ionized and, despite the short mean-free-path, radiation can efficiently transport energy away from the core. Past ~0.7 R_⊙, however, the lower temperatures allow atoms (particularly of the heavier elements) to form, which can more readily absorb the outgoing radiation; the increased opacity reduces the efficiency of radiative energy transport and steepens the negative temperature gradient, making the plasma convectively unstable – plasma blobs that rise and expand/cool adiabatically are hotter (and thus less dense) than their surroundings, remaining buoyant and rising further. Within this layer, termed the *convective zone*, such instability leads

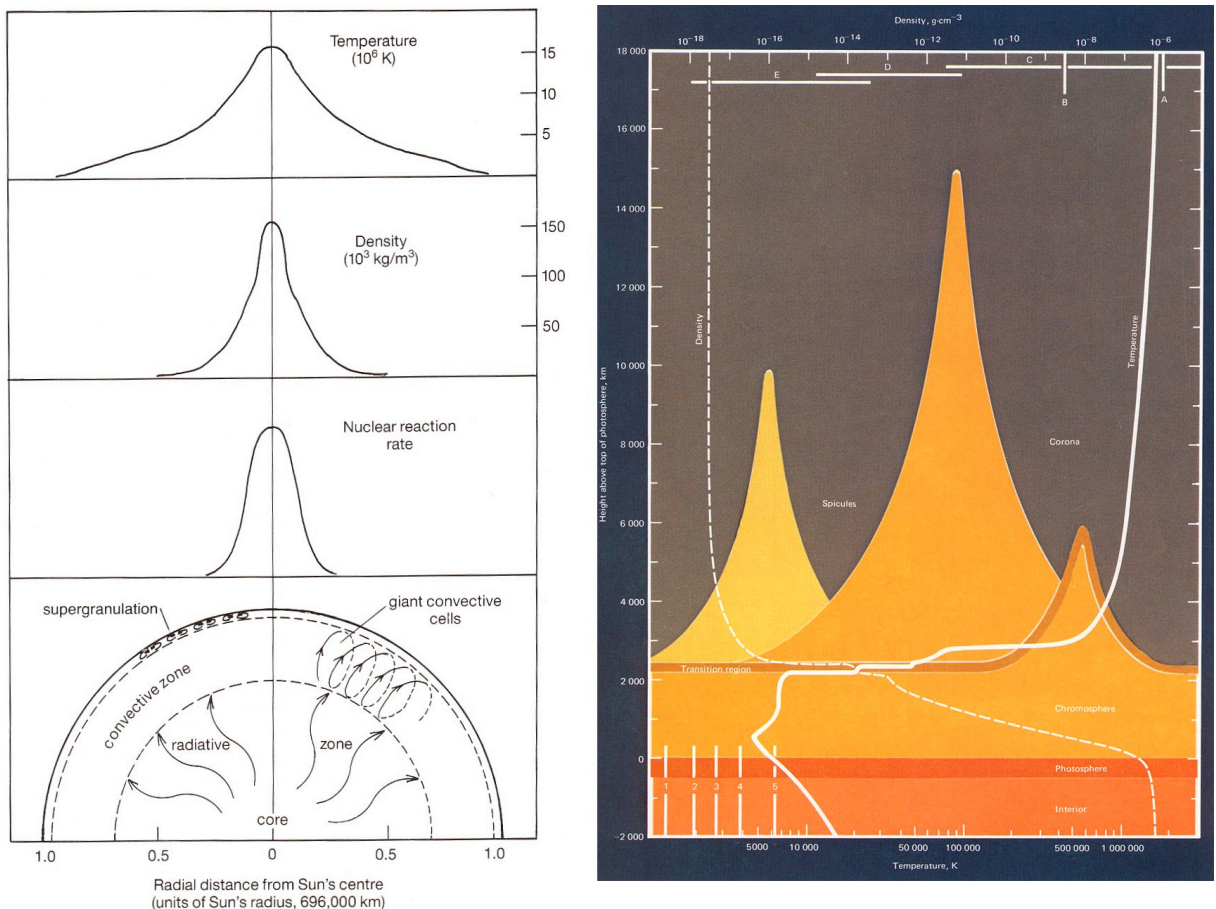

**Figure 1.2** – [left] Temperature and density as a function of radial distance from Sun center (from Phillips [1992]). [right] Temperature and density variation with altitude above the photosphere. (Image credit: NASA)

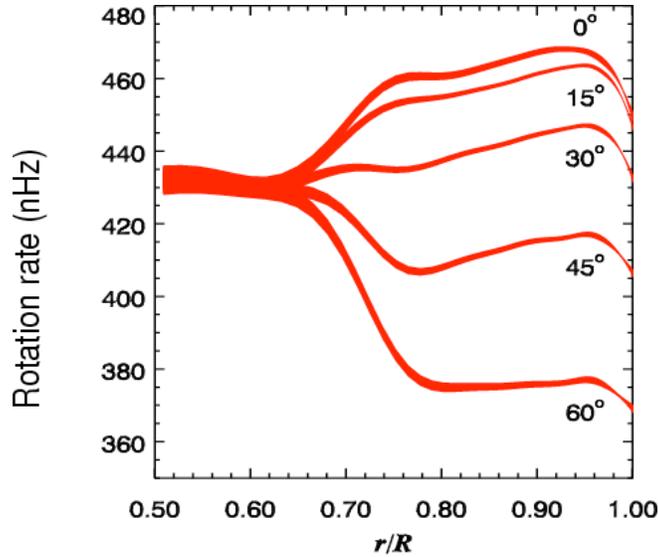

**Figure 1.3** – Rotation rate versus depth from the photosphere (as a fraction of the solar radius), cut across various latitudes; the *tachocline* is at ~0.66 R⊙ (Image credit: NSF/NSO)

to the formation of large-scale convection cells that carry the thermal energy the rest of the way to the surface.

As the atmospheric density decreases, so does its opacity; at the top of the convective zone, the density is low enough that outward-emitted photons are unlikely to interact with the ambient material and can escape the Sun. Here, the energy convectively transported from the radiative zone is released as photons in a thin (~few hundred km) layer termed the *photosphere*, which has an effective temperature of ~5800 K and defines the visible surface of the Sun (Figure 1.2, right). The tops of the convection cells from the convective zone – where the hot plasma rises, releases its energy, cools, and sinks – manifest as turbulent "super-granules" that perturb the photosphere. About 500 km above this, although the density continues to decrease, the temperature actually begins to increase with altitude through a ~2000 km-thick layer termed the *chromosphere*, reaching ~20,000 K at the top of the layer. (The exact source of this heating, which would seem to defy the second law of thermodynamics, is still the subject of much debate, although leading candidates include plasma waves and/or heating from magnetic reconnection [see §1.3] in nano-flares). At these temperatures, hydrogen and helium become partially ionized, which, combined with the low density, makes radiative energy loss inefficient compared to the energy input; this further steepens the positive temperature gradient and results in a ~2 order of magnitude increase in temperature (and a corresponding decrease in density) over a very thin (~100 km) layer called the *transition region*. Above this lies the outermost layer of the atmosphere, the hot and tenuous *corona*, which extends out into interplanetary space.

While the photosphere and all higher layers can actually be seen, the interior zones are not observable directly. However, motions within the interior generate oscillations that *are* visible on the surface, and observations of these spherical harmonics (known as helioseismology), coupled with knowledge of atomic physics, thermodynamics, and fluid mechanics, can be used as probes to determine the specific characteristics of the core, radiative, and convective zones.

## 1.2 Solar activity

The Sun does not rotate as a rigid body, but rather experiences differential rotation, wherein the rotation period is a function of radial distance and polar angle. The surface rotation period is determined observationally by tracking visible photospheric features; in the sidereal (fixed relative to distant, thus "slow-moving," stars) reference frame, it is ~24.5 Earth days at the equator, increasing to ~34.2 days at the poles. Models based on helioseismologic data suggest that this differential rotation extends inwards, with the latitude-dependent period also changing with depth down to the radiative zone, which is thought to rotate as a rigid body (Figure 1.3).

### 1.2.1 The Sun's magnetic field

At the interface between the uniformly-rotating radiative zone and the differentially-rotating convective zone, known as the *tachocline*, the large-scale shear in plasma flow is thought to cause a dynamo effect which continuously generates the Sun's magnetic field. With a few exceptions (discussed below), the solar plasma can generally be reasonably approximated as a perfectly-conducting fluid which thus obeys the principles of ideal magnetohydrodynamics (MHD). This zero-resistivity approximation immediately leads to the "frozen flux condition" wherein the magnetic field lines are fixed to and move together with a given fluid volume element, as changes (in time or space) in the external (i.e. dynamo) field induce diamagnetic currents in the plasma that generate a secondary magnetic field such that the total field within a given fluid element remains constant.

The relative strengths of the external magnetic pressure (or, equivalently, the field energy density) and the plasma thermal energy density (and hence the magnetic pressure of the induced diamagnetic field) perpendicular to the external field determine whether the field or the plasma dominate the large-scale motion; it is convenient to define the plasma parameter β which relates these two, in cgs units:

$$\beta = \frac{n k_B T}{B^2/8\pi} \tag{1.1}$$

where $B$ is the external magnetic field strength (in Gauss), $k_B$ is Boltzmann's constant, and $n$ is the number density (in cm$^{-3}$) for a plasma with temperature $T$ (in K). (β is sometimes defined with a factor of 2 to account for both electrons and ions, as done in Figure 1.4.)

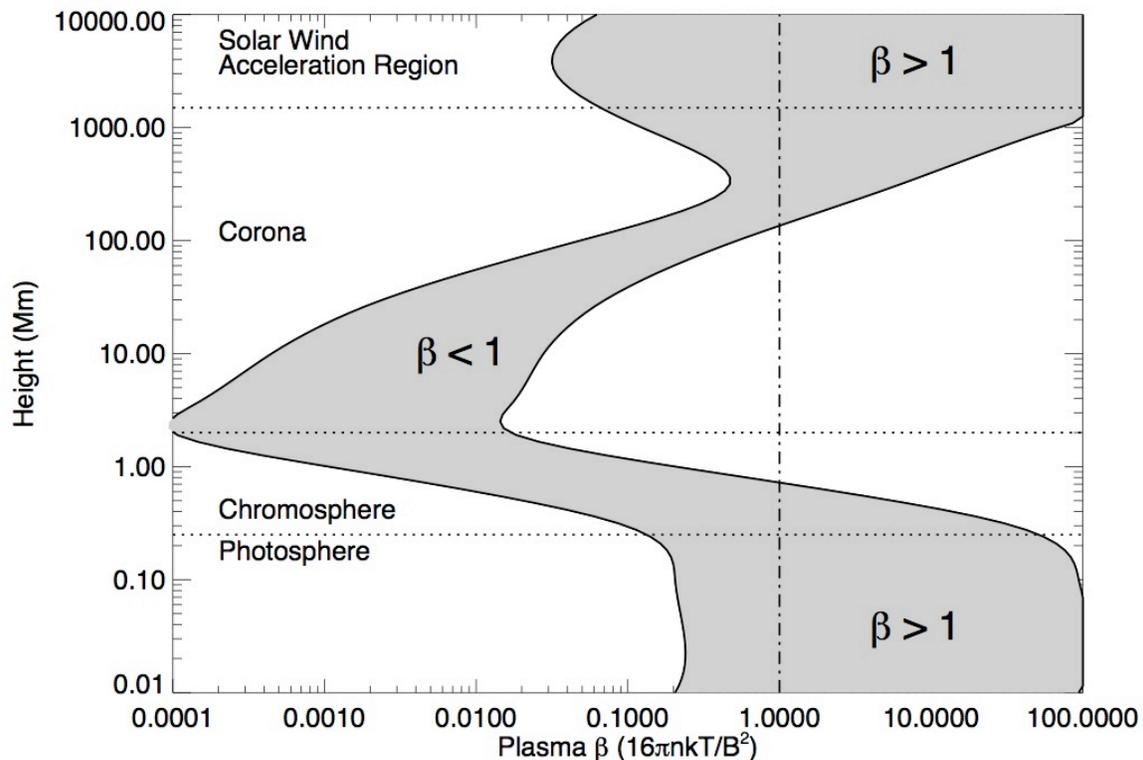

**Figure 1.4** – Plasma β variation with height above the photosphere, for field strengths between 100 G and 2500 G (from Aschwanden [2005]).

The most commonly accepted model describing the time evolution of the magnetic field is that of Babcock (1961) and Leighton (1964, 1969), shown schematically in Figure 1.5. In the convective zone, photosphere, and much of the chromosphere, the densities and temperatures are high enough that β > 1 – the plasma pressure dominates and field lines (or "flux tubes") are dragged along with plasma motion. Thus, differential rotation "winds up" the magnetic field lines, gradually converting the initial poloidal (longitudinal) field to a largely toroidal (azimuthal) configuration. The magnetic pressure is increased in areas where the field lines become more concentrated; this can be enhanced by convective and/or turbulent motion of the plasma, which can further kink the field lines and/or twist them into rope-like structures. Because these processes are adiabatic, the total pressure must remain constant – the increased magnetic pressure results in a lower plasma pressure, increasing the buoyancy of the flux tubes (compared to the surrounding plasma) and causing them to emerge from the surface, forming observable phenomena such as sunspots, loops, or "open" field lines (which are not truly open, as this would violate Maxwell's equations, but are so-called because they extend well into interplanetary space).

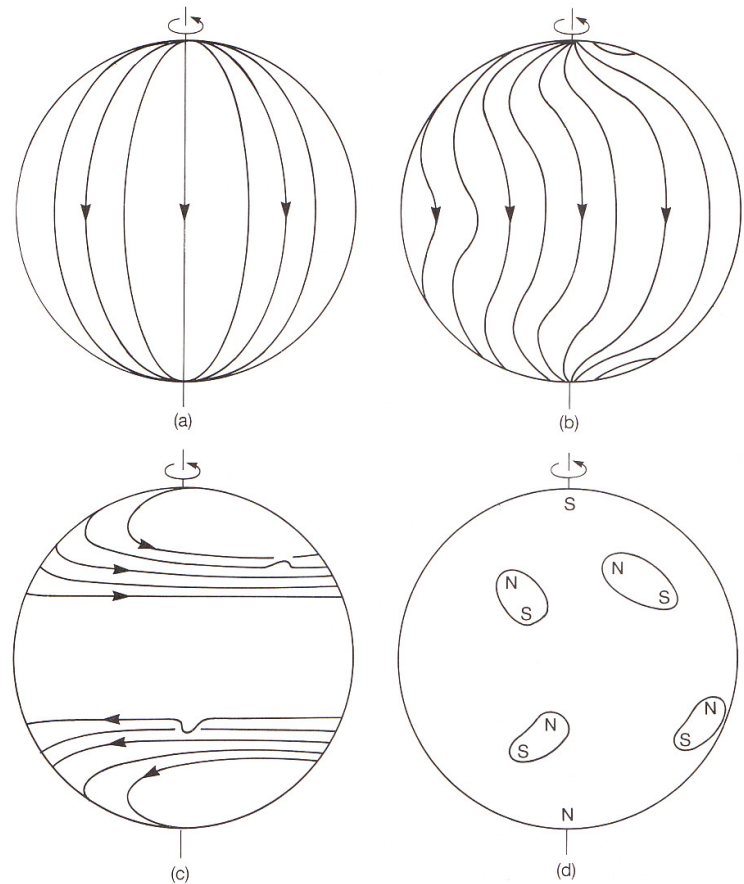

**Figure 1.5** – Schematic representation of the Babcock-Leighton model; an initially-poloidal field [a] is wound up by differential rotation [b], eventually leading to a largely-toroidal configuration [c]. The leading and trailing zones of emergent flux regions in each hemisphere have, respectively, equal and opposite polarity of that hemisphere's polar field [d]; meridional flow of the emergent flux (not shown) weakens and eventually reverses the polar field (from Phillips [1992]).

Although these emergent flux regions can have complicated shapes, in general they have leading (rotationwards) and trailing zones; the leading zone, which is closer to the equator, has the same field polarity as that emerging from the pole of its respective hemisphere, while the trailing zone has the opposite polarity. The leading zone brings the toroidal field equatorwards while the trailing zone neutralizes (and ultimately reverses) it polewards; over time, continued differential rotation thus "unwinds" the field, leading eventually to a poloidal field with a polarity reversed from previously. In this way, the field strength/complexity and polarity cycle over a period averaging 11 and 22 years, respectively.

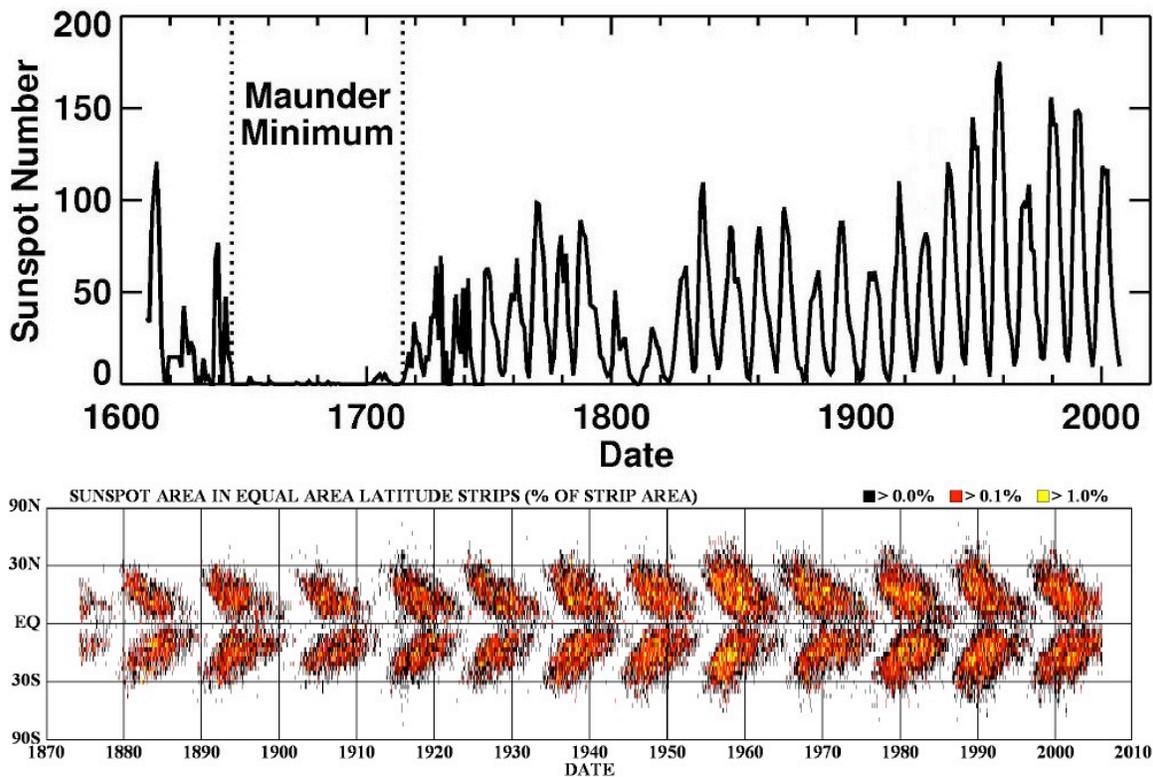

**Figure 1.6** – [top] Yearly-averaged sunspot number from 1610 to 2008. [bottom] The so-called "butterfly diagram" showing how new sunspots appear closer to the equator as the solar cycle progresses. (Image credit: NASA)

### 1.2.2 Sunspots and active regions

Emergent flux regions often manifest as sunspots, which are visible as dark regions on the solar disk. Field strengths within sunspots are often thousands of Gauss, compared to only tens of Gauss in "quiet" regions; the higher magnetic pressure suppresses convective motion within the sunspot umbra (the dark central region) and also lowers the plasma pressure, both of which result in a lower plasma temperature within the sunspot, causing it to appear darker than the surrounding area. As the solar cycle progresses per the Babcock-Leighton model, new sunspots typically appear closer to the equator than previous ones, starting from ~40-50° latitude early in the cycle to ~5-10° latitude near the end (Figure 1.6). The number of sunspots generally increases as the global field winds up (increasing the field strength), then decreases as it unwinds; the sunspot number is therefore often used as a proxy for the solar activity cycle as a whole. While the average cycle period is 11 years, individual cycles may vary from ~9 to 13 years; the amplitude of the cycle, and thus its average activity level, can vary greatly, including such extremes as the "Maunder Minimum" wherein very few sunspots occurred for ~6 entire cycles.

Sunspots always appear within larger regions of emergent flux called "active regions," although not all active regions contain sunspots. Active regions are characterized by high magnetic field fluxes and often-complicated field configurations, with zones of opposite polarity in close proximity to, and often intertwined with, one another. These zones are "footpoints" of emerging flux tubes that, because of their buoyancy and the nearly vertical density (and thus pressure) gradient of the transition region, can balloon high (typically ~1-10 Mm) into the corona to form loops; because of the corona's low β, plasma in the loops remains relatively confined and can be at different temperatures compared to the surrounding atmosphere. In their minimum-energy state, the loops follow a potential configuration, where the magnetic field can be expressed as the gradient of a scalar function ($\nabla \times \mathbf{B} = 0 \implies \mathbf{B} = \nabla \psi$). Turbulent and convective motion on the surface moves the footpoints around, causing loops to shear and twist into a so-called "non-potential," higher energy state. The kinetic energy of photospheric surface motion is thus stored as increased magnetic energy within the complex field configurations in the corona. Eventually, the field must reach a critically-unstable point, whereupon the stored energy is released in the form of a solar flare.

### 1.3 Solar flares

The first recorded observation of a solar flare was made in 1859 by Sir Richard Carrington and, independently, by Richard Hodgson. Using conventional optical telescopes, they observed a brief but intense visible-light brightening within a large sunspot group, which appeared suddenly and then faded over the subsequent few minutes. Since then, flares have been shown to emit strongly across the entire spectrum, not just in visible light (indeed, "white light" brightenings are actually fairly rare). Emission in Hα – a 656-nm excitation line (cf. §2.1.3) of neutral hydrogen – was first observed by Charles A. Young in 1870; bright radio emission from flares at centimeter (microwave) and meter wavelengths was discovered in the 1940s, first by Appleton & Hey (1946) and followed by many others. The space age extended observing capabilities to frequencies normally blocked by Earth's atmosphere and heralded the discoveries of intense flare emission in ultraviolet (UV), extreme ultraviolet (EUV), X-rays (cf. §1.4), and gamma rays. (Henceforth, "soft" X-ray [SXR] emission generally refers to ~0.1-20 keV photons, while "hard" X-ray [HXR] emission refers to photons above ~20 keV.)

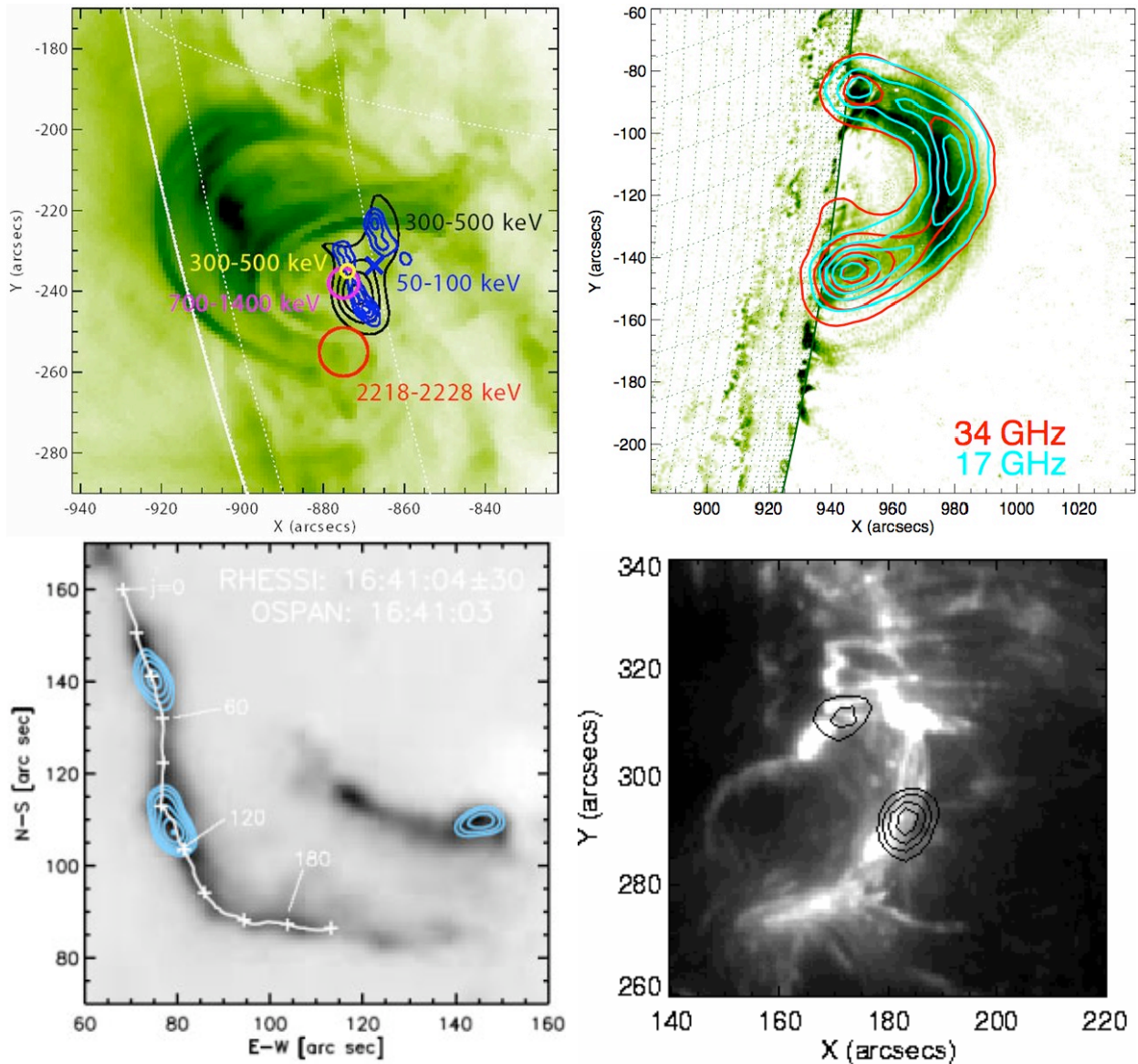

**Figure 1.7** – Images of flare emission at various energies: [top left] RHESSI X-ray and gamma-ray emission contours overlaid on TRACE 195 Å EUV emission (primarily from thermally-excited Fe XII and XXIV lines) for the 2002 July 23 flare (from Hurford *et al.* [2003]); [top right] NoRH radio emission contours overlaid on TRACE 195 Å emission for the 2002 Aug 24 flare; [bottom left] RHESSI 25-50 keV contours overlaid on (reverse-color – darker is brighter) OSPAN Hα emission (from neutral chromospheric hydrogen) for a 2005 May 13 flare (image credit: RHESSI Science Nuggets); [bottom right] *Yohkoh* HXT HXR emission contours overlaid on TRACE 1600 Å emission for the 2000 Nov 24 flare (image credit: *Yohkoh* Science Nuggets).

Flare emission is observed at all altitudes, from the chromosphere to the corona, and even out to interplanetary space. As can be seen in the examples in Figure 1.7, Hα emission is primarily chromospheric and often appears as two long ribbons, especially for large flares in complex active regions. UV and EUV excitation lines, which are emitted by atoms in various states of ionization (§2.1.3) and are characteristic of hot thermal plasma (§1.4), are observed from loop-like structures in the corona and transition region; SXR emission is also observed from the loops, and is often brightest at the looptops. HXR emission is generally observed from the loop footpoints that lie along the Hα ribbons, though it has also been observed at or above the SXR-emitting looptops; for very energetic flares, gamma-ray emission is also observed from the footpoints. Radio emission is observed from both the footpoints and the loops, as well as from much higher in the corona and, at decreasing frequencies, out to interplanetary space.

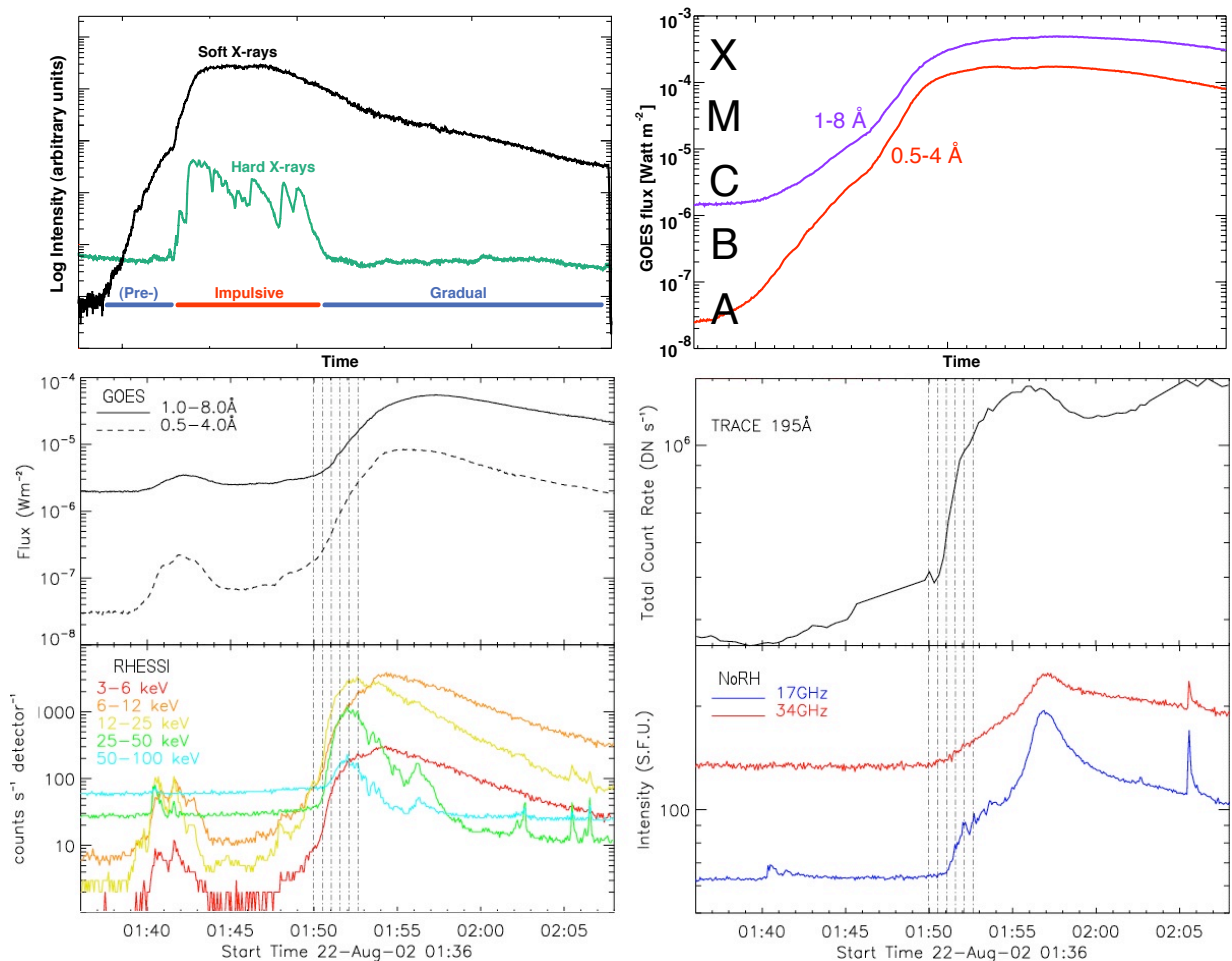

**Figure 1.8** – [top left] Schematic lightcurves exemplifying the pre-impulsive, impulsive and decay/gradual flare phases, identified by the behavior of the X-ray (and other wavelength) emission. [top right] Example lightcurve showing the "GOES class" flux-based flare classification scheme; the flare shown has GOES-class X4.8. [bottom] Lightcurves from GOES, RHESSI, TRACE, and NoRH showing the time evolution of flare emission at various energies/wavelengths for an event on 2002 Aug 22 (adapted from Bain & Fletcher [2009]).

Flares are highly dynamic, and the evolution of a flare is thus observed via its radiative output, as shown in Figure 1.8. During the initial impulsive (or flash) phase, the radiation in all wavelengths increases, often by multiple orders of magnitude; HXR, microwave, and (when observed) white light emission exhibit a burst-like behavior, varying significantly on short timescales and often with multiple successive bursts, while the UV, EUV, and SXR emission tend to vary more smoothly. The impulsive phase typically lasts up to tens of minutes and is then followed by a decay (or gradual) phase, where the bursty HXR and microwave flux is typically negligible and the UV, EUV, and SXR fluxes decay smoothly over time; the decay phase can last for hours, depending on the power and intensity of the flare. Some flares also exhibit an early, pre-impulsive phase marked by a gradual emission increase, especially in X-rays.

Because radiation is the most readily and immediately observable signature of a flare, it lends itself well as a means of categorizing them. Multiple flare classification schemes exist, depending on the type of radiation observed, but the one most commonly used over the past ~30 years is that of "GOES class," which ranks flare magnitude based on the maximum 1-8 Å (~1.6-12.4 keV) SXR flux measured during the flare by photometers on the *Geostationary Operational Environmental Satellite* (GOES) spacecraft. The GOES class is a logarithmic scale wherein the arbitrary letter designations A, B, C, M, and X represent the exponents of SXR fluxes of $10^{-8}$ through $10^{-4}$ W/m² (as measured at the spacecraft), respectively, with a corresponding number signifying the mantissa (e.g. M3.6 = $3.6 \times 10^{-5}$ W/m²). The SXR continuum observed by GOES is generally dominated by thermal emission processes (§1.4) and, as with the UV/EUV line emission, indicates the presence of hot, thermal plasma.

The HXR and gamma-ray continuum emission above ~20 keV is entirely dominated by electron-ion bremsstrahlung (§2.1.1), a much-studied and well-understood physical process. (The coronal magnetic field is not strong enough to allow production of X-ray synchrotron emission, and the free-bound continuum [§2.1.2] drops off rapidly with energy.) For most flares, this continuum is well-fit by a power-law and can be inverted (cf. Brown 1971) to yield the emitting electron spectrum, itself a power-law. Assuming a thick-target model, where Coulomb collisions dominate the electron energy losses, only ~$10^{-5}$ of the total electron power goes into radiating bremsstrahlung, suggesting that the total energy in energetic electrons (cf. equation [2.6]) is often of the order of $10^{32}$-$10^{33}$ ergs, tens of percent of the total estimated flare energy budget (Lin & Hudson 1971, 1976; Emslie *et al.* 2004, 2005). In the corona, the magnetic field is the only source of sufficient energy, suggesting a direct link between the magnetic field and flare energy release (a relationship hypothesized even before X-ray observations were available [Giovanelli 1946]); indeed, the free (non-potential) magnetic energy in an active region is also typically $10^{32}$-$10^{33}$ ergs (e.g. Yang *et al.* 1983; Emslie *et al.* 2004), suggesting that flares very efficiently convert stored magnetic energy into the kinetic energy of accelerated particles.

The various forms of emission that are enhanced during a flare can thus be interpreted as signatures of energetic particles and/or hot thermal plasma, and observations of such emission therefore serve as measurements of the underlying particle populations. The accelerated electrons will emit bremsstrahlung (cf. §2.1.1), spanning from radio waves to X-rays or gamma rays, as they travel through the ambient plasma; their energy spectrum and the characteristics (e.g. temperature, ionization state, elemental abundances, etc.) of the ambient plasma directly determine the photon spectrum that is subsequently observed. If the magnetic field is sufficiently strong, the accelerated electrons will also emit radio synchrotron radiation, and plasma waves produced in the corona by the accelerated electrons can also result in radio emission. The HXR and gamma-ray bremsstrahlung continuum spectra, as well as radio synchrotron spectra, indicate that flares routinely accelerate electrons to energies of hundreds of keV, sometimes up to tens of MeV. Radio measurements from the upper corona and beyond show that the energetic electrons exist not just within closed loops but also along open field lines that reach into interplanetary space (in some cases, they can reach Earth, where they can be directly detected). The hot, thermally-ionized atoms in the ambient plasma will emit UV, EUV, and X-ray excitation lines (cf. §2.1.3), and free thermal electrons will also emit SXR bremsstrahlung with a characteristic spectral shape; the temperature, density, elemental make-up, and ionization balance of the plasma all directly affect the resulting observed spectra, which indicate that plasma temperatures range from ~1-30 MK for most flares, but can approach ~50 MK for the most powerful flares.

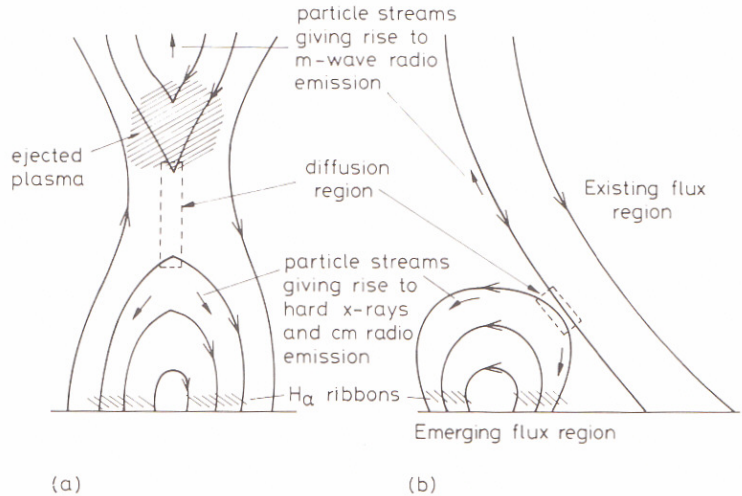

**Figure 1.9** – Cartoons depicting two scenarios in the "standard flare model:" reconnection of [a] open field lines or [b] emergent & existing flux (from Phillips [1992]).

Spectroscopic and imaging observations of flares, from radio waves to gamma rays, can thus serve as probes of the various physical processes occurring within flares and provide a test-bed for theoretical models of those processes. The prevailing theory describing the onset and evolution of flares is the so-called "standard flare model," although this is somewhat of a misnomer as the model is greatly simplified and there are other competing models, as well – the specific details of these models are still the subject of heated debate.

In the standard model, coronal magnetic field lines of opposing polarity are brought into close proximity with each other – this can be caused by, for example, field motion due to convective currents on the solar surface, or when new magnetic flux emerges into a region of existing flux, as depicted in cartoon form in Figure 1.9. The opposing field lines are separated by a sheet of perpendicular current (required by such a configuration). Although the frozen-flux condition would normally forbid a change in the magnetic topology, plasmas do have some resistivity; near the current sheet, the resistivity enables the magnetic field to diffuse through the plasma, allowing the field lines to reconnect (as depicted schematically in Figure 1.10) and altering the magnetic topology. (The rate of diffusion when considering only collisional resistivity is orders of magnitude smaller than required to explain flare energy release [Parker 1963], thus other mechanisms must also contribute to increase the diffusion rate.)

Although the mechanisms are still not well understood, within the reconnection region, the energy released by the magnetic reconfiguration is thought to be transferred to plasma heating, waves, bulk motion (plasma outflows), and particle acceleration. Magnetic tension pulls the newly-reconnected field lines into more potential, lower-energy states, and the subsequent com-pression of the flaring loop can add further kinetic energy to the plasma. Accelerated particles stream away from the reconnection region along the field lines, either out into interplanetary space or down to the footpoints, where they deposit their energy via collisions in the dense chromosphere. The sudden input of energy heats the chromospheric material quickly to temperatures of up to ~10-20 MK, resulting in "chromospheric evaporation" wherein the heated material rises, often explosively, to fill the flaring loop. Super-hot plasma, with temperatures of ~30-50 MK, is also commonly observed; its origins are the subject of this dissertation.

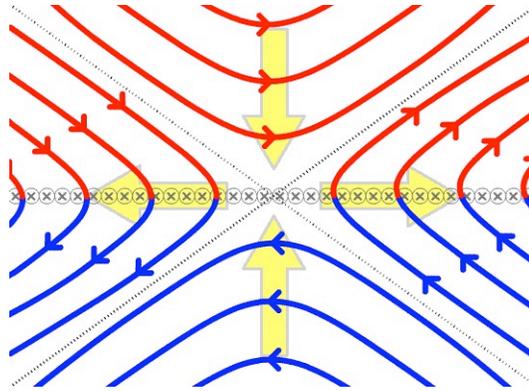

**Figure 1.10** – Schematic of magnetic reconnection in one possible configuration; oppositely-oriented field lines (red and blue) are pushed together (inflowing arrows), e.g. by the mechanisms shown in Figure 1.9. Because of the finite resistivity of the plasma, the oppositely-oriented lines can diffuse through the plasma and reconnect across the separatrices (gray lines), forming field lines of a different configuration (joint red/blue lines); magnetic tension then pulls these newly-connected lines away from the reconnection region (outflowing arrows). (Image credit: Wikipedia)

While highly idealized, this theoretical model explains, at least hypothetically, the origins of the measurable radiation observed over the last ~150 years. The electrons accelerated on open field lines create Langmuir waves in the coronal plasma as they escape to interplanetary space, which interact through plasma processes to generate meter- and decimeter-wave radio emission; relativistic electrons will also emit synchrotron radiation at centimeter wavelengths. Electrons accelerated downwards emit (so-called "non-thermal") bremsstrahlung X-rays and gamma rays as they interact in the chromosphere. Thermally-ionized atoms in the hot (~1-20 MK) plasma will emit UV and EUV excitation lines; at temperatures above ~10 MK, the hot plasma will also emit X-ray lines and (so-called "thermal") bremsstrahlung extending up to ~30-40 keV.

*1.4 X-ray observations of thermal plasma in solar flares*

The thermal plasma in flares is often considered a secondary product, resulting from heating due to collisional energy loss of accelerated particles. Indeed, this connection is suggested by the so-called Neupert effect (Neupert 1968; Dennis & Zarro 1993), where during the impulsive phase, the SXR time profile is often observed to be roughly proportional to the time-integral of the microwave or HXR flux. An example of this is shown in Figure 1.11, in which the time-derivative of the GOES SXR time profile and the RHESSI HXR time profile show good agreement in their peaks and valleys, though not in the relative normalization of those peaks. Deviations from an ideal Neupert effect suggest that energy deposition by energetic particles is not necessarily the only process that can heat plasma – ohmic (Joule) losses from currents, interac-tions with plasma waves, and the reconnection process itself may all contribute to flare heating. Indeed, during the decay phase of large flares, when the HXR emission – and hence the inferred accelerated particle population – is negligible, the plasma temperature nevertheless drops more slowly than would be expected from naïve calculations of radiative and conductive cooling (e.g. Moore *et al.* 1980; Veronig *et al.* 2002a, 2002b), suggesting that there is additional heat input to the plasma during this time, despite the apparent lack of accelerated particles. Additionally, not all flares exhibit a well-defined Neupert effect (e.g. Veronig *et al.* 2002b), and some flares even appear to be almost entirely thermal (e.g. Kobayashi *et al.* 2006). Studying the evolution of the thermal plasma can therefore provide insight into how the plasma is heated and, ultimately, into how energy is released and transported during a flare. Since X-rays can only be emitted by particles with energy ≳1 keV, SXR and HXR observations are a direct probe of both hot (≳10 MK) plasmas and accelerated particles.

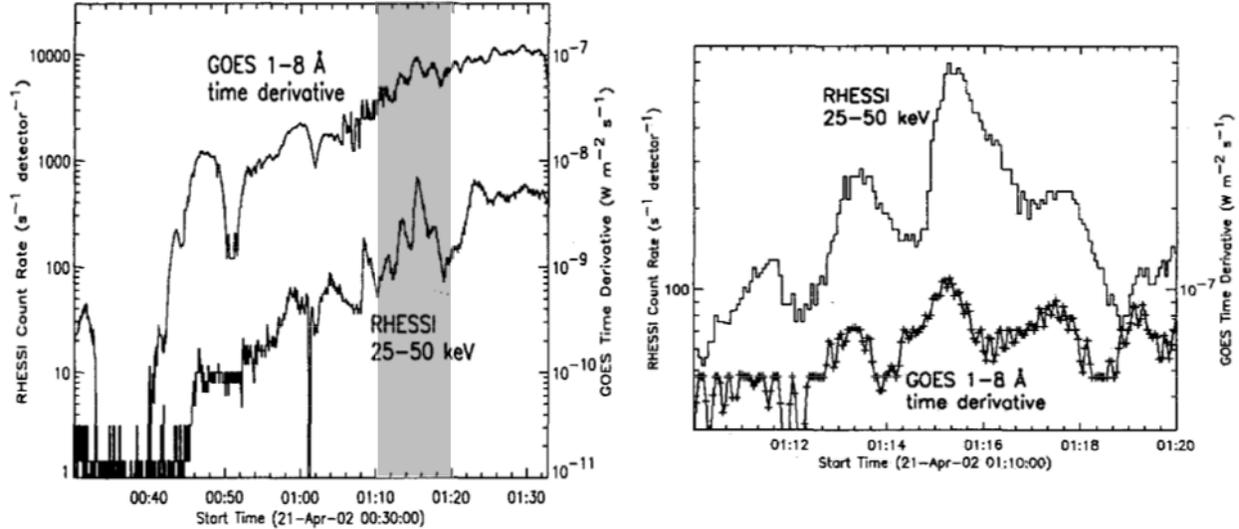

**Figure 1.11** – Lightcurves exemplifying the Neupert effect, wherein the SXR time profile is observed to behave similarly to the time-integral of the HXR time profile. [left] Time-derivative of GOES 1-8 Å SXR time profile, compared to the RHESSI 25-50 keV HXR time profile; there is general agreement in the overall and fine-structure behavior, though no actual correlation of values. [right] Close-up view of the shaded region from left. (Adapted from Dennis *et al.* [2003].)

The first observation of SXRs from solar flares was made by Chubb *et al.* (1957), using a Geiger counter on-board a balloon-launched rocket. Although the detector had no spectral resolution, they were able to deduce a rough energy spectrum by convolving the measured counts in the detector's ~3-9 Å (~1.4-4.1 keV) passband with the energy-dependent X-ray attenuation (§2.2) of the overlying atmosphere as the rocket gained or lost altitude. This, of course, relied on the accuracy of the model atmosphere and an assumption that the incident photon spectrum was static throughout the duration of the rocket flight. Nevertheless, the SXR spectrum derived from this albeit crude method could be interpreted as thermal bremsstrahlung from a ~4-6 MK plasma and gave the first indications that flares could heat plasma well above the ambient coronal values.

This has been well-verified by the routine SXR measurements made by GOES. In addition to the 1-8 Å flux from which the flare magnitude (GOES class) is derived, GOES also measures the 0.5-4 Å (~3-25 keV) flux. With knowledge of the instrument's response to X-rays and the (overly broad) assumption that the emission is from an isothermal plasma (see §2.1), the ratio of these two fluxes yields a temperature for the emitting plasma. GOES measurements have shown that ~10 to ~20 MK thermal plasmas are ubiquitous in flares. The GOES-derived temperature is often quoted as *the* temperature of the flare plasma, but because this relies on only two data points and many assumptions, it is really only a gross approximation. Indeed, observations of simultaneous Hα, UV/EUV, and X-ray emission (e.g. Figure 1.7) indicate that a wide range of temperatures, from ~10,000 K to tens of MK, is present in flares; it is therefore much more realistic to consider that flares contain multiple thermal populations at different temperatures, or even with a broad and continuous distribution of temperatures. Precise, high-resolution broadband spectral observations, such as from RHESSI (below), are therefore required to accurately identify and characterize the multi-temperature hot plasma in flares; in Chapter 4, we present such observations and show that the X-ray-emitting thermal plasma is well-described by two distinct populations: the traditional ~10-20 MK GOES plasma and a separate, much hotter component with temperatures of up to ~45 MK.

X-ray excitation lines from thermally-ionized atoms also support a multi-thermal picture. First detected from flares by Neupert *et al.* (1967) using a crystal spectrometer on-board the OSO III satellite, X-ray lines have been routinely observed with similar spectrometers on numerous subsequent missions. Since the ionization balance of a thermal plasma is closely tied to its temperature, so, too, are the fluxes in various ionic excitation lines; measurements of these lines are thus a sensitive indicator of the plasma temperature. As with the GOES measurements, temperatures inferred from line spectra (in particular, from Fe XXIII through XXV) showed that ~10-20 MK plasmas were a common feature of nearly all solar flares, regardless of the flare intensity, although as with GOES, X-ray lines can provide only a few discrete temperature measurements. Although measurements of line emission from the most highly-ionized solar atoms (Fe XXV and Fe XXVI) hinted that temperatures above ~30 MK may also be present in flares, confirmation of these "super-hot" plasmas from continuum observations was possible only after high-resolution broadband spectrometers (see below) became available.

The first HXR observations of the solar flare continuum were by Peterson & Winckler (1959), who used a balloon-borne Geiger counter to observe at 100-500 keV, but with no energy resolution. Chubb *et al.* (1960) then observed down to ~20 keV with a rocket-borne instrument with a coarse energy resolution of ~10 keV FWHM. They were able to fit a power-law to their data and interpreted it as the first observation of non-thermal bremsstrahlung from a solar flare. Since then, satellite experiments have routinely made HXR flare observations, but with only two exceptions (discussed shortly), all of these instruments used scintillators or other detectors that had a coarse energy resolution, and binned their data even more coarsely to improve statistics, yielding a $\Delta E/E$ of ~25% to ~133% for their observations. The spectra were typically fit as power-laws (Kane *et al.* 1980) and interpreted as non-thermal thick-target bremsstrahlung from accelerated electrons, with the consequence that the inferred electron population (see §2.1.1) often contained a large fraction – tens of percent – of the total flare energy, on the same order as the total energy available from the magnetic field (Lin & Hudson 1971, 1976), and that the number of energetic electrons was of the same order as the total number of electrons available in the flaring coronal loop; it was difficult to explain how flares could so efficiently accelerate so many electrons to high energies, and how the electron population could be replenished sufficiently quickly to explain the duration of the emission. However, the coarse resolution of the observations also allowed acceptable fits to an isothermal bremsstrahlung continuum from plasmas with tempera-tures of ~100 MK to ~1 GK (e.g. Crannell *et al.* 1978; Elcan 1978). Because thermal electrons are in collisional equilibrium and therefore do not (on average) experience collisional energy losses, far fewer electrons with far less total energy are required to produce the same HXR continuum as from non-thermal electrons (Brown & Smith 1980); this very-high-temperature interpretation therefore eliminated the need to explain efficient particle acceleration and replenishment, but it was difficult to explain how the plasmas could reach and sustain such high temperatures.

In 1980, Lin *et al.* (1981) obtained the first high-resolution (~2 keV FWHM) HXR observations from flares using cryogenically-cooled planar germanium solid-state detectors (GeDs; see §2.3.2), to precisely measure the HXR continuum and determine whether it was thermal or non-thermal, at what energies, and with what parameters. Their measurements revealed that the continuum above ~33 keV was indeed a power-law with a sharp break, consistent with non-thermal bremsstrahlung rather than with emission from ≳100 MK thermal plasmas. Below ~33 keV, however, the precise measurements revealed an isothermal continuum with temperatures of up to ~34 MK (Figure 1.12); this component decreased exponentially with energy, with an *e*-folding of ~2 keV. Previous instruments, which had coarse energy resolutions of ≳10 keV FWHM in the ~20-40 keV range over which this hot component dominates the emission, were entirely incapa-ble of resolving the steeply-falling spectrum to precisely measure the isothermal temperature. These observations were the first positive measurements of plasma temperatures well above the usual ~10-20 MK that had been previously measured through continuum observations; the ≳30 MK temperatures were quickly dubbed "super-hot" (e.g. Hudson *et al*. 1985; Lin *et al*. 1985) for this reason.

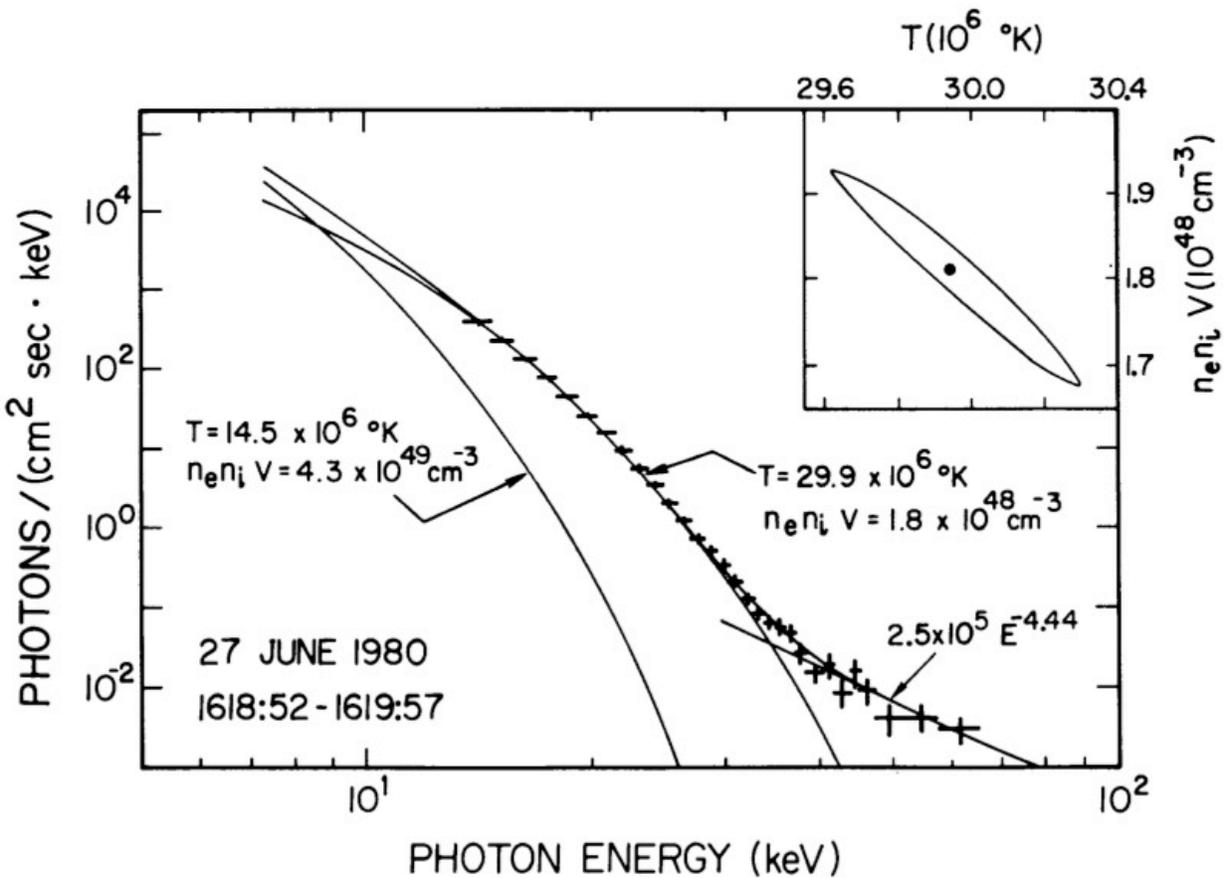

**Figure 1.12** – The first high-resolution X-ray spectrum obtained with cryogenically-cooled germanium detectors (hashes); the model fit (solid lines) indicates the first observation of super-hot (T > 30 MK) thermal plasma (from Lin *et al.* [1981]).

*Hinotori*, a Japanese spacecraft with a suite of solar-observing instruments, launched a year later. Its Soft X-ray Spectroscopy (SXS) achieved ≲1 keV resolution, although it was only able to observe the solar spectrum from ~2-12 keV (sometimes up to ~17 keV). Nevertheless, model fits to the spectrum were consistent with super-hot temperatures for some of the larger flares. Supporting observations with the Bragg Crystal Spectrometer (BCS) of ratios of Fe XXVI to Fe XXV line intensities were also consistent with >30 MK temperatures (Tanaka 1987). *Yohkoh*, the next-generation Japanese spacecraft, also flew a BCS and obtained similar line ratios to support super-hot temperatures (Pike *et al*. 1996). The Hard X-ray Telescope (HXT) showed that the high-energy emission came primarily from flare loop footpoints, while the Soft X-ray Telescope (SXT) showed that the ~10-20 MK plasma indeed filled the flare loop (Masuda 2002); the time profiles from BCS and SXT often showed a Neupert effect (e.g. McTiernan *et al*. 1999). Combined with *Hinotori* and *Yohkoh* BCS measurements of Doppler shifts in the excitation line profiles (e.g. Antonucci 1989), these measurements provided good evidence to support the picture of chromospheric evaporation as the source of ~10-20 MK plasma. However, SXT could not distinguish super-hot plasma from the ~10-20 MK plasma in the flare loops. In some flares, HXT also observed a source at or above the SXR loops seen by SXT (e.g. Masuda 2002), but with HXT's coarse energy resolution, it could not be definitively determined whether these sources were non-thermal or super-hot.

Finally, in February of 2002, the *Reuven Ramaty High Energy Solar Spectroscopic Imager* (RHESSI) was launched, and it remains in operation after over 8 years (see Chapter 3). Compared to previous missions, RHESSI offers significantly improved capabilities, observing down to ~3 keV and up to ~17 MeV with better spectral resolution (~1 keV FWHM), and imaging over the entire energy range with angular resolution down to ~2 arcsec below ~100 keV and ~35 arcsec for gamma rays, along with moveable attenuators to enable these precise measurements over a dynamic range of ~$10^7$ in flare intensity, allowing a single instrument to observe the dynamics from microflares to the largest X-class events. With its powerful observing capabilities, RHESSI provides the richest data set for studying both thermal flare plasma and non-thermal accelerated particles, and has finally begun to yield clues as to the origins of super-hot plasmas. With its high spectral resolution, RHESSI is ideally suited for observations of the steeply-falling continuum from super-hot plasmas, and has shown (see Chapter 4) that super-hot plasma can exist from the very beginning of the flare, during a period when no HXR footpoint emission is visible and hence when chromospheric evaporation is unlikely to occur. Spectra and imaging suggest that, while the ~10-20 MK plasma likely does result from chromospheric evaporation, the super-hot plasma is most likely heated directly at the looptop. Additionally (see Chapter 5), the existence of super-hot plasma appears dependent on flare magnitude, with super-hot temperatures measured in most X-class flares, but in very few M-class flares.

# Chapter 2: X-Rays

## 2.1 X-Ray Production

We cannot observe *in situ* the physical processes in flares that so efficiently accelerate particles and heat thermal plasma to tens of MK. However, energetic electrons interacting with ions produce electromagnetic radiation (Figure 2.1), including in the X-ray band, and the nature of that radiation yields information about its production. Thus, observations of X-ray emission from flares can provide insight into the underlying physical phenomena.

### 2.1.1 Free-free (bremsstrahlung) emission

When a free electron experiences a close encounter with an ion, it is de-accelerated, losing kinetic energy and emitting a photon (equivalent to the energy lost) in the process; if the electron remains free after scattering (hence a "free-free" interaction), the resulting radiation is termed *bremsstrahlung*. In flares, this is the dominant form of emission at X-ray energies. Excellent reviews of this process can be found in Jackson (1998), Tandberg-Hanssen & Emslie (1988), or Haug & Nakel (2004).

For long-range interactions, the bremsstrahlung process can be treated semi-classically; the incident electron loses only a small fraction of its energy and the resultant photon is typically in the radio or visible wavelengths. X-ray photons are produced by short-range collisions, however, where quantum effects become significant and the semi-classical approximation is insufficient to fully describe the electron-ion bremsstrahlung cross-section. There is no general closed-form solution to the Dirac wave equation for an electron in a Coulomb field, so approximations must still be made in order to derive analytic solutions for the cross-section in various limits; Koch & Motz (1959) compiled a comprehensive tabulation and discussion of such formulae.

The simplification most commonly employed is the Born approximation, whereby the inci-dent and scattered electron wave functions are taken as plane waves and perturbed to first order, yielding the well-known Bethe-Heitler formula (Bethe & Heitler 1934). In the non-relativistic limit, often used for X-rays below ~100 keV, this reduces to:

$$\frac{d\sigma_B}{dk} = \frac{16}{3}\alpha Z^2 r_0^2 \frac{1}{kp_i}\ln\left(\frac{p_i + p_f}{p_i + p_f}\right) \quad (\mathrm{cm}^2 / m_e c^2) \qquad (2.1)$$

where $\alpha \approx 1/137$ is the fine-structure constant, $Z$ is the ion atomic number, $r_0$ is the classical electron radius, $k$ is the photon energy in units of $m_e c^2$, and $p_i$ ($p_f$) is the initial (final) momentum of the electron in units of $m_e c$; or, in units more convenient for our purposes (cf. Brown 1971):

$$\frac{d\sigma_B}{d\varepsilon} = \frac{7.9 \times 10^{-25} Z^2}{\varepsilon E}\ln\left(\frac{1 + \sqrt{1 - \varepsilon / E}}{1 - \sqrt{1 - \varepsilon / E}}\right) \quad (\mathrm{cm}^2 / \mathrm{keV}) \qquad (2.2)$$

where $E$ is the initial electron energy and $\varepsilon$ is the photon energy, both in keV. The accuracy of this approximation decreases rapidly with increasing $E$ and yields significant (>10%) relative error even for mildly relativistic electron energies of a few deka-keV (Haug 1997), thus it is preferable to use the full cross-section (Koch & Motz 1959, eqn. 3BN) for most studies of hard X-ray emission.

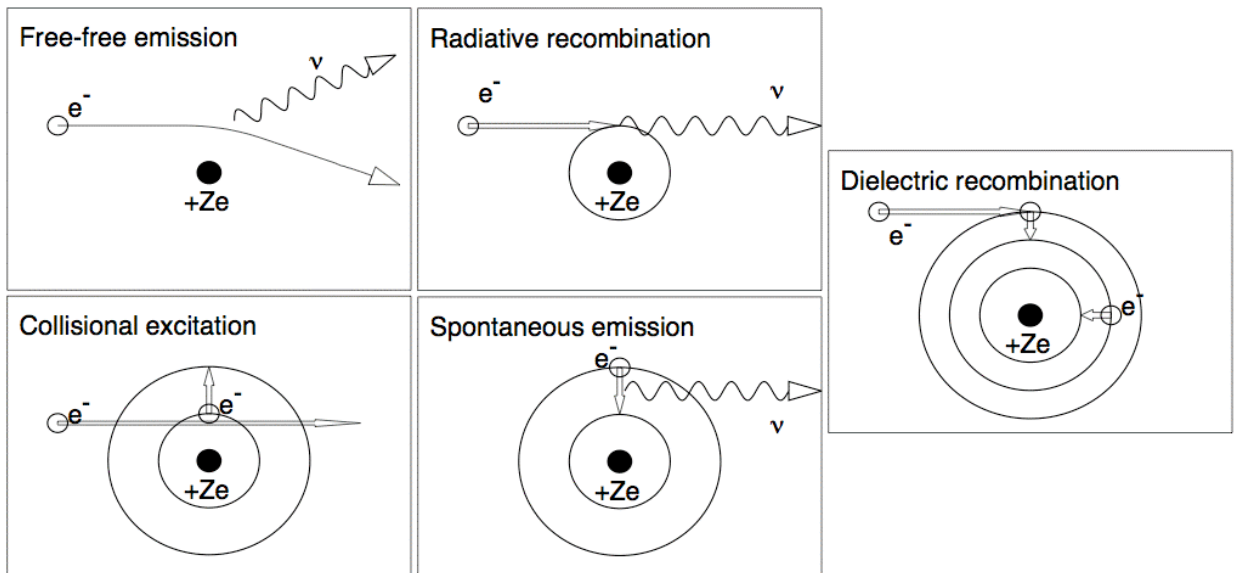

**Figure 2.1** – Schematic representations of X-ray emission mechanisms (adapted from Aschwanden [2005]).

Because the Born approximation neglects the effect of the ion's Coulomb field on the electron wave function, it becomes increasingly inaccurate as the photon energy approaches its maximum of the initial electron energy; as $p_i$ - $p_f \rightarrow 0$ and thus $\varepsilon \rightarrow E$, equations (2.1) and (2.2) go to zero, while the true bremsstrahlung cross-section remains finite. For steeply-falling photon spectra, as are generally observed in solar flares, most of the photons at energy $\varepsilon$ are produced by electrons of energy not much greater than $\varepsilon$ (Holt & Cline 1968), so this effect can be important in interpreting observed spectra. Elwert (1939) derived a correction factor to approximately account for this effect, and in practice, the Bethe-Heitler cross-section (whether full or in the non-relativistic limit) is generally multiplied by this factor to improve accuracy over X-ray and gamma ray energies (Pratt & Tseng 1975).

Given an electron spectrum, the bremsstrahlung cross-section allows calculation of the emitted photon spectrum. Conversely, by inverting the cross-section, one can deduce the instantaneously-emitting electron spectrum from a photon spectrum. For a power-law photon spectrum, equation (2.2) can be inverted analytically (Brown 1971), but the full cross-section (with or without the Elwert factor) cannot be; it can, however, be inverted numerically, allowing an empirical determination of the parent electron spectrum, with quantifiable uncertainties, directly from the observed flare photon spectrum (Johns & Lin 1992). Because such inversion can utilize the full cross-section, the results are not limited to specific energy domains nor are they dependent upon any *a priori* assumptions about the emitting electron population. As with any inverse method, however, small uncertainties in the photon spectrum can be greatly magnified by the inversion, thus the applicability is often limited by statistics.

Because of this, and because the photon spectrum (especially below ~10 keV) contains significant non-bremsstrahlung contributions (discussed below), most flare spectral analyses – including those presented in this thesis – utilize forward modeling, whereby a model electron spectrum is assumed and its parameters varied until the calculated photon spectrum best fits the observations. Forward modeling is, of course, inherently dependent upon the initial assumptions about the electron spectrum; as such, the models generally fall into two regimes (Figure 2.2):

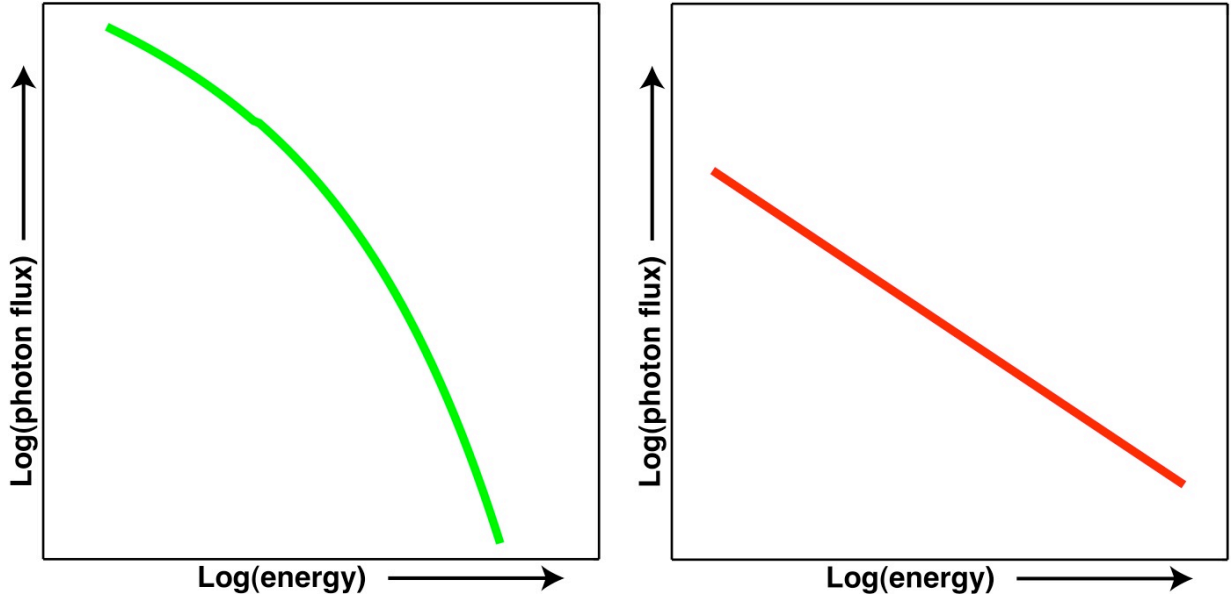

**Figure 2.2** – Schematic representation of bremsstrahlung spectra from a thermal (left) or non-thermal power-law (right) electron population.

*1) Thermal* – here, the electron population is assumed to be Maxwellian (i.e. in thermal equilibrium) with temperature $T_e$. If $T_e$ is sufficiently high (generally above a few MK, whereby $k_B T_e$ is generally above ~0.1 keV), the electrons will have sufficient energy to produce X-ray bremsstrahlung as they interact with the ions (which are assumed stationary, as even if $T_i = T_e$, the average ion thermal velocity is smaller than that of the electrons by the square-root of the electron-ion mass ratio, and thus can be effectively taken to be zero). Since even at extremely high temperatures of ~100 MK, the average electron energy $k_B T_e \lesssim 10$ keV, the non-relativistic approximation is often used; then, combining an isothermal Maxwellian distribution with equation (2.2) and integrating over all electron energies yields a photon spectrum of the form:

$$I(\varepsilon) \propto Z^2 Q g(\varepsilon, T_e) \frac{\exp(-\varepsilon/k_B T_e)}{\varepsilon \sqrt{T_e}} \quad (\text{photons}/\text{s}/\text{keV}) \tag{2.3}$$

The emission measure (luminosity per unit emitting volume) $Q \equiv \int_V n_e^2 dV$, where $n_e$ is the electron density within the emitting volume $V$; for spatially-integrated measurements, the density is often assumed uniform, whereby $Q$ reduces simply to $n_e^2 V$. With the simple cross-section of equation (2.2), the Gaunt factor $g$ is an analytic, slowly-varying function of order unity; when the full cross-section and/or Elwert correction are used, the Gaunt factor is calculated numerically (e.g. Itoh *et al.* 2000).

Thus, by fitting equation (2.3) to an observed spectrum and varying $T_e$ and $Q$, the best-fit source electron temperature and emission measure (and hence density, if $V$ is known or assumed) can be determined, subject to the model assumptions. Alternatively, rather than assuming an isothermal distribution, one may define the differential emission measure $Q_{DEM}(T_e) \equiv \frac{d}{dT_e} \int_V n_e^2 dV$ and turn equation (2.3) into an integral over temperatures; by assuming a parametric form for $Q_{DEM}(T_e)$ (i.e. for $dQ/dT_e$), its parameters can be determined via fitting of $I(\varepsilon)$. If other observations are present to constrain $Q$ (or $n_e$), then $Z^2$ (or, for a plasma with multiple elemental species, the abundance-averaged $\overline{Z^2}$) can be determined; in practice, for the measured elemental abundances in the solar corona, this is usually approximated as a constant $\overline{Z^2} \approx 1.4$.

*2) Non-thermal* – here, the electron population is anything non-Maxwellian, but is most often assumed to be a power-law (or broken power-law). To avoid infinities, a low-energy cutoff or rollover is imposed, and the spectral index (the negative power-law exponent) is taken to be no less than 1; a high-energy cutoff is sometimes imposed, as well. Using equation (2.2) for the cross-section, the resulting photon spectrum above the low-energy cutoff is also a power-law; below the cutoff, the spectral index is not constant, but decreases gradually with decreasing energy and can thus often be approximated as a power-law, as well. By fitting a model photon spectrum to the observations, the electron spectral index and cutoff value(s) can be determined.

In a non-thermal population, the bremsstrahlung-producing electrons have energies much greater than those in the target/interaction region, unlike in a Maxwellian where the emitting and target electrons are the same population; Coulomb collisions are usually assumed to be the dominant form of energy loss for non-thermal electrons (e.g. in the thick-target approximation, below). Consequently, depending on the column density of the target traversed compared to the initial electron energy, the instantaneous electron spectrum may differ substantially from the originally-injected spectrum. This is described by the Coulomb energy loss equation, which, in the non-relativistic approximation (cf. Brown 1971, 1972; Lin 1974), is:

$$\frac{dE}{dt} = -\frac{Kn_e v}{E} = -\frac{\kappa n_e}{\sqrt{E}} \tag{2.4}$$

and quantifies the energy lost over time via collisions of an electron of energy $E$ (in keV) and corresponding velocity $v$ (in cm/s) traveling through a medium of number density $n_e$; the constants $K \equiv 2\pi e^4 \Lambda \approx 2.6 \times 10^{-18}$ cm$^2$ keV$^2$ for a fully-ionized hydrogen plasma (the Coulomb logarithm $\Lambda \approx 20$ for the solar corona) and $\kappa = K\sqrt{10^{11} \cdot 2e/cm_e} \approx 4.9 \times 10^{-9}$ cm$^3$ keV$^{3/2}$ s$^{-1}$. If we take $E = E_0 e^{-t/\tau_c}$ then we immediately obtain the energy-dependent collisional loss timescale $\tau_c(E) = E^{3/2}/\kappa n_e$. If we consider the column density through which the electron travels over time, $N = n_e vt$ (in cm$^{-2}$), then we may also write (cf. Brown 1972):

$$\frac{dE}{dt} = -\frac{Kn_e v}{E} = -\frac{K \, dN/dt}{E} \quad \Rightarrow \quad \frac{dE}{dN} = -\frac{K}{E} \quad \Rightarrow \quad E^2 = E_0^2 - 2KN \tag{2.5}$$

whence $N_c \equiv E_c^2/2K$ is the maximum column density through which *all* of the electrons in a power-law spectrum with low cutoff energy $E_c$ can traverse without being stopped.

This gives rise to the two limiting cases most often considered: *thin-target*, where $N \ll N_c$ (e.g. for low-density targets such as in the corona, or when the observational timescales are much shorter than $\tau_c[E_c]$) such that the energy loss due to Coulomb collisions is negligible, whence the emitting electron spectrum can be assumed static over the observation time and the instantaneous spectrum is not significantly changed from the injected one; and *thick-target*, where $N \gg N_c$ (e.g. for high-density targets such as the chromosphere, or when the observational timescales are long compared to $\tau_c[E_c]$), such that the electrons are assumed to lose *all* of their energy to collisions. Then, in the thin-target case, the electron spectrum inferred from *any* observed photon spectrum is exactly the instantaneously-emitting spectrum (averaged over the observation time), with no assumptions necessary; if one does assume complete collisional energy loss, then the injected electron spectrum may also be inferred, subject to the thick-target assumption. For a given injected power-law electron spectrum, both the thin- and thick-target photon spectra will be power-laws, with their spectral indices differing by 2 (Brown 1971). The analytic relationship between a power-law electron spectrum and the resulting power-law photon spectrum under the thick-target interpretation with the Bethe-Heitler cross-section (equation [2.2]) allows the total energy deposited via Coulomb collisions by the non-thermal electrons to thus be directly estimated from the photon spectrum, as the power $P$ in non-thermal electrons above a low cutoff energy $E_0$ is (cf. Lin 1974; Lin *et al.* 2001):

$$P(> E_0) = 9.5 \times 10^{24} \gamma^2 (\gamma - 1) \beta(\gamma - \tfrac{1}{2}, \tfrac{3}{2}) A E_0^{-(\gamma - 1)} \text{ erg/s} \tag{2.6}$$

where $A$ and $\gamma$ are the normalization and spectral index of the *photon* power-law, and $\beta(m,n)$ is the beta function. By observing the evolution of the non-thermal photon spectrum over time, one can then determine the total energy in non-thermal electrons.

Electron-electron bremsstrahlung also occurs in flares, but the interaction has no dipole moment and is described only by the quadrupole moment; below ~1 MeV, it is dominated by the dipole moment interaction of electron-ion bremsstrahlung. Thus, while potentially significant for relativistic electrons, electron-electron bremsstrahlung can be ignored for electron energies below ~200 keV (Haug 1975, Kontar *et al.* 2007).

### 2.1.2 Free-bound (radiative recombination) emission

Rather than scattering in a bremsstrahlung process, an energetic electron may instead be captured into a bound state of the target ion (hence a "free-bound" interaction). Unlike in bremsstrahlung, where a given electron energy $E$ can yield a photon of any energy $\varepsilon \leq E$, a photon emitted from radiative recombination has energy equal to the full electron kinetic energy plus the binding energy of the bound state, thus $\varepsilon > E$; moreover, since the ion bound states are discretized, $\varepsilon$ can take only discrete values above $E$. With a continuum of electron energies, such as in a Maxwellian distribution, the emitted photons form the free-bound continuum. This continuum is not entirely smooth, but rather includes sharp edges corresponding to the edges in the electron capture cross-section of the target ions; the number and energies of the edges depend on the elemental abundances and ionization balance of the target/interaction region.

The capture cross-section drops off rapidly with energy (edges notwithstanding), thus in flares, the free-bound continuum is primarily important for thermal populations and is generally ignored for non-thermal models. While it is important to consider the free-bound contribution to the continuum when interpreting thermal spectra, free-free emission dominates the X-ray continuum above ~10 keV and ~10 MK; the fractional contribution of the free-bound component decreases with increasing photon energy and with increasing plasma temperature (White *et al.* 2005).

### 2.1.3 Bound-bound (excitation line) emission

If a bound electron is energized into an excited state – through, for example, a collision with a free electron or absorption of an ambient photon – it will emit a photon as it spontaneously decays back to the ground state. Photons emitted in this manner are at discrete energies, equal to the difference in binding energies between the two states, and form narrow excitation line features in the spectra (e.g. Figure 2.3). Dielectronic (3-body) recombination – whereby a free elec-tron is captured into a high excited state of an ion and excites an already-bound electron, forming a doubly-excited ion that then decays – also contributes significantly to the line spectra (Burgess 1964); although technically a free-bound interaction, the resulting emission is nevertheless discrete.

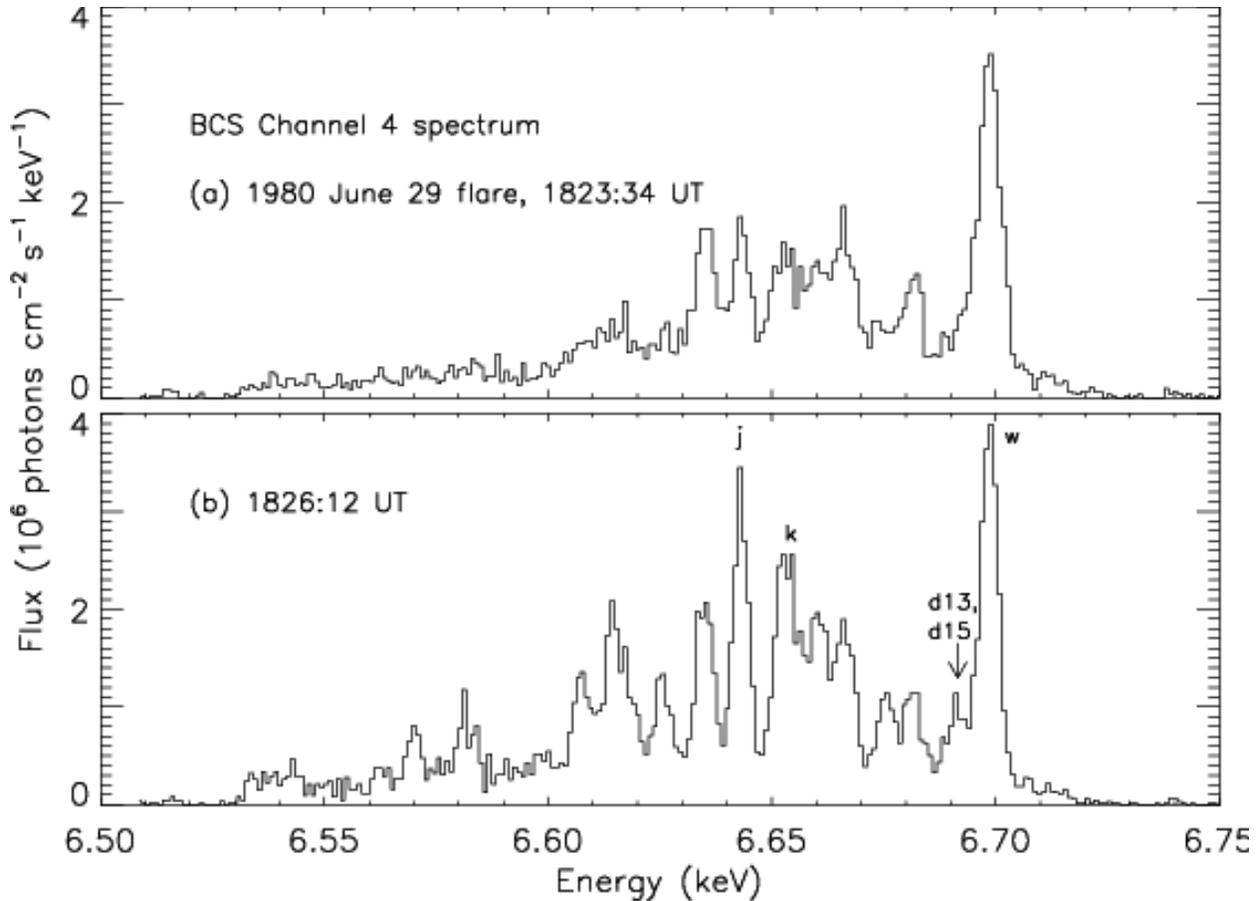

**Figure 2.3** – X-ray spectrum from the BCS instrument on SMM for a flare on 1980 Jun 29, showing Fe XXV excitation lines and Fe XXIV dielectronic recombination lines, among others; ratios of these lines are a diagnostic of the plasma temperature (from Phillips [2008]).

For solar conditions, various ionization states of Si, Ca, S, Fe, and Ni produce lines in the X-ray range; the latter two, in high (hydrogen- and helium-like) ionization states, can contribute significantly to the ~6-8 keV flare spectrum (Phillips 2004) and are of particular importance for studies of super-hot thermal plasmas (see Chapter 4). Because the solar atmosphere – the corona and chromosphere in particular – is optically thin to X-rays (Ohki 1969), both the ionization and excitation processes are collisionally-dominated. In thermal (i.e. collisional) equilibrium, the ionization balance is determined by temperature; thus, the ratios of lines from different ionization states of the same atom are a direct measure of the electron temperature. At a measured temperature, the absolute line fluxes then provide either the emission measure or elemental abundances, if the other quantity is known or assumed.

## 2.2 X-ray interactions in matter

Regardless of the means of production, an X-ray photon must interact with a detector in order to be measured. The photon may also (or instead) interact with various overlying material before it reaches the detector. Knowledge of these interactions is important in ensuring a good understanding of the instrument response (see §3.2.3) and, in turn, of the observations.

Below ~1 MeV, two processes dominate X-ray interaction with matter: photoelectric absorption and Compton scattering. The cross-sections for these interactions depend on the photon energy and the target material, or more specifically, on its composition and density; as exemplified in Figure 2.4, the cross-sections are usually represented as a mass attenuation coefficient $\mu/\rho$, whereby the photon flux $I$ after interaction with a material of mass density $\rho$ and length $\ell$ is given by $I = I_0 \exp\left[-(\mu/\rho)\rho\ell\right]$ (thus, the flux interacting within the material is $I_0 - I$). Photoelectric absorption dominates the cross-section at low energies, e.g. below ~55 keV for Al, a material commonly used as a passive attenuator (see §3.2.1), and below ~150 keV for Ge, the detector material discussed below; in general, photoelectric absorption dominates to higher energies with higher-$Z$ materials.

The photoelectric absorption process is essentially a time-reversed free-bound interaction: an incident photon is absorbed by an atom, ejecting an electron from a bound state; the ejected electron has kinetic energy equal to the photon energy minus the binding energy of the bound state. As with the free-bound interaction cross-section, the edges in the photoelectric absorption cross-section arise from the discretization of the atomic bound state energies, whereby a photon with energy just above an edge has an increased probability of interaction since it can eject an electron from one more shell than could a photon with energy below the edge. For X-ray energies, the electron is typically ejected from an inner shell, e.g. the K ($n = 1$) or L ($n = 2$) shells; the va-cancy left by the ejected electron is quickly filled by an electron from a higher-energy bound state or by a free (but generally low-velocity) electron, resulting either in the emission of a characteristic excitation line X-ray photon (as in a bound-bound interaction) or the ejection of a low-energy bound electron, termed an Auger electron. In macroscopic materials, the secondary X-ray photon is usually subsequently photoelectrically absorbed by nearby atoms, kicking out its own photoelectron; as such, by measuring all ejected electrons, the entire photon energy can be recorded, hence most detectors rely on photoelectric absorption as their primary interaction mechanism. In some cases, the secondary photon can escape the detector, which will contribute to the instrument response (see §3.2.3).

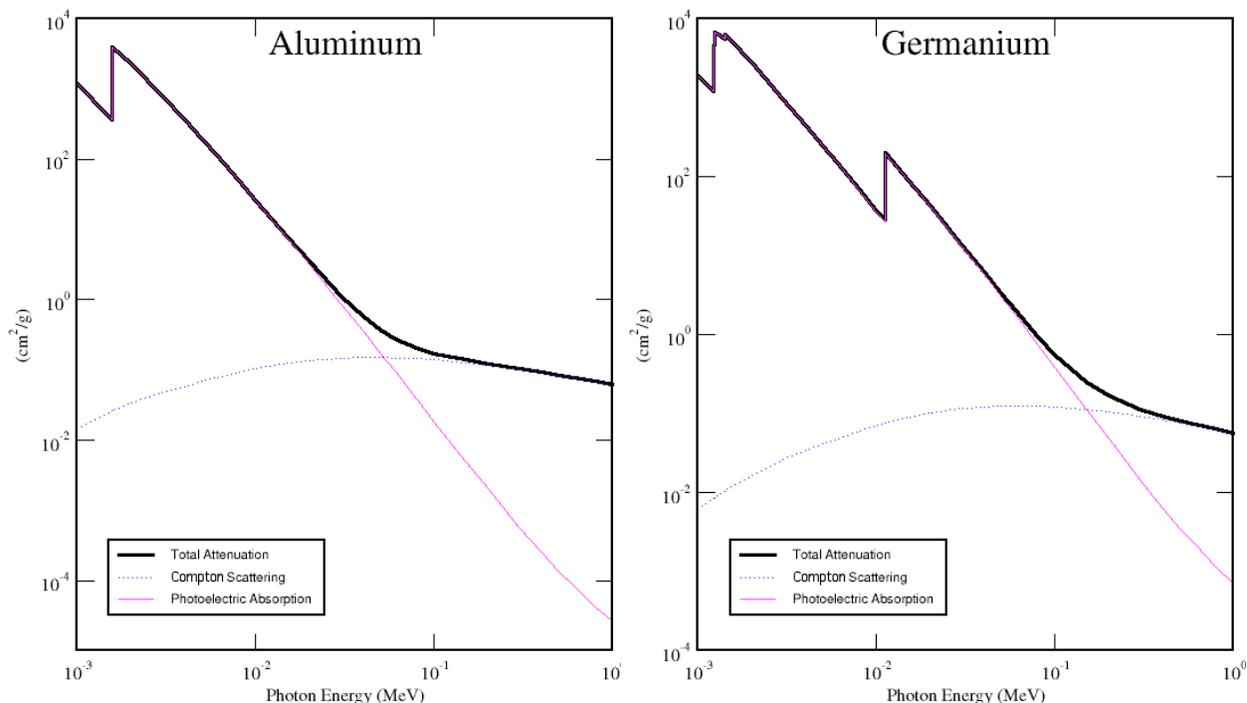

**Figure 2.4** – Mass attenuation coefficients, including the contributions from photoelectric absorption and Compton scattering, for Al and Ge. (Image credit: NIST XCOM)

Compton scattering dominates the interaction cross-section at higher energies. In this process, an incident photon elastically scatters off of an electron (whether free or bound), transferring some of its momentum and energy to the electron. The amount of energy lost is a function of the scattering angle; in a forward scatter, the photon loses relatively little energy, while a large-angle or backward scatter (e.g. ~90° to 180°) results in significant energy loss. If a photon scatters within overlying material before being detected, and/or if it scatters within the detector and then escapes, only some of the incident photon energy will be recorded; this must be accounted for within the instrument response, although for X-rays below ~100 keV and a Ge detector with overlying Al material, Compton scattering within the spacecraft is only of minor importance and only for photons with energies above ~55 keV which can forward-scatter within the Al material (losing a small fraction of energy there) and still reach the detector.

Compton scattering can also occur *outside* the spacecraft, thus the photon spectrum incident upon the spacecraft may differ from the spectrum originally emitted by the source. For solar flares, X-rays can suffer backward Compton scatter off of the dense photosphere, which, unlike the corona, is *not* optically-thin to such radiation. Being composed primarily of hydrogen and helium, the photosphere's average $Z$ is ~1.2 and Compton scattering dominates the interaction cross-section above ~10 keV; the backscattered flux can be significant from ~10-100 keV (above ~100 keV, the incident photon has penetrated too deeply for the scattered photon to escape), and the spacecraft-incident photon spectrum is therefore the sum of the originally-emitted spectrum and the Compton-backscattered component, called "albedo" (e.g. Bai & Ramaty 1978). To recover the original spectrum, it is therefore necessary to compensate for the albedo; although not technically part of the instrument, this can be included within the instrument response (see §3.2.3).

*2.3 X-ray detection*

The essence of a detector is to turn an interaction event into a measurement; Knoll (2000) offers a comprehensive reference of detector technology and implementation. The first observations of X-rays from solar flares were made with Geiger-Müller counters, which measured photon fluxes but not their energies. Other simple detectors, called ionization chambers, were routinely used for basic flare observations (on board the GOES satellites) even until very recently; they measured the integrated energy deposited by the interacting photons but not the actual photon fluxes. For either instrument, the detected photon energies were known only to within the upper and lower limits of the detectors' energy ranges, determined by the column-density of the detector and the preceding absorbing material, respectively.

Proportional counters improve observations by enabling measurement of the photon energy. In a gas proportional counter, a voltage is applied across the detector and, when an incident pho-ton ionizes the fill gas, the primary electrons and ions further ionize the gas as they migrate towards the electrodes, multiplying the original (small) charge which is subsequently measured as a charge pulse; in a scintillator (either inorganic or gas), the incident photon induces the emission of fluorescence photons, which are subsequently detected by a photomultiplier tube and also measured as charge pulses. In both cases, the number of secondaries (electron-ion pairs or fluorescence photons) – and hence the measured charge pulse – is directly proportional to the energy of the incident photon. However, for X-ray observations, the choice of detector material and configuration is important: inorganic scintillators offer a good response up to gamma ray energies but have fairly coarse energy resolution (e.g. Figure 2.5) because of the inefficiencies associated with creating and measuring the fluorescence photons, while gas proportional counters (including gas scintillators) offer good (~keV) energy resolution and low-energy response but are inefficient at high energies due to their relatively low column density. At the other extreme, Bragg crystal spectrometers offer very high (~eV) energy resolution but only over a very narrow (~keV) energy range, and thus are primarily suited for detailed observations of excitation lines – including their precise shapes – rather than of continuum emission.

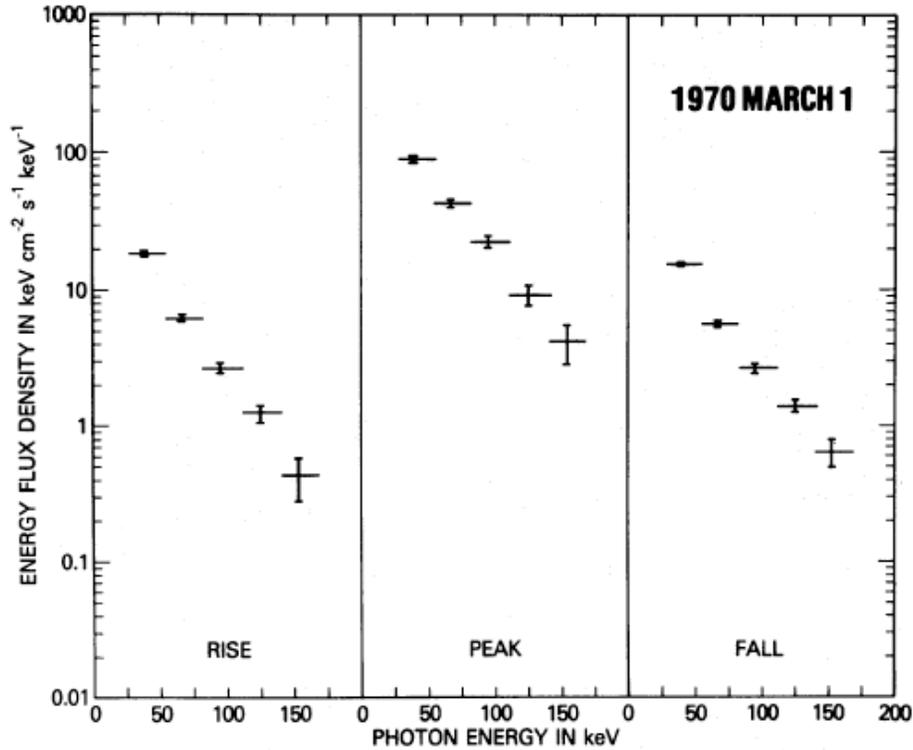

**Figure 2.5** – X-ray spectra from the HXR spectrometer on OSO-5, a CsI(Na) scintillator; the coarse energy resolution ($\Delta E/E \approx 60\%$ at ~100 keV) made it difficult to distinguish between thermal and non-thermal spectra (from Crannell *et al.* [1978]).

Solid-state detectors use semiconductors as the interaction medium, instead of gas or scintillation material. The detector material and geometry can be chosen to tailor the energy response from SXRs to gamma rays and to provide an energy resolution of a few keV or better across the entire range of detectable energies. Unlike in scintillators, where photon measurement requires the intermediate steps of creation and detection of fluorescence photons, semiconductor detectors instead convert incident photons into charges that can be directly measured to determine the photon energy. The charge carriers are produced proportionately per ~3 eV of deposited energy, compared to the ~100 eV or more required per detected fluorescence photon in a scintillator; the counting statistics are further improved by a multiplier known as the Fano factor, typically ~0.1, because the charge carriers are not produced independently of one another. Semiconductor detectors thus offer a significantly better intrinsic energy resolution than inorganic scintillators; at low energies, the resolution can be dominated by electronic noise, depending strongly on the geometry (see below), but is typically a few keV or better across all energies for the configurations often used for high-resolution broadband flare observations.

### 2.3.1 Semiconductor detectors – general properties

In a semiconductor, the atoms form a crystalline lattice and share electrons in covalent bonds. In their ground state, the electrons completely fill the valence band, but with some energy input above a minimum value (called the band gap, which is typically ~1-2 eV; for Si and Ge, two common detector materials, it is ~1.1 and ~0.67 eV, respectively), they can be excited into the conduction band where they are effectively free charge carriers; the vacancies now left in the valence band – known as holes – behave as quasi-particles and are also mobile, acting as positive charge carriers. If a voltage is applied across the semiconductor, the electrons and holes will migrate to the anode and cathode, respectively, where they can be measured as a charge pulse (usually through a pre-amp).

When an X-ray photon is photoelectrically absorbed in the detector, the resulting photoelectron travels through the semiconductor bulk, exciting electrons and holes into the conduction band. As with all proportional counters, the number of charge carriers is proportional to the energy of the photoelectron; the resulting current intensity is thus a direct measurement of the incident photon energy. The proportionality factor is larger for a smaller band gap as less energy is required to excite the electron-hole pairs, resulting in more charge carriers, increased counting statistics, and therefore better energy resolution. (However, a smaller band gap also enhances thermal excitation of charge carriers into the conduction band, which generates a leakage current and necessitates cooling of the detector to reduce the noise from this process.) For Si, one electron-hole pair is created for every ~3.7 eV of energy deposited by the incident photon; for Ge, with its smaller band gap, only ~3 eV is required per pair. In contrast, on the order of ~100 eV or more is required to produce each fluorescence photon in an inorganic scintillator due to the various inefficiencies in the multiple steps required in detection (the detector fluorescence yield, collection of the fluorescence photons, and the quantum efficiency of the photomultiplier tube). Thus, a fully-absorbed ~10 keV photon will release only ~100 fluorescence photons in an inorganic scintillator but ~3000 electron-hole pairs in a semiconductor detector, resulting in significantly-improved counting statistics. The counting uncertainty is further improved beyond the $\sqrt{n}$-behavior suggested by naïve Poisson statistics because the multiple electron-hole pair creations are not independent events; the empirically-determined ratio of the actual to the expected variance, known as the "Fano factor," is ~0.1 for Si and Ge.

We can therefore quantify the intrinsic FWHM energy resolution due to counting statistics, measured at photon energy $E$, as $W_i = (2.35)\sqrt{F\varepsilon E}$ for a detector material with Fano factor $F$ and electron-hole pair-liberation energy $\varepsilon$. However, the total energy resolution will also include contributions from the electronics noise $W_e$, as well as other noise sources $W_x$ within the detector, such as thermal leakage current or charge trapping in a damaged detector; these are added in quadrature, and thus the total FWHM is $W_t = \sqrt{W_i^2 + W_e^2 + W_x^2}$ (cf. Knoll 2000). The electronics noise $W_e$ depends on the detector capacitance and thus is determined largely by the detector geometry, though other factors (e.g. the time constants of the charge pulse shaper amplifiers downstream from the pre-amp) will also contribute. Since $W_i$ decreases with energy, below a threshold energy where $W_i < W_e$, the FWHM resolution is electronics-limited and relatively constant; a lower detector capacitance will reduce $W_e$, which can be $\lesssim 100$ eV for low-capacitance pixelated Si detectors. While in principle $W_x$ can be made nearly negligible, e.g. by cooling the detector to reduce thermal noise and by optimizing the charge-pulse integration time to ensure full charge collection, in practice this component can become important, especially for large-volume detectors that have been radiation damaged (see below) such that charge carriers can be trapped – and therefore not measured – for timescales on the order of the charge integration time. For low energies, however, $W_e$ will still usually dominate the energy resolution.

Any real semiconductor contains some amount of natural impurities, and may also have had impurities (known as dopants) added deliberately. The impurity (or dopant) atoms occupy normal lattice sites but contribute a different number of valence electrons than do the bulk atoms; donor (acceptor) impurities have more (fewer) valence electrons and thus readily contribute free electrons (holes) to the conduction band. Semiconductors are termed n-type (p-type) if they are dominated by donor (acceptor) impurities and hence when the majority charge carriers are electrons (holes); the excess charge carriers from the impurities raise the conductance of the material and, when a voltage is applied across the detector, contribute to a steady leakage current that overwhelms any true signal from an absorbed photon.

To combat this, the excess carriers must be depleted from the detector bulk; this is accomplished via the p-n junction, a single piece of semiconductor that is p-type on one side and n-type on the other. The large concentrations of majority carriers on either side of the boundary (electrons for n-type, holes for p-type) attract each other and diffuse across the junction, neutralizing one another. This creates a depletion region with no free carriers, where the charges of the impurity/dopant atoms are no longer screened, forming a space charge that grows away from the junction until the subsequent potential drop is sufficient to prevent further diffusion; the potential gradient is shallower, and hence the depletion region larger, for lower intrinsic carrier concentrations (i.e. lower impurity levels). By applying a reverse bias voltage that sufficiently increases the potential drop (and thus enhances the space charge), the depletion region can be enlarged so as to encompass the entire semiconductor thickness; the voltage at which this is achieved is termed the depletion voltage, which is smaller for higher purity material. Detectors operated at (above) the depletion voltage are termed fully (over-) depleted. Over-depletion minimizes the detector capacitance – reducing the electronic noise – and makes the electric field more uniform over the detector volume – optimizing the charge collection efficiency – thereby maximizing the energy resolution, especially at low energies where electronic noise dominates.

While the reverse bias prevents diffusion of the majority carriers across the junction, it *attracts* minority carriers across, which can result in a reverse leakage current as electrons (holes) are injected from the cathode (anode). This can be minimized by providing a heavily-doped p+ or n+ layer on the side of the same type – the heavy doping effectively suppresses injection of the minority carriers across the layer.

Hence, a practical semiconductor detector is generally made from the highest-purity n-type (or p-type) material available. One surface is then heavily doped to create a a thin p+ (n+) layer, forming the p-n junction – this layer also serves directly as the cathode (anode) and is termed the p+ (n+) rectifying contact; it is also the blocking contact. The opposing side is heavily doped to create a thin n+ (p+) layer that serves as anode (cathode) and blocking contact. This configuration achieves depletion at a level far beyond the purity of the bulk material; except for thermal excitation and other small effects, the only charges traveling across the junction are those liberated by absorption of a photon, allowing a precise measurement of the photon energy.

### 2.3.2 Semiconductor detectors – germanium

Si, Ge, CdTe, and $Cd_{1-x}Zn_xTe$ (CZT) are the most common semiconductors used for detectors. Si is inexpensive, widely available, easily manufactured, and offers the possibility of room-temperature operation (though its performance is significantly improved when cooled to reduce thermal noise). However, even with the highest currently-available purity levels, the depletion voltage for detectors thicker than a few mm approaches the breakdown voltage of the crystal; because of Si's relatively low $Z$, this thickness limitation restricts the energy range over which photons can be efficiently absorbed, making Si-based detectors inefficient for observations of X-rays above ~30-50 keV. CdTe and CZT also require no cooling and their higher $Z$ yields more efficient absorption of higher-energy X-rays, but their limited hole mobility necessitates complex electrode configurations to ensure efficient charge collection and thus achieve good energy resolution; this consideration restricts detector thicknesses to only ~1 cm or less, limiting their effective energy range to below ~200 keV.

While more expensive than Si, Ge's lower melting point allows for easier refinement and purification; high-purity germanium (HPGe) is available with impurity levels low enough to allow detector thicknesses (i.e. cathode-anode distances) of a few cm. Because of Ge's small band gap (~0.7 eV), HPGe detectors must be cooled to minimize the thermal leakage current and maximize the detector sensitivity, with the best performance achieved at cryogenic temperatures such as ~77 K, readily accomplished with liquid nitrogen. To achieve a large detector volume, thereby maximizing absorption efficiency even up to MeV energies, the planar configuration is eschewed in favor of a closed-ended coaxial configuration, generally oriented such that most photons are incident along the axial direction (though photons from any direction may be detected if the detector is unshielded). The small inner bore lowers the detector capacitance, improving the energy resolution compared to a planar configuration of similar thickness. Regardless of whether the detector is n-type or p-type, the rectifying contact is placed on the outer surface rather than the inner bore as this maximizes the field strength near the outer radius; most of the area, and hence most of the photon interactions, is at larger radii, so this contact configuration maximizes the efficiency of charge collection for the majority of events. The p+ contact is typically created via boron implantation, while the n+ contact is made via lithium diffusion.

Detectors operated in space are subject to high-energy particle radiation. For Earth orbit in particular, cosmic rays and energetic protons from Earth's radiation belts are of particular importance. These high-energy particles damage the crystal lattice and behave as implanted impurities which reduce the inherent purity of the bulk material, raising the depletion voltage, which over time can exceed the breakdown voltage of the crystal; in such a case, the detector can no longer be operated in a fully-depleted mode, reducing the active volume over which measurements can be made. In HPGe, radiation damage tends to produce acceptor sites that act as hole traps, slowing or stopping holes from reaching the cathode, thereby degrading the energy resolution due to poor or incomplete charge collection. This can be mitigated via careful choice of contact configuration: since most photon interactions take place at large radii, placing the cathode on the outer surface minimizes the average hole travel distance and, if this is the rectifying contact, also takes advantage of the stronger electric field and thus faster charger collection. Hence, the optimal choice for a space-borne coaxial HPGe detector is an n-type bulk with p+ outer contact. Radiation damage may be somewhat reversed by annealing the detector at ~100° C; this temperature is sufficient to repair some of the lattice damage without significantly affecting the overall structure.

**Chapter 3: RHESSI**

*3.1 Overview*

The *Reuven Ramaty High-Energy Solar Spectroscopic Imager* (RHESSI) is the current generation instrument for solar X-ray and gamma-ray observations. It was designed to further our understanding of the relationship between the rapid release of magnetic field energy, particle acceleration, and flare heating by using large-volume HPGe detectors (§3.2) and rotation-modulation collimation imaging (§3.3) to precisely measure solar flare emission from ~3 keV to ~17 MeV with high resolution spectrally (down to ~1 keV FWHM), spatially (down to ~2 arcsec FWHM), and temporally (down to tens of ms).

The detailed mission overview is provided by Lin *et al*. (2002). Funded under the NASA Small Explorer (SMEX) program, RHESSI's design, construction, and operation were and are principally led by the RHESSI team, headed by Principal Investigator Robert P. Lin, at the Space Sciences Laboratory at the University of California, Berkeley. RHESSI was launched into nearly-circular, 38°-inclination, low Earth orbit in February 2002 with a nominal mission lifetime of 2 years. Despite some performance degradation over time (§3.2.2), as of May 2010, it continues to successfully operate and remains the only solar instrument with its observing capabilities.

RHESSI is a solar-pointed spinning spacecraft with a nominal rotation period of ~4 sec; the instrument itself comprises much of the spacecraft bus and is shown in detail in Figure 3.1. The spectrometer uses cryogenically-cooled HPGe detectors (§3.2) and charge-sensitive pulsed-reset preamplifiers to observe down to ~3 keV and up to ~17 MeV with a spectral resolution down to ~1 keV FWHM at X-ray energies (limited by electronics noise, cf. §2.3.1). Moveable attenuator discs that reduce incident SXR flux, along with on-board pulse pileup rejection (§3.2.1), enable accurate spectroscopy over the wide dynamic range from microflares to the largest X-class flares. Coupled with a bi-grid rotation-modulation collimator (RMC) imager assembly (§3.3) using eight tungsten (W) and one molybdenum (Mo) grid-pairs, this allows imaging over the entire energy range with an angular resolution down to ~2 arcsec below ~100 keV (~35 arcsec for gamma rays) and a temporal resolution down to ~2 seconds (half the spacecraft spin period, for full imaging detail) or even tens of ms (for very basic images). Each individual detected photon is tagged with its measured energy and time of arrival (with μs precision), then stored in on-board memory and transmitted to the ground without further processing, preserving as much raw information as possible to provide wide latitude in data analysis and in-flight calibration improvements. The data is typically accessed and analyzed using routines written in Interactive Data Language (IDL) and distributed as part of the open-source *SolarSoftWare* (SSW) package[1].

*3.2 Spectroscopy*

Because the RMC imaging implementation (§3.3) does not require the detectors themselves to have any spatial resolution, the spectrometer design could be optimized to maximize spectroscopic accuracy and precision.

---

[1] http://www.lmsal.com/solarsoft/sswdoc/index_menu.html

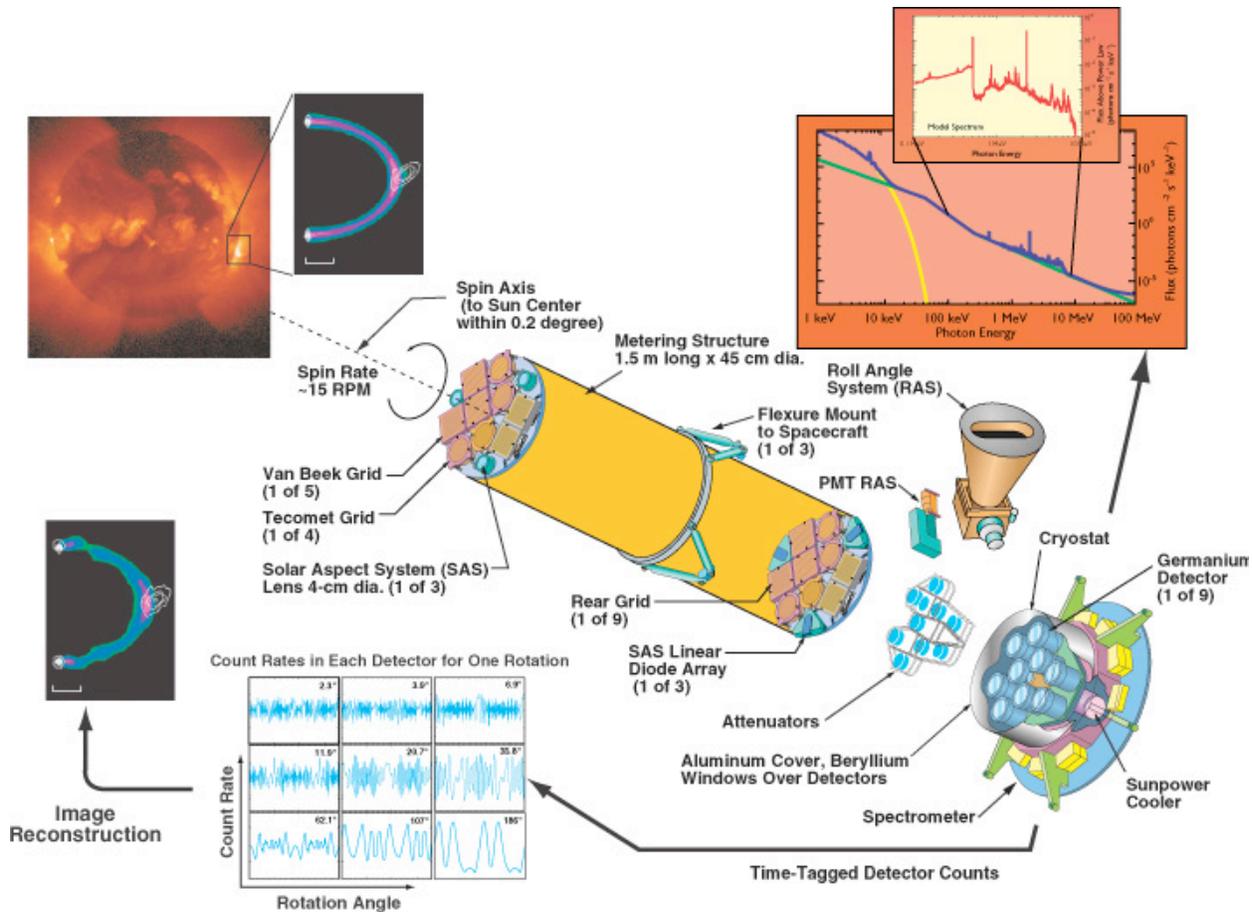

**Figure 3.1** – Exploded view of the RHESSI instrument, including the RMC imaging grid assembly and spectrometer (from Lin *et al.* [2002]).

### 3.2.1 Spectrometer – detectors, electronics, and attenuators

The RHESSI spectrometer is described in detail by Smith *et al.* (2002). It is composed of nine closed-ended, slightly n-type, coaxial HPGe detectors (see §2.3.2), each ~7 cm in diameter and ~8.5 cm tall. The detectors (GeDs) are all housed in a single Al cryostat and held at cryogenic temperature (~77 K) using a Stirling-cycle mechanical cooler. A thin Be entrance window above each GeD effectively attenuates solar photons below ~5 keV. The front and side surfaces of the GeDs have a thin boron-implanted layer that serves as the p+ cathode and rectifying contact; the inner bore has a thin lithium-diffused layer that serves as the n+ anode. When a GeD is fully depleted (typically at ~2500 V), the electric field configuration (see Figure 3.2) is such that the front and rear portions of the crystal are electrically segmented; the anode is discontinuous at the segmentation point, providing separate front and rear contacts from which signals can be extracted independently, allowing a single GeD to be operated as two separate detectors. X-rays (primarily below ~200 keV) have shallow penetration in Ge and are measured primarily in the front segment, while higher-energy gamma rays penetrate more deeply and are measured primarily in the rear segment.

By absorbing most of the ≲200 keV X-rays that dominate flare emission by many orders of magnitude over higher-energy photons (e.g. Figure 4.3, which shows a 7-order-of-magnitude decrease in photon flux over only ~100 keV), the front segments serve as natural attenuators for the rears and allow the rear segment live time to remain high for precise gamma-ray spectroscopy. The background due to ambient radiation and cosmic ray interactions is reduced in the front segments because of their smaller active volume, thereby improving the sensitivity of X-ray measurements. The front and rear segments may be anti-coincidenced to reject undesirable simultaneous events, e.g. from non-flare sources such as cosmic rays or from solar photons Compton-scattering between segments (whence their full energy may not be accurately measured). Currents induced in the front segments by charge motion in the rears (during charge collection after a photon interaction) may be recorded as spurious counts; these can also be rejected via anti-coincidence, though this is typically only important for observations below ~6 keV.

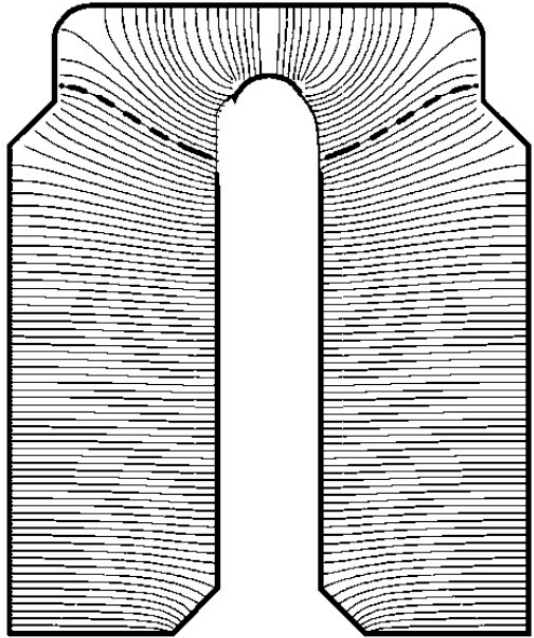

**Figure 3.2** – Schematic cross-section of one RHESSI GeD, showing the segmented inner anodes and outer cathode (bold lines) and the electric field lines between them; the dotted lines shows the segmentation boundary, where field lines intersect with either the front or rear anode, respectively (from Smith *et al.* [2002]).

The signal from each segment is fed into a charge-sensitive pulsed-reset preamplifier that accumulates the charges collected during each photon event on a feedback capacitor until reaching an upper threshold, whereupon it is reset to baseline by a transistor pulse. The output signal of the preamp is therefore the step-function increase in capacitor voltage from each measured photon. This signal is then split and sent in parallel through two amplifiers: a fast triangular-pulse shaper amp (~800 ns pulse width) used for timing and rough energy measurement, and a slow shaper amp (~8 μs peaking time) for precise spectroscopy. In the front segments, the long shaping time is chosen to optimize electronic noise while maintaining a high counting rate; in the rears, it ensures that all charges are collected from the full crystal volume regardless of the location of charge deposition. To eliminate spurious events due to electronic noise, both shapers include low-level discriminators (LLDs) that pass events only above a certain energy; the fast and slow LLDs are typically set to ~7 and ~3 keV for the front segments, respectively, and to ~20 keV for the rears. The base-to-peak amplitude of the slow shaper output pulse is fed through an analog-to-digital converter (ADC) to digitize the photon energy measurement (with 8192 chan-

nel = ~1/3 keV precision in the fronts, and ~1/3-2 keV precision in the rears), which is then recorded in the on-board solid-state memory along with the detection time (with µs precision) and other diagnostic information.

If two photons interact in a segment within the slow shaping time, the second event "piles up" on the first and distorts the energy measurement of one or both photons. This problem is mitigated using the fast pulse timing for automatic on-board rejection of piled-up events. If two events are separated by less than the ~8 µs slow peaking time, they would be read as a single higher-energy event by the ADC and therefore both are discarded. If the second photon arrives after the first pulse peak has been sampled, the first event is retained as its energy has already been accurately measured; if the slow shaper has not yet returned to baseline, the second pulse is still discarded as its peak measurement would be distorted. However, if the two photons arrive within the ~800 ns fast pulse width, they are indistinguishable from a single event and neither can be discarded; events with energies below the fast LLD threshold don't generate a fast pulse at all and can therefore pile up with other events (of any energy), also without rejection. It is possible to compensate for both effects in ground software during spectral analysis (see §3.2.3).

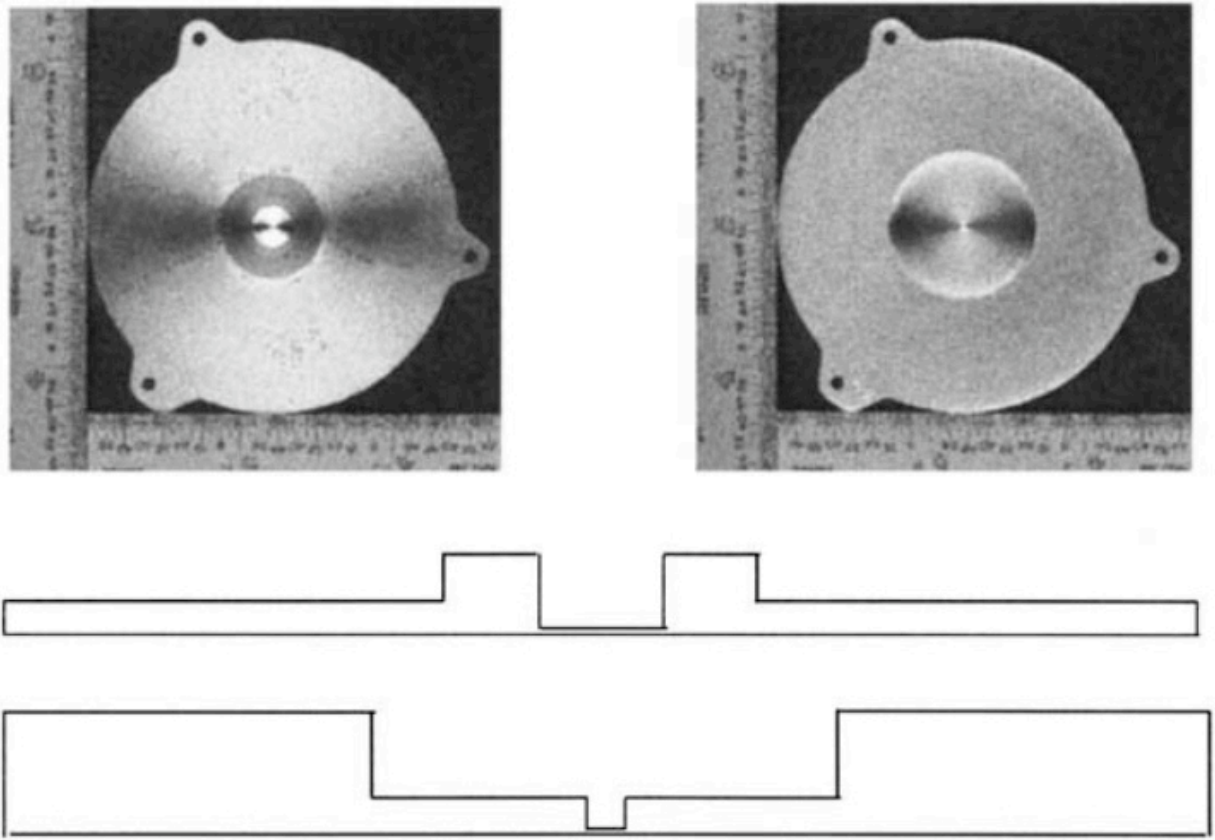

**Figure 3.3** – Photos and schematic cross-sections of the thin (left photo; upper schematic) and thick (right photo; lower schematic) attenuators. The attenuation is averaged over the entire disc area; in the A1 (thin-only) state, there are 3 annular regions of differing thickness contributing to the average, while for the A3 (thin+thick) state, there are 5 such regions. In both states, the center region is thinnest, and is where almost all of the low-energy counts will be measured (from Smith *et al.* [2002]).

Because of the finite ADC processing time and pileup rejection, the number of counts measured is less than the number actually generated in the detector. The fraction of measured-versus-generated counts is the detector live time (or, conversely, the fraction missed is the dead time) and is recorded along with the photon energy and time; in general, the live time goes down as the incident count rate increases. Knowledge of the live time is important for accurate spectroscopy, both to reconstruct the true incident count rate and to optimize the software pileup correction.

Directly above the cryostat lie two frames, each holding nine aluminum discs – thin and thick attenuators – that can be moved in front of the nine GeDs to reduce incident SXR flux. Because flare emission is strongly dominated by lower-energy photons, attenuating the SXR flux reduces detector dead time and pulse pileup (albeit also reducing the relative sensitivity at low energies), maintaining accurate spectroscopy for higher-energy photons even during high incident flux. The centers of the attenuators are thinned (as shown in Figure 3.3) to preserve some low-energy response – the transmitted flux is reduced by $1/e$ below ~15 and ~25 keV for the thin and thick attenuators, respectively (Figure 3.4); consequently, the detector resolution below ~10 keV is somewhat improved in attenuator states A1 and A3 (to ~0.75 keV FWHM, compared to ~1 keV nominally), as the low-energy photons are then detected only at the center of the GeDs, where the electric fields are strongest, charge travel distance is smallest, and charge collection is fastest. Although all attenuators of a given type (thin or thick) move together, the thin and thick attenuators can be inserted independently. In practice, only three of four modes are used, in order of increasing attenuation: A0 – no attenuators; A1 – thin attenuators only; and A3 – thin and thick attenuators engaged. The on-board computer automatically engages the appropriate attenuators as the detector-averaged live time drops below specified thresholds (~92% for A0, ~90% for A1). When the live time remains above 99% for ~4 minutes, the most recent attenuator is disengaged. If the now-unattenuated rates are still sufficiently high so that the live time falls below the threshold, the attenuator is reinserted, and the process repeats; the data during these successive attenuator changes should generally be omitted during spectral analysis.

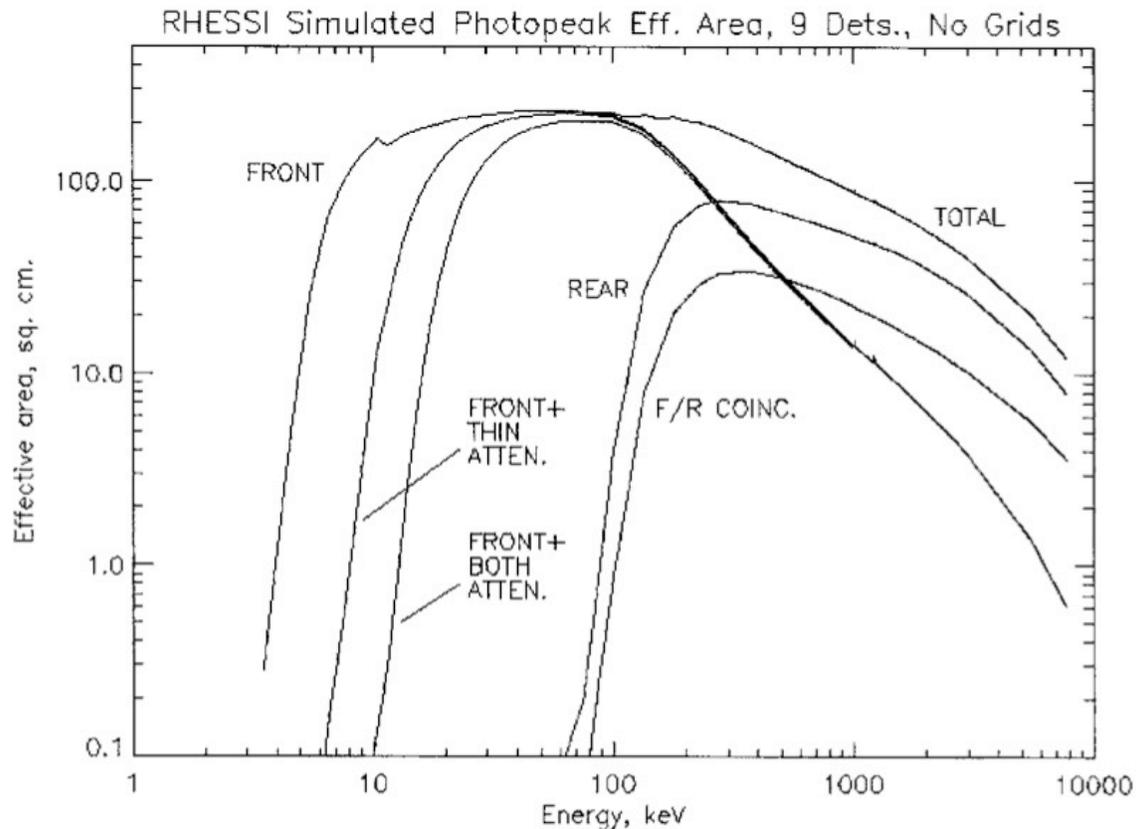

**Figure 3.4** – Simulated total effective area (physical detector area times transmission fraction of the overlying material) versus energy for various states; note the significant decrease at low energies when attenuators are engaged (from Smith *et al.* [2002]).

### 3.2.2 Radiation damage and annealing

In space, the GeDs are susceptible to radiation damage from energetic particles. Specifically, RHESSI's orbital altitude and inclination take it through zones of high magnetic latitude and through the South Atlantic Anomaly (SAA). The cryostat is unshielded, so in these zones, the GeDs are subject to a large flux of energetic protons from the Van Allen belts. The particle interactions in the detector not only contribute very high count rates that can overwhelm a true solar signal (hence why events during SAA passage are rejected entirely), they also contribute to radiation damage as they dislocate Ge atoms from the crystal lattice. The disordered regions behave like acceptor sites that, over time, become sufficiently numerous so as to dominate the resident donor impurities and change the detector bulk from slightly n-type to p-type. The inner n+ anode therefore becomes the rectifying contact; the depletion region grows from the inside outwards, and the higher depletion voltage required by this configuration results in weak or zero electric field strength at larger radii, reducing the detector's active volume. This problem is compounded by the need to periodically decrease the bias voltage for some detectors to combat arcing (the exact cause of which is unknown), further reducing the active volume. The disordered regions also act as hole traps that, coupled with the weaker electric field, cause significant degradation in the energy resolution due to poor/incomplete charge collection; this is especially important for studies of gamma ray lines, but is negligible (below the electronic noise contribution) for X-rays below ~100 keV.

RHESSI's design includes the capability of annealing the GeDs to correct some of the radiation damage. Resistors in the cryostat can heat the Ge to restore dislocated Ge atoms back to their lattice positions and/or break up some of the disordered regions to decrease their trapping efficiency. However, heating the crystal increases the mobility of the Li ions in the inner contact, allowing them to diffuse away from the original implantation layer. This diffusion blurs the discontinuity between the inner anodes, requiring larger bias voltages to further concentrate the electric field to maintain segmentation. The larger voltage may increase the leakage current and, after sufficient Li diffusion, the required bias can exceed the limits of the on-board high-voltage power supply, eliminating the ability to segment the detector. Additionally, the anneal temperature is limited to ~100° C to avoid heat-related shrinkage of the aluminized-mylar thermal blankets within the cryostat.

To quantify the allowable anneal time, we performed a test anneal (in collaboration with A. Y. Shih and D. M. Smith) on a spare GeD at the Space Sciences Laboratory, identical to the ones selected for flight. Since the detector performance can only be evaluated at cryogenic temperatures, the test was performed in iterative steps by annealing the detector for 3-4 days, re-cooling it and determining the segmentation voltage and energy resolution, and repeating. As expected, the test showed an increase over time in the voltage required for segmentation; after ~21 days, the segmentation voltage approached 5000 V, the limit of the on-board power supply. The diffu-sion of the lithium contact from annealing did not appear to have a significant detrimental effect on the energy resolution. Since the spare GeD was not radiation-damaged, the test could not predict the exact performance improvements expected from an in-flight anneal.

Radiation damage had noticeably degraded the RHESSI detector performance by early 2006, significantly reducing the active volume in the rear segments and thus compromising gamma ray observations. Because of the limited anneal lifetime and the potential complications, the anneals must be carefully managed; the first anneal occurred only in late 2007. The procedure maintained an anneal temperature of ~90° C for 7 days and successfully repaired some of the damage, but the subsequent detector performance was restored only to ~mid-2005 levels. By 2010, detector performance had fallen to pre-anneal levels; a second anneal was performed in April 2010, this time at ~100° C for 10 days. At the time of writing, the resolution and active volume appear to have been restored to ~early-2005 levels, while operating voltages for most of the detectors have been restored to at or near their initial values after launch.

### 3.2.3 Data handling for spectral analysis

The minimal on-board processing of the photon data maximizes the flexibility for data analysis. Since photons are precisely tagged with their energy and arrival time, they can be binned entirely arbitrarily to produce spectra optimized for the analyst's needs, whether to maximize sensitivity, resolution, statistics, etc. However, because of attenuation, dead time, and various other factors, the observed count spectrum is not identical to the incident photon spectrum, and any analysis must compensate for the effect of the instrument. Much of this is (or can be) done automatically via the RHESSI SSW software (Schwartz *et al.* 2002).

The photon energy is recorded digitally and corresponds to a channel number; the detector gain and offset map that channel number to the more useful units of energy. The gain and offset can drift slowly over time; this is easily monitored by measurements of known background lines, typically done automatically every few days. However, data during flares shows apparent front-segment offset variations on much shorter timescales (minutes to seconds), which appears to be directly related to the detector live time (Figure 3.5). The exact mechanism for this is unclear, but is thought to be due to a slow recovery time in the baseline restorer for the slow (spectroscopy) shaper amp, leading to insufficient baseline restoration during high count rates which therefore adds a small amount to the measured energy of each photon at these times. This cannot be adequately monitored using background lines because such lines have poor statistics, require long integrations, and are completely dominated by flare emission during these periods of high dead time. Tests during flares, assuming certain spectral models, suggest that this shift is a linear offset rather than a multiplier, and is generally small ($\leq 1/3$ keV at the highest count rates); thus, its effect on continuum studies is relatively minor, but it must be carefully considered when studying the Fe and Fe/Ni line complexes (see Chapter 4) and during in-flight calibration (see Appendix A).

To get a true measure of the incident flux, the data must also be corrected for live time to account for counts that were missed due to pileup rejection or during the ADC processing time. The counter live time is recorded with the photon energy and time (one measurement per 3 photons [Curtis *et al.* 2002]), thus a spectrum can be live time-corrected simply by dividing the measured fluxes by the average live time over each time bin. (This is only well-defined when the total flux is assumed to change on timescales longer than the time bin width, although an error will affect only the overall normalization, not the spectral shape.) Measured spectra also ex-hibit "dropouts" where no counts are measured for a given period regardless of the overall live time. These are thought to be caused by a cosmic ray interaction within the detector electronics, resulting in a dead period as the electronics reset. Dropouts occur up to once every few seconds and last for ~10-100 ms each time, but they can be identified by a distinct electronic signature; in practice, they are treated as dead time and incorporated into the live time correction (Smith *et al*. 2002).

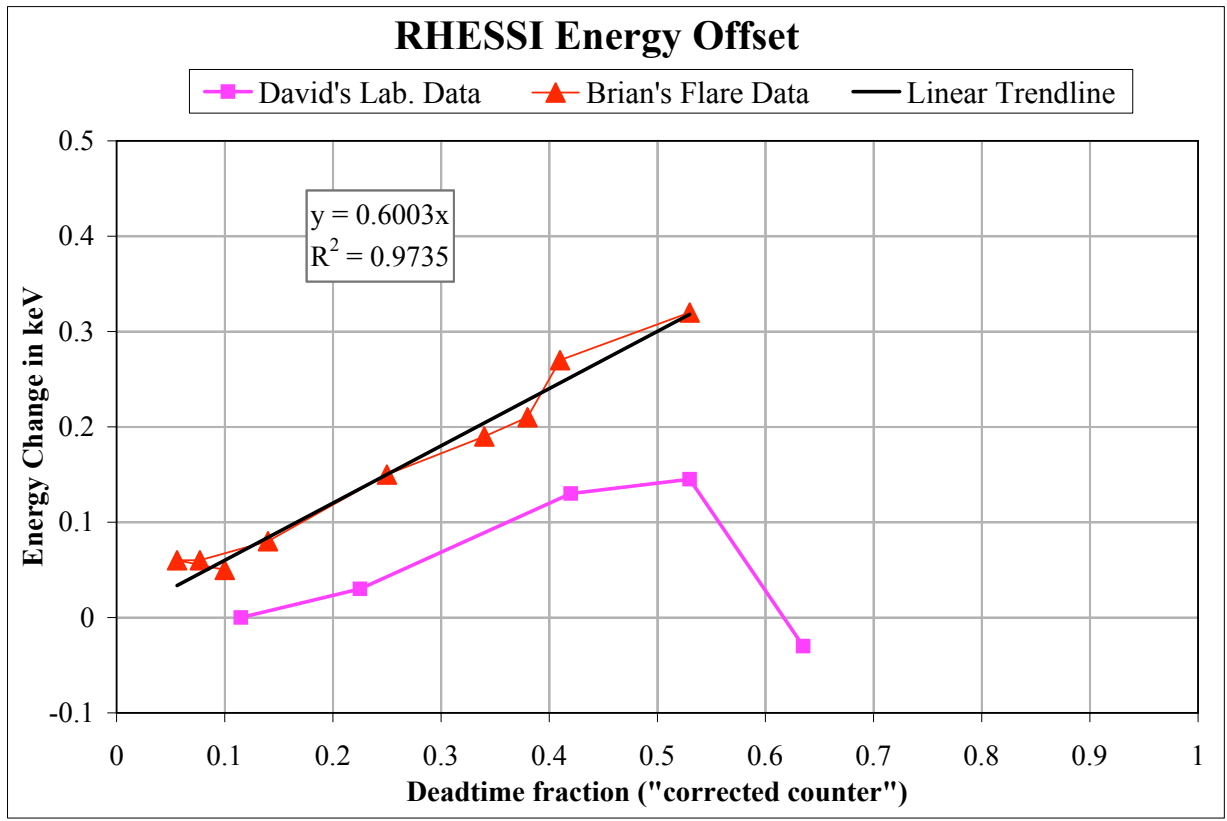

**Figure 3.5** – Measurements of the apparent offset as a function of dead time based on lab and flare data (fuchsia squares & red triangles, respectively); a linear fit (black line) is also shown for the latter. Lab measurements are based on the $^{55}$Fe radioactive decay line at ~6 keV; flare measurements are based on the Fe-line complex at ~6.7 keV. (B. R. Dennis & D. M. Smith, private communication)

The spectra must also be corrected for count decimation. To prevent the on-board memory from filling too quickly during high count rates, the flight computer employs a progressive decimation scheme whereupon $N$-1 of every $N$ counts below an energy $E$ are automatically discarded, with $N$ and $E$ functions of the memory fill level, attenuator state, and segment (front or rear); for example, at the lowest decimation level for the front segments, 50% ($N$ = 2) of counts below ~7 keV are rejected, while at the highest level, ~94% ($N$ = 16) of counts below ~100 keV are rejected. Since the decimation scheme is exact, the original count flux can be precisely reconstructed by multiplying the measured count rate below energy $E$ by $N$ (and, in practice, this can be folded into the live time correction). However, decimation does reduce the effectiveness of spurious count rejection via anti-coincidence of the front and rear segments (cf. §3.2.1) as the *N*-1 decimated counts are never recorded and thus cannot be used for anti-coincidence.

Putting the data into usable (energy, time) units now allows compensation for the effects of the physical instrument – known as the instrument response – including the detectors and attenuators. It is usually convenient to represent the linear static and quasi-static (changing over timescales that are long compared to the analysis times) components of the detector response numerically as a Detector Response Matrix (DRM) that maps incident photon energies to observed count energies, or, in the inverse sense, that maps observed counts to linear combinations of incident photons. The actual parameters of the DRM – the quantification of the various physical effects described by the matrix – depend on the specific detector (and segment) and attenuator state, and are determined via numerical mass modeling (Monte Carlo simulations), pre-flight ground calibration, and in-flight calibration, along with situational specifics such as the emission source's location in the sky.

For a given photon energy, the photopeak efficiency (cf. Figure 3.4) is the percentage of incident photons that are recorded as counts with that same energy, within the detector resolution; this is represented by the main diagonal of the DRM, and components of the response which affect primarily the photopeak efficiency are termed "diagonal" components. These include:

*1) Grid transmission* – the slats in the two RMC imaging grids (§3.3) are opaque below certain energies and generally reduce the incident X-ray flux by ~75%. However, this number depends on the photon energy, as the grids' X-ray transparency increases with energy, and on the angular distance of the source from the spacecraft spin axis (the "off-axis position"), as internal shadowing due to the grids' finite thickness and (for higher-energy photons) the larger effective column density become important for off-axis distances that are a significant fraction of the grid field-of-view (FOV). These variations are generally negligible for X-rays below ~100 keV, where the grid transmission is thus essentially a constant ~0.25.

*2) Attenuator, blanket, & Be window transmission* – apart from the grids, all of the other material overlying the GeDs also attenuates incident photon flux. This is determined simply by the material mass attenuation coefficient and thickness, per §2.2. The material thickness is not uniform – deliberately so for the attenuators – and since the detectors are monolithic and have no position sensitivity, the transmission factor is integrated over the entire detector area to obtain the weighted-average transmission as a function of energy (and, since they are movable, the attenuator state). These response components are especially important below ~100 keV for attenuator states A1 and A3, as the transmission fraction decreases exponentially with energy. For the rear segments, the attenuation due to the front segments must also be considered.

*3) Low-level discriminator* – the slow shaper LLD prevents counts below its threshold energy from being recorded, but the threshold is not entirely sharp, allowing some fraction of near-threshold-energy counts to pass; the threshold spans ~1 keV and therefore must be considered when using fine energy bins. Although for most of the GeDs, attenuation entirely dominates over the LLD contribution to the photopeak efficiency, the LLD threshold for GeD #7 was set to ~7 keV to combat noise due to a loose front-segment electrode, and the LLD threshold for GeD #8 was raised to ~6 keV in mid-2006, also to reduce noise (from an as-yet unidentified source); the LLD contribution is thus not negligible for these detectors. (GeD #2 suffered an apparent breakdown in the crystal early in the mission and must be operated at reduced bias to avoid arcing; it failed to segment through most of the mission and has poor energy resolution. Its LLD threshold was set to ~20 keV, but it is not used for spectroscopy at any energy. Following the second anneal in April 2010, both GeDs #7 and #2 showed remarkable recoveries and their front-segment LLDs were reset to the nominal ~3 keV.)

In flares, although the photon flux typically rises exponentially with decreasing energy, these three factors combine to yield a photopeak efficiency that drops even faster; the measured count flux actually decreases below a peak energy (~6 keV for A0, ~10 keV for A1, and ~18 keV for A3), and the counts below this energy become quickly dominated by other than the diagonal response. Contributions to the response that map an incident photon energy to a different measured count energy further reduce the photopeak efficiency and are termed "off-diagonal" components; they include:

*1) Energy resolution* – the spectroscopic precision (number of ADC channels times the gain) exceeds the detector resolution, and it is thus possible to choose energy bin widths that oversample the resolution by a factor of ~3, as would be desirable when studying narrow line features such as the Fe and Ni lines. In doing so, the nominal photopeak efficiency smears out from the diagonal to the adjacent sub- and super-diagonal chords; the "diagonal" response instead becomes a diagonal band matrix, with the band width determined by the degree of oversampling. While technically off-diagonal, this response component is "quasi-diagonal" in that a diagonal response can be easily recovered by using somewhat coarser energy bins, which is not the case for the following contributions.

*2) K-escape* – when photoelectrically absorbed (§2.2) in Ge, photons above ~11 keV will primarily liberate a photoelectron from the K shell, often causing the emission of a ~10-keV fluorescence photon. If the interaction region is close to the detector surface, the fluorescence photon has a non-negligible probability of escaping the detector before being reabsorbed; the measured count energy is therefore decreased by ~10 keV (the energy of the escaping fluorescence photon) from that of the incident photon. This phenomenon is increasingly important for lower-energy photons (down to the K-shell binding energy), as they tend to have shallow penetration; because of the exponentially-decreasing diagonal response, K-escape counts are the dominant contribution to the count spectrum at low energies. (Incident photons below ~11 keV will interact with the L shell, but the L-shell binding energy is only ~1.4 keV and the fluorescence yield is low, so escape of L-shell photons is negligible.)

*3) Compton scattering* – before reaching the detector, incident photons may Compton scatter within any of the overlying materials (grids, attenuators, etc.). The diagonal response (photopeak efficiency) already accounts for large-angle scatters whereupon the scattered photon misses the detector and is not counted; if a forward-scattered photon *is* detected, however, it will have lost some fraction of its energy during the Compton scatter. Similarly, if a photon scatters within a detector segment and subsequently escapes before being completely absorbed, it will deposit only part of its energy. Both processes contribute to the off-diagonal response. Since the Ge, Mo, and W (the grid material) interaction cross-sections are dominated by photoelectric absorption up to ~170, ~200, and ~500 keV, respectively, the Compton contribution is vanishingly small for ≲100 keV photons. The contribution from Al and Be (the cryostat window) are not necessarily negligible, but in practice the counts at a given energy are completely dominated by either the diagonal response or by K-escape, which is the dominant energy-loss mechanism relevant for this study.

*4) Albedo* – photons from the flare source are not emitted only towards the observer, but also towards the solar surface. Because the solar atmosphere is primarily H and He, Compton scattering rather than photoelectric absorption dominates the interaction cross-section (cf. §2.2); the dense photosphere is optically-thick to X-rays and there is a non-negligible probability that downward-directed X-rays will be Compton-backscattered from the photosphere towards the observer at Earth. The observed spectrum will therefore be the sum of the original flare spectrum emitted towards the observer and this albedo component. Although this scattering does not occur within the spacecraft and thus is not technically an instrumental effect, it is still a linear process just as if it did occur within the spacecraft; it is essentially an off-diagonal modification to the original spectrum, thus it can be treated as a *de facto* part of the DRM. The primary factors on which this component depends are the heliocentric angle of the flare (a larger angle, i.e. closer to the limb, yields less albedo) and the directivity of the flare photons (a higher flux beamed towards the photosphere yields a higher fractional contribution from albedo). The actual parameters of the albedo response, as incorporated into the DRM, are given by Kontar *et al*. (2006).

*5) Pulse pileup* – although most piled-up counts are automatically rejected, some will get through the on-board veto circuit; this is especially important during high counting rates when the probability of pileup within the ~800 ns fast shaper time resolution is non-negligible. Because flare bremsstrahlung spectra decrease steeply with energy (typically exponentially or as a power-law, per §2.1.1), pileup can significantly distort the spectral shape at higher energies, as even a small fraction (e.g. ~1%) of the counts at energy $E$ can yield pileup at energy *2E* comparable to the diagonal response. Pileup is the dominant off-diagonal contribution at energies above the count-rate peak energy (but below energies where Compton scattering becomes significant); however, since pileup is inherently a non-linear effect that depends on the exact shape of the spectrum, it cannot be included in a linear DRM, and must be handled separately in the analysis.

With the exception of those that are functions of the electronics, all of the above response components (specifically: grid transmission, attenuation, K-escape, Compton scattering, and albedo) assume that the incoming photons are incident along the axis, i.e. from the Sun. For such spectral analysis to be well-defined, the non-solar background must either be negligible or subtracted from the data before analysis; above ~100 keV, or at any energy for small flares, the non-solar background can be significant compared to the flare emission, and proper subtraction (see Appendix B) is therefore important.

### 3.3 Imaging

While X-ray imaging can be achieved directly via grazing-incidence focusing optics, such optics cannot efficiently focus photons above ~60-80 keV even with current technology, and the angular resolution above ~20 keV is currently limited to ~7 arcseconds; even below ~20 keV, achieving arcsecond-scale angular resolution for SXRs requires precise control of the telescope pointing, long focal lengths, and position-sensitive detectors with fine spatial resolution. Fourier techniques using shadow masks (see below) allow imaging over a wide energy range, from SXRs to gamma rays, and provide arcsecond-scale resolution with much shorter focal lengths. Single-plane coded-aperture masks are one such technique, but require finely-pixelated detectors to achieve good angular resolution. Instead, RHESSI uses the bi-grid RMC, which essentially decouples the imaging resolution from the detector performance, enabling the use of large-volume monolithic detectors and thereby allowing simultaneous optimization of both imaging and spectroscopic performance. The imager is discussed in detail by Hurford *et al*. (2002).

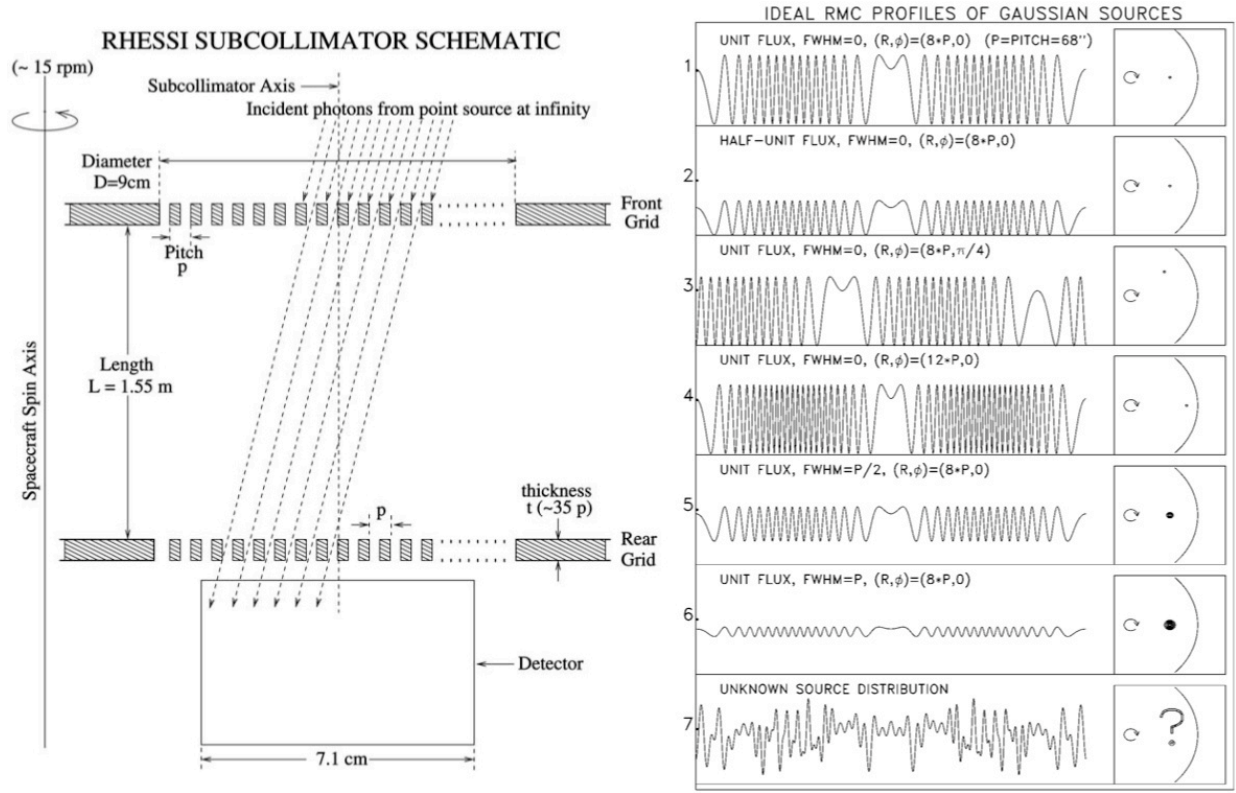

**Figure 3.6** – [left] Schematic representation of the RHESSI RMC: the top grid casts a shadow on the bottom grid; rotation of the grids changes the photon angle of incidence and thus the position of the shadow on the bottom grid, thereby changing the fraction of photons passing through to the detector. The measured lightcurve is thus modulated in time. [right] Idealized lightcurves for Gaussian sources showing how spacecraft rotation modulates the observed source intensity for various source parameters; changing the source brightness, rotation angle, radial position, or size affects the maximum intensity, modulation phase, frequency, or amplitude, respectively (from Hurford *et al.* [2002]).

### 3.3.1 Fourier imaging

The bi-grid RMC imager uses multiple slitted grid-pairs at different pitches to image sources through the concept of shadow masking, as illustrated in Figure 3.6. A point source shining through a grid-pair casts a specific shadow on the detector; as the source moves perpendicular to the slit direction, the amount of light on the detector is modulated. This modulation is effectively the Fourier transform of the 1-D (perpendicular) source profile, converting 1-D spatial frequencies (the Fourier components) into temporal frequencies (the lightcurve modulation). Each grid pitch (the slit-to-slit distance) samples a different 1-D spatial frequency, and varying the grid[2] azimuthal orientation (e.g. by rotating the spacecraft) samples the source from different angles, thereby measuring the frequencies in the 2$^{nd}$ dimension and providing multiple Fourier components from a single grid pitch (provided that the pitch is small enough such that the modulation period for a given rotation angle is small compared to the overall spin period). A half-rotation of the spacecraft is sufficient to measure all available Fourier components. The angular resolution of each grid is defined by the ratio of half the grid pitch to the grid separation (*p/2L* per Figure 3.6).

---

[2] Henceforth, "grid" is taken to be synonymous with "grid-pair."

Changes in the source size, position angle, off-axis distance, or intensity change the modulation profile in predictable ways. Thus, accurate knowledge of the spacecraft pointing is required for image reconstruction. This is achieved on RHESSI by a high-bandwidth solar aspect system (Zehnder *et al.* 2003) which allows the pointing to be known to sub-arcsecond accuracy for every photon recorded by the spectrometer; the pointing stability therefore need be accurate to only a few arcminutes. Then, by combining multiple Fourier components, one can then reconstruct the source image via the inverse Fourier transform – converting the spatial frequencies into an intensity map. The imaging performance is relatively insensitive to changes in the grid separation, grid tilt, or to translational shifts between the grids; the critical alignment requirement is that of the relative horizontal twist between the grids, which for RHESSI need be accurate to only ~1 arcminute.

Fourier imaging was successfully implemented on *Yohkoh* with the HXT, which used 64 non-rotating bi-grid collimators (2 orthogonal grid-pairs at each of 32 different pitches) to measure 32 complex Fourier components, imaging with ~8 arcsec resolution in the ~2-100 keV range; and on *Hinotori* with the SXT modules, which used 2 bi-grid RMCs to measure ~200 Fourier components, imaging with ~28 arcsec resolution in the ~20-40 keV range. RHESSI improves on both designs by using 9 bi-grid RMCs to measure ~1100 Fourier components, imaging with an angular resolution down to ~2 arcsec below ~100 keV and down to ~35 arcsec for gamma rays.

### 3.3.2 RHESSI grids

RHESSI uses 9 different grid pitches, varying from ~2.8 mm for grid 9 down to ~.034 mm for grid 1, spaced by root-3 between each grid. Grids 2 through 9 are composed of W and are completely opaque to X-rays below ~300 keV; because of manufacturing limitations, grid 1 is composed of Mo and is completely opaque only below ~100 keV. In order to maintain a nominal ~1° FOV, the grid thickness increases proportionately to its pitch (though grids 7 and 8 are thinner, to improve gamma ray sensitivity, and consequently have a larger FOV). This configuration provides an imaging angular resolution down to ~2 arcsec for the finest grid (below ~100 keV); for gamma rays, only grids 6 and 9 are sufficiently opaque to allow imaging, yielding a ~35 arcsec resolution.

### 3.3.3 Image reconstruction

The measured Fourier components can be directly inverted to obtain a "back-projection" of the source, which provides a general idea of the source intensity and morphology but includes considerable artifacts due to the nature of the discrete Fourier transform. For a single grid pitch (and complete Fourier sampling, i.e. some multiple of a half-rotation), the back-projection of a point source is the radial *sinc* function *A×sin(r)/r* (Figure 3.7), which has significant positive and negative sidelobes. (The finite angular resolution of the grid determines the width of the main lobe, and hence the observed width of the point source; this is known as the "point-spread function" or PSF.) Because of the root-3 factor between successive grid pitches, the sidelobes from all the grids tend to cancel each other out. Nevertheless, the relatively sparse sampling of the Fourier components (including the complication that grids 2 and 7 couldn't be used for imaging at low energies throughout nearly all of the mission, due to their noise) yields back-projection images that still retain noticeable artifacts (e.g. areas of negative intensity). These artifacts can be removed approximately via various algorithms such as CLEAN (described below), although such methods include specific assumptions that must be considered when interpreting the images.

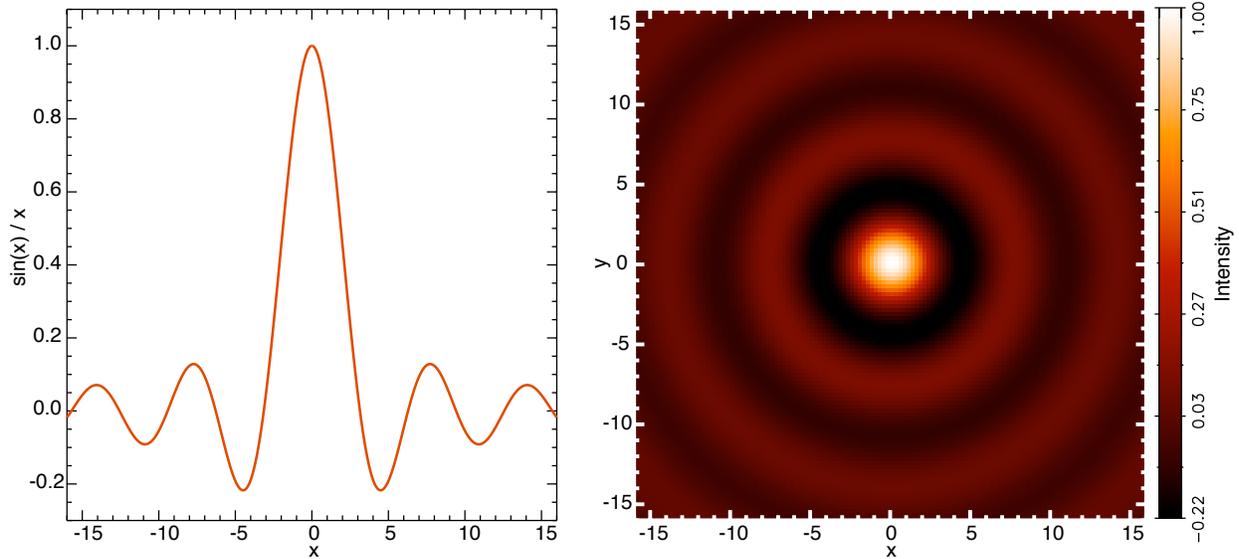

**Figure 3.7** – The *sinc* function (*sin[x]/x*) plotted in 1D [left] and 2D [right], normalized to unit intensity; the back-projection of a point source using only one grid looks like the 2D image, with the spatial scale set by the grid's angular resolution. Note the significant positive and negative sidelobes.

Because the Fourier components are not sampled with uniform density (in frequency space), the back-projection images can be biased by the sampling density. Images can be reconstructed assuming a "natural" weighting, whereby the Fourier components from each grid are equally weighted, or with a "uniform" weighting, whereby the components are weighted by their spatial frequency. Because uniform weighting emphasizes the finer grids, it has a smaller grid-averaged PSF and yields images with finer spatial resolution; it is most appropriate for complex sources with small-scale morphological variations, where the finer resolution is a benefit. However, the higher weighting of the fine grids also emphasizes their artifacts and noise, thereby decreasing the overall signal-to-noise of the image compared to natural weighting and introducing potentially spurious sources that must be carefully considered during analysis. Natural weighting does not suffer from this issue and therefore has increased sensitivity; it is appropriate for simple compact sources, or for large/diffuse sources where a higher weighting of the finer grids would be a hindrance, but its larger PSF can smear out small details in complex sources. Since it is difficult to know, *a priori*, the specific attributes of an observed source, it is often practical to image the source using both weighting methods (Figure 3.8).

CLEAN is a secondary algorithm applied to back-projection images in order to remove the sidelobes introduced by the Fourier transform. It works by assuming that the true source, with arbitrary morphology, is a collection of point sources (one at each pixel) that have been smeared by the grid-averaged PSF; the back-projection map is therefore the collection of true point sources, convolved with the PSF. CLEAN begins with the back-projection and selects the pixel with the highest absolute intensity, calculates its PSF (assuming it were a point source), and sub-tracts it from the map; it then iterates this process, selecting the brightest pixel each time, until the newly-selected pixel has a negative intensity; since negative intensity is unphysical, this pixel cannot be part of the source and must be representative of the background level. The final CLEAN map is then composed of the chosen pixels convolved with the PSF (the "clean components"), along with the residuals from the back-projection. CLEAN is computationally light and preserves the source flux, but by its nature is biased towards point/compact sources and tends to yield "rounder" images.

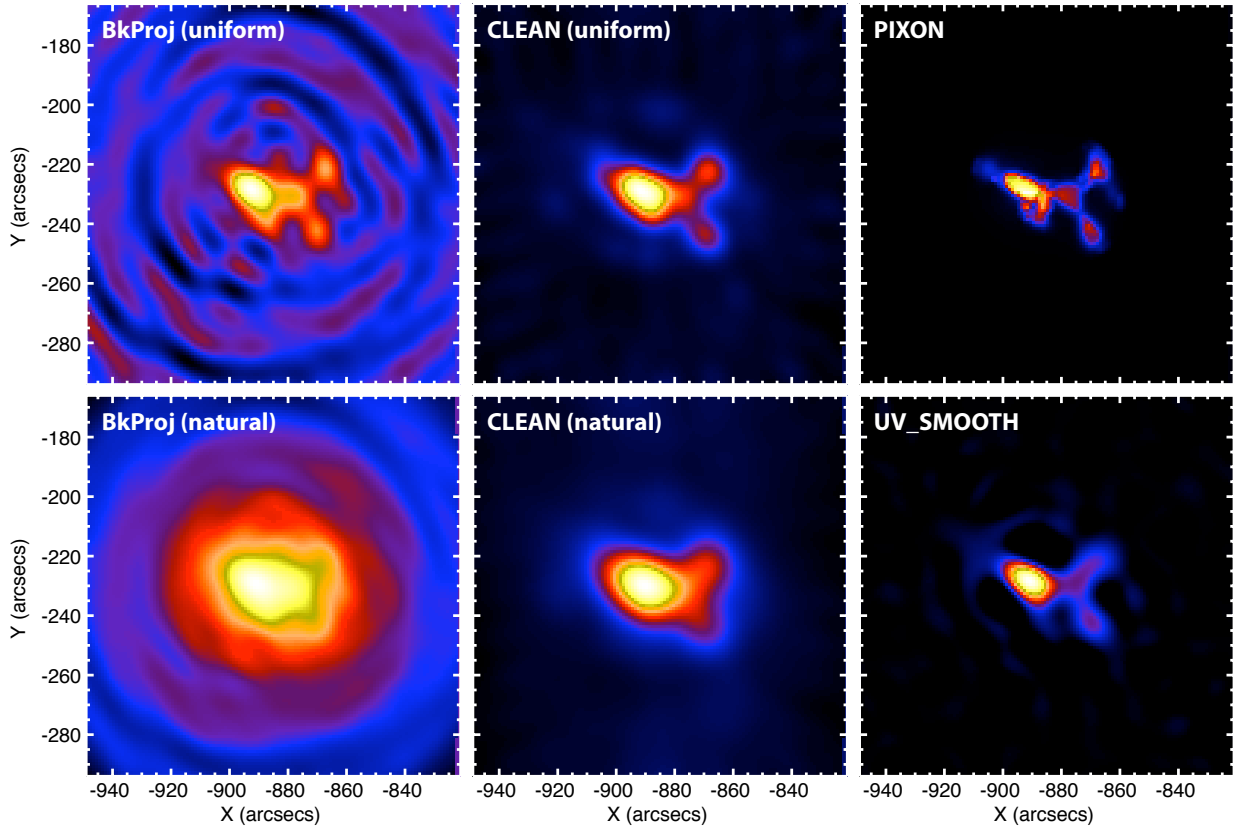

**Figure 3.8** – Comparison of image reconstruction using back-projection, CLEAN (using both uniform and natural weighting), PIXON, and UV_SMOOTH for the 25-50 keV emission during 2002 Jul 23 00:30-00:31 UT, using grids 3-9 (excluding 7). The strong sidelobes evident in back-projection are removed by the other algorithms; note the suppression of fine structure when using natural weighting.

Rather than starting from the back-projection, other imaging reconstruction algorithms (e.g. PIXON, MEM-VIS, Forward Fit) work by starting from the modulated lightcurves directly and iteratively fitting a source configuration, finding the "minimal" source (with various initial assumptions) that best fits the observed lightcurves. Such algorithms are beyond the scope of this dissertation.

### 3.3.4 Visibilities

The Fourier components can be represented identically as complex numbers of amplitude and phase; after the instrumental response has been removed, the calibrated numbers are called "visibilities." The same back-projection map can be built from visibilities as from the lightcurve, although with much greater speed since visibilities can be directly inverted with the Fast Fourier Transform. More importantly, however, visibilities are both linear – they can be scaled, added, or subtracted – and have quantifiable uncertainties; this makes them useful for more advanced source reconstruction algorithms. Specifically, models can be forward-fit to the observed visibilities to yield best-fit sources (given initial assumptions) with quantifiable errors. Because they are complex numbers, visibilities may also be interpolated to obtain inferred visibilities between the relatively sparse measured ones; this concept is exploited by the UV_SMOOTH algorithm (Massone *et al*. 2009) to reconstruct images with very few assumptions and vastly reduced artifacts compared to other algorithms.

### 3.3.5 Data handling for imaging analysis

Routines exist within the RHESSI SSW software that provide all of the above functionality automatically. Because of the flexibility of RHESSI data, images can be accumulated over arbitrary energies and times, using any subset of the grids. With a 1° FOV, the grids observe the entire Sun and images can therefore be made for arbitrary source locations and with arbitrary FOVs (up to 1°). Since the images are mathematically reconstructed, the pixel scaling (in arcseconds) may also be chosen arbitrarily; optimally, the pixel scale should oversample the angular resolution by a factor of a few, to achieve smooth images while minimizing extraneous processing.

The freedom to choose arbitrary energies for image reconstruction enables the technique of imaging spectroscopy for spectral analysis of spatially-separated sources. Here, images are created over successive energy bins using otherwise identical algorithms and parameters. One or more source regions are chosen, and the count flux within those regions is determined as a function of energy; the spectrum can then be analyzed as any other (§3.2.3). This allows spectral analysis of specific source regions, not possible with traditional spatially-integrated spectroscopy. It has the additional benefit that non-solar background is automatically subtracted by the imaging process, since it is not modulated by the grids and therefore does not contribute to the reconstructed image flux. However, because the spectra are now a Level 2 product (derived from an already-derived product, i.e. the image), the signal-to-noise is decreased compared to traditional spatially-integrated spectra, and therefore coarser energy and/or time bins are required to improve statistics.

**Chapter 4: Super-hot Thermal Plasma in the 2002 July 23 X4.8 Flare**

Since its launch in February 2002, RHESSI has observed over 15,000 flares in total, with over 50 of those having GOES-class of X1.0 or higher; these large flares are the most likely to reach super-hot temperatures (see §5.1). The X4.8 event on 2002 July 23 (hereafter "Jul 23") was the first RHESSI-observed flare where thermal bremsstrahlung emission from super-hot plasma was definitively measured; its high fluxes, long duration, near-limb position, and host of supporting observations make it an ideal candidate for a detailed study of the evolution of super-hot flare plasma. Jul 23 also marked a number of other unprecedented observations, including the first discovery of pre-impulsive-phase non-thermal coronal emission and the first gamma-ray line spectroscopy and imaging; because of this latter milestone, an entire issue of the *Astrophysical Journal Letters* (volume 595, issue 2) was devoted solely to the observational results of this flare, providing extensive context for interpretation of our analysis.

*4.1 Flare overview*

Jul 23 reached a GOES class of X4.8 (Figure 4.1), corresponding to a 1-8 Å SXR energy flux of $4.8 \times 10^{-4}$ W/m$^2$ measured at Earth; assuming isotropic emission from the flare source, this represents an instantaneous luminosity of $\sim 1.1 \times 10^{26}$ ergs/s, or $\sim 3 \times 10^{-8}$ of the total solar luminosity, from the SXRs alone. The flare occurred on the southeast limb at $\sim$S13° E72°, corresponding to a heliocentric angle of $\sim$73°; it was observed nearly broadside, removing much of the observer projection and allowing good spatial separation of emission at various altitudes in the atmosphere. Although microwave emission had been present for nearly an hour prior to the flare (White *et al.* 2003), the flare onset – marked by rising X-ray emission – occurred at $\sim$00:18 UT. Hα, microwave, and HXR emission – all corresponding primarily to non-thermal processes (see §1.3 and §2.1) – peaked at $\sim$00:28-00:31 UT, while the SXR emission – corresponding to thermal bremsstrahlung – peaked at $\sim$00:31-00:33 UT in RHESSI (6.3-7.3 keV) and at $\sim$00:35 UT in GOES (1-8 Å). Except for a brief ($\sim$3-min) burst around $\sim$00:49 UT, the HXR emissions decayed to near-background levels by $\sim$00:45 UT; the SXR emissions decayed to 10% of their peak value by $\sim$01:25 UT, but did not reach pre-flare background levels until after 04:00 UT.

The X-ray observations divide the flare naturally (cf. Lin *et al.* 2003) into three distinct phases. The *pre-impulsive* rise phase, from $\sim$00:18 to $\sim$00:26 UT, is marked by a gradual, fairly smooth increase in both SXR and HXR emission. This is a particularly intriguing phase (see §5.2) as the HXR emission appears to be primarily non-thermal but is dominated by an extended ($\sim$22" diameter) coronal source, without significant footpoints in HXRs or other wavelengths. Co-spatial radio emission is also consistent with non-thermal emission (White *et al.* 2003), indicating the presence of a significant non-thermal electron population within the corona that either remains trapped or escapes, but does not reach the chromosphere. This behavior is not commonly observed (likely due to instrumental sensitivity), and suggests potentially substantial energy contained in the accelerated electron population (see §4.4 and §5.2).

Following this gradual rise is the *impulsive* phase, from $\sim$00:26 to $\sim$00:43 UT, which is marked by multiple intense bursts of HXR emission from compact footpoint sources with spectra that are well-fit by doubly-broken power-laws, consistent with non-thermal electron bremsstrahlung (see §2.1.1); gamma-ray emission is also present and extends beyond $\sim$7 MeV, indicating significant ion acceleration. The temporal variations in the HXR fluxes of the footpoints track each other to within seconds (Krucker *et al.* 2003), suggesting that they anchor the same flare loop; the motion of the HXR sources indicates that reconnection is occurring in the corona, forming new (larger) loops along which particles are accelerated, causing the apparent footpoint motion. The SXR emission during this phase is dominated by an extended (~30" diameter) coronal source whose spectrum is primarily thermal; this source is at super-hot (≳30 MK) temperatures (see §4.4) for nearly the entire impulsive phase duration, until ~00:42 UT. This source is elongated and appears to be *above* the loop connecting the footpoints.

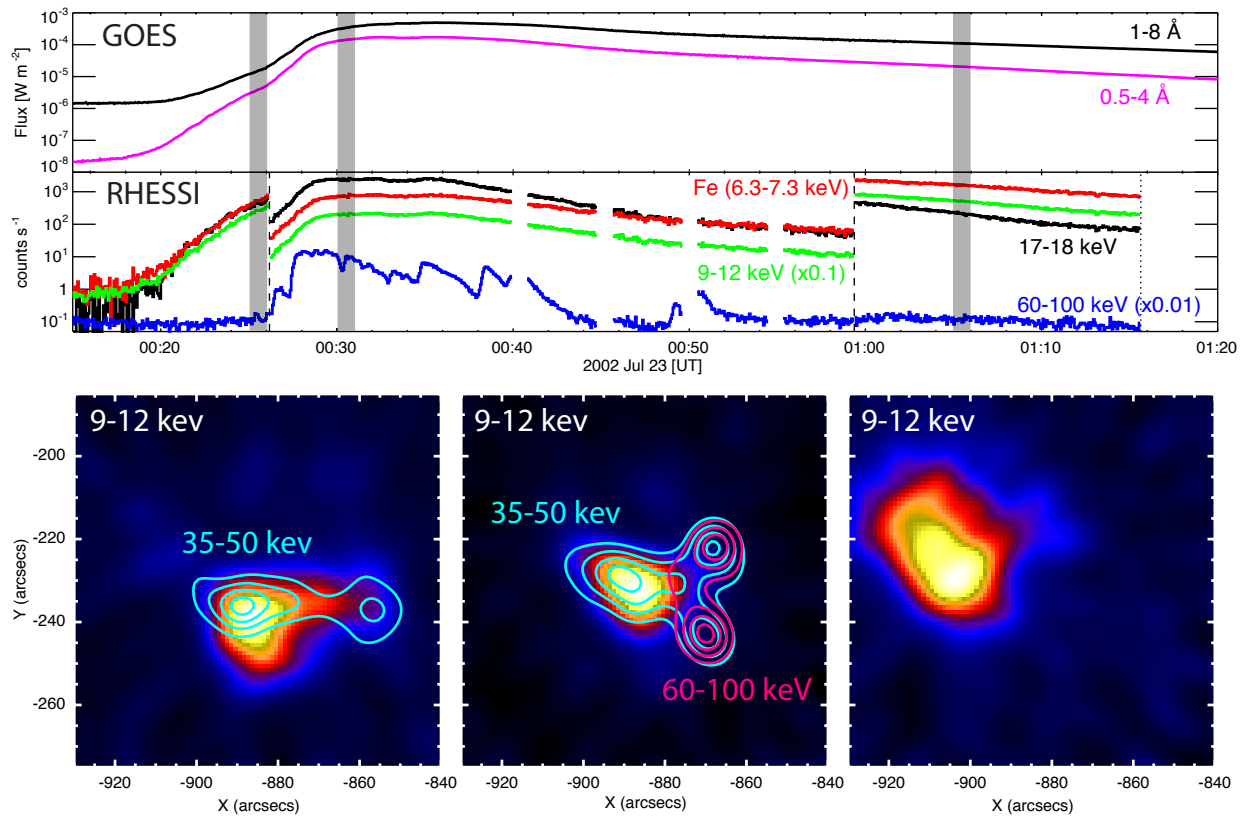

**Figure 4.1** – [top] Time profiles of the GOES and RHESSI X-ray observations for the 2002 July 23 X4.8 flare in various energy bands. The dotted vertical line represents spacecraft nighttime. Dashed vertical lines represent RHESSI attenuator transitions (A1→A3→A1); interstitial transitions have been excised. Shaded bars represent the imaging times. [bottom] RHESSI 9-12 keV images at the selected times, overlaid with the (30%, 50%, 70%, 90%) contours of 35-50 keV and 60-100 keV emission.

During this phase, the flare exhibits a very rough Neupert effect (cf. Figure 4.2), in that the time profile of the GOES or RHESSI SXR flux (e.g. in the ~6.3-7.3 keV band) is similar to that of the time integral of the HXR flux (e.g. in the ~60-100 keV band). However, the Neupert effect isn't cleanly established here: the time-integral of the HXR flux rises faster than does the SXR flux, especially after the initial HXR peak, when the SXR flux appears to flatten and remain nearly constant for some time. However, the choice of energy bands for such a comparison is rather arbitrary; it is more physically appropriate (cf. Lee *et al.* 1995; Veronig *et al.* 2002a) to compare the thermal energy (see §4.2.2) with the time-integral of the power in non-thermal elec-trons (cf. equation [2.6]). Nevertheless, the similar rise and plateau times of the SXR and time-integrated HXR emission hints at some (not necessarily causal) link between the super-hot plasma and the accelerated electrons, but the deviation from a clean Neupert effect plus the apparent location of the SXR emission above – rather than at – the looptop suggests that the super-hot plasma may result from processes other than the traditional pictures of chromospheric evaporation and heating by non-thermal electrons.

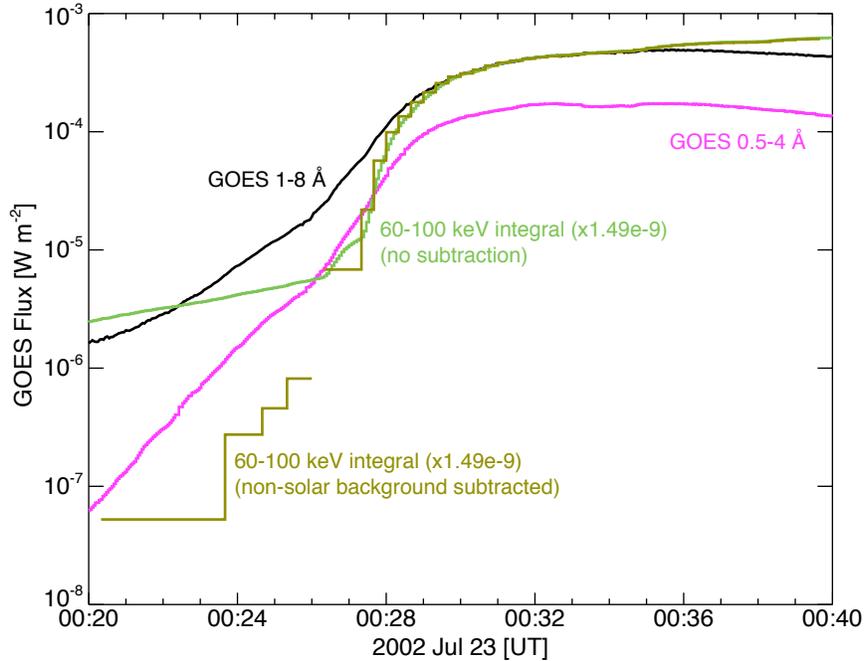

**Figure 4.2** – Evaluation of the Neupert effect during the Jul 23 flare: GOES SXR light-curves are overlaid with the (scaled) time-integral of the RHESSI 60-100 keV observations, including or omitting non-solar background. (The step-like behavior in the subtracted curve is from time binning used for background subtraction.) The time profiles show similarity, but no actual agreement.

The *decay* phase, from ~00:43 UT onwards (until the SXR emission returns to background levels after ~04:00 UT), shows no appreciable HXR emission above ~25 keV, except for a brief burst from ~00:49 to 00:52 UT; the thermal SXR emission decays slowly and smoothly throughout this phase, and the coronal source generally expands higher into the corona over time. TRACE observations in the 195 Å passband, dominated by excitation lines from thermal ions with peak formation temperatures of ~2 and ~20 MK (Fe XII and XXIV, respectively), also show the thermal flare loops growing and the thermal source moving higher into the atmosphere.

### 4.2 Methodology and analysis

RHESSI spectra for Jul 23 typically showed (Figure 4.3) an intense SXR continuum that decreased rapidly with energy and, during the impulsive phase, a relatively hard HXR continuum extending to MeV energies. Throughout the flare, two narrow line features were also observed at ~6.7 and ~8 keV; these are the Fe and Fe/Ni line complexes, comprised of numerous excita-tion lines (both spontaneous de-excitation and dielectronic recombination; see §2.1.3) of highly-ionized Fe and Ni. The primary ionization states contributing to the line complexes are the helium-like Fe XXV and Ni XXVII, which have peak formation temperatures (cf. Figure 4.4) of ~40 and ~50 MK, respectively (Mazzotta *et al.* 1998), and thus their presence in the spectrum is a clear indication of hot plasma.

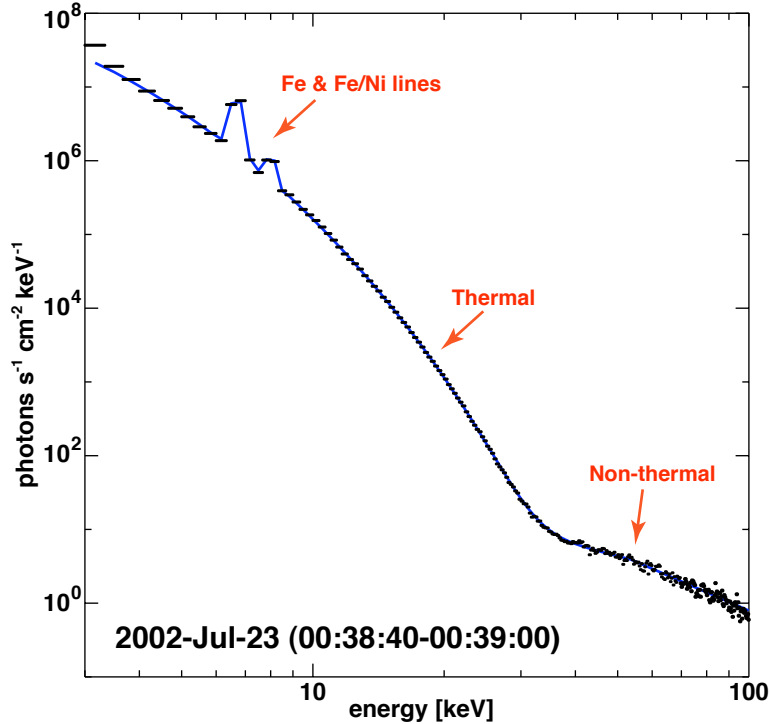

**Figure 4.3** – A typical photon spectrum observed by RHESSI during Jul 23, showing a rapidly-decreasing SXR continuum, an HXR continuum, and the Fe & Fe/Ni line complexes. The SXR continuum is exponential, typical of thermal emission; the HXR continuum is a power-law, typical of non-thermal bremsstrahlung.

At temperatures of ~20-50 MK, the individual lines cluster within an energy interval of only ~0.15 keV, which RHESSI cannot resolve; the instrumental response smears the individual lines into two quasi-Gaussian features. Nevertheless, the integrated flux of each complex, like that of the individual lines, is strongly dependent on temperature; Phillips (2004) used the CHIANTI atomic X-ray spectral database (ver. 5.2; Landi *et al.* 2006) to predict the emissivity (flux per unit emission measure) of the Fe and Fe/Ni line complexes as a function of continuum temperature, assuming an isothermal source (Figure 4.5). The ratio of the Fe and Fe/Ni line fluxes is also temperature-dependent, but is independent of the source emission measure and nearly independent of the elemental abundances (K. Phillips 2005, private communication); the line complexes may thus be used as diagnostics of the plasma temperature, separately from the bremsstrahlung continuum (see §4.4).

To investigate the evolution of the super-hot plasma, we analyzed time-series spectra to measure and track its temperature and emission measure. This was done using forward modeling, where a model incident photon spectrum is assumed and convolved with the instrument response to obtain a model observed spectrum, which is then compared with the true observations;

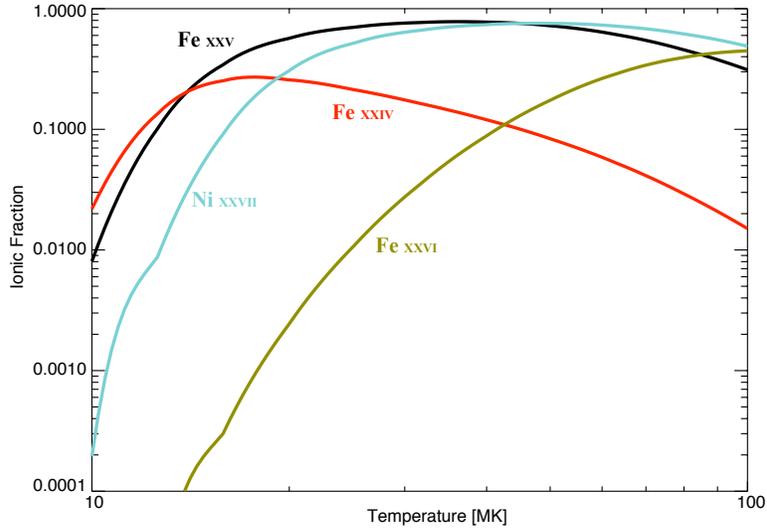

**Figure 4.4** – Ion population (as a fraction of the total population for that element) versus temperature for Fe XXIV-XXVI and Ni XXVII, for an isothermal plasma in equilibrium (based on Mazzotta *et al.* [1998]). Fe XXV and Ni XXVII are the primary contributors to the Fe & Fe/Ni line complexes observed by RHESSI.

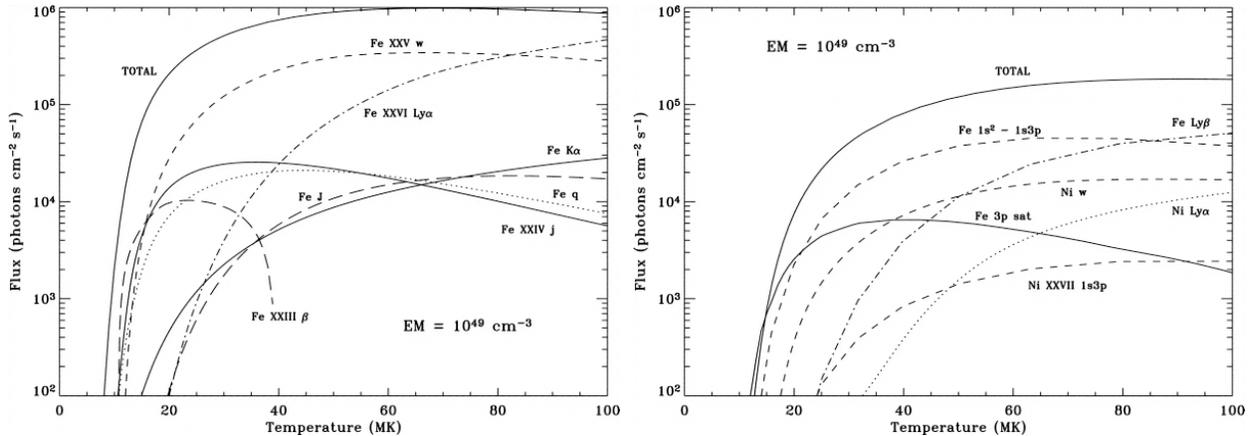

**Figure 4.5** – Integrated fluxes of the Fe [left] and Fe/Ni [right] line complexes centered at ~6.7 and ~8 keV, respectively, and their primary constituent lines, as a function of isothermal temperature. The ratio of the lines is strongly temperature-dependent (from Phillips [2004]).

the model parameters are varied until a good fit is obtained.

### 4.2.1 Instrument calibration

It was initially noted that the default instrument calibration parameters yielded significant differences in the model fits before and after a change in the attenuator state (see §3.2.1), e.g. from A3 to A1. If the true incident spectrum is changing slowly – a reasonable assumption during the late decay phase – then these differences must be due primarily to disagreements in the instrument response parameters (see §3.2.3) used for each respective attenuator state. Below

~6 keV, the model spectra in either attenuator state were inconsistent with observations from other instruments, e.g. GOES, suggesting an additional shared error in the response. Hence, to maximize the accuracy of spectral analysis, it was necessary to first improve upon the default calibration parameters. The calibration procedure and results are summarized below; for a full discussion with supporting figures, see Appendix A.

To simplify the analysis by avoiding detector cross-calibration issues, we used only a single GeD, detector G4, which had the best nominal front-segment resolution (~0.98 keV FWHM) based on ground calibration. In either the A1 or A3 attenuator states, the actual resolution was found to be ~0.75 keV FWHM at ~7 keV (§A.2.4), determined by analysis of the Fe line complex, since with either attenuator inserted, the low-energy photons are then detected in the center of the GeD where the electric field is strongest and charge collection fastest (see Figure 3.2).

To establish a baseline attenuator calibration, RHESSI A1 spectra for a few flares were compared with simultaneous observations from the *Solar X-Ray Spectrometer* (SOXS), a Sun-observing silicon *p-i-n* detector on an Indian satellite (Jain *et al.* 2005). The usable overlap range of RHESSI A1 and SOXS observations is ~6-12 keV; above ~12 keV, SOXS suffers from uncorrected pulse pile-up that distorts the spectral shape (R. Schwartz 2006, private communication), while below ~6 keV, RHESSI's direct sensitivity in the A1 state is too small – its response is dominated by off-diagonal contributions that overwhelm the direct flux to which SOXS is still sensitive. Within this range, however, the incident photon spectra inferred from RHESSI and SOXS agreed to within ~5-10% (Figure A.1), well within the uncertainties of the SOXS instrument response.

Thus, the default attenuator response for the A1 state was taken as confirmed. The final transition between the A3 and A1 states, during the decay of Jul 23, was used to cross-calibrate the A3 attenuator response against the A1 baseline (§A.2.1), as well as to optimize the software pile-up correction (§A.2.2) – the thick attenuator contribution to the DRM and the pile-up parameters were varied until good agreement was achieved between the inferred spectra from the A1 and A3 observations. Below ~5 keV, where the A1 and A3 attenuation of direct photons is greater than $10^5$ and $10^7$, respectively, the response is dominated by the K-escape contribution (§3.2.3), which appeared to be underestimated – even using the improved calibration from above, the observed count fluxes at ~4-5 keV in both attenuator states were significantly higher than would be expected from the inferred ~14-15 keV incident photon flux using the default K-escape calibration; the K-escape contribution was therefore increased (§A.2.3) to obtain good agreement and eliminate this discrepancy.

### 4.2.2 Spectral analysis with forward-modeling

For spectral analysis, we accumulated the observed 3-100 keV counts from detector G4 using 1/3-keV energy binning – the nominal instrumental channel width – and 4-second time bins – the spacecraft spin period – for the entire period of ~00:20 to ~01:15 UT on 2002 July 23. Multiple time bins were then grouped into 20- to 240-second intervals to optimize counting statistics. At each interval, we initially fit the spectrum from ~10-100 keV with a photon emission model consisting of a super-hot isothermal continuum and a non-thermal continuum, convolved with the instrument response. The free model parameters (see below) were optimized by iterative chi-squared minimization using the *Object Spectral Executive* (OSPEX) package using the improved calibration, with the dynamic pileup and energy offset corrections (§3.2.3) applied at each interval; for full details on OSPEX and the forward-modeling procedure, see Appendix B.

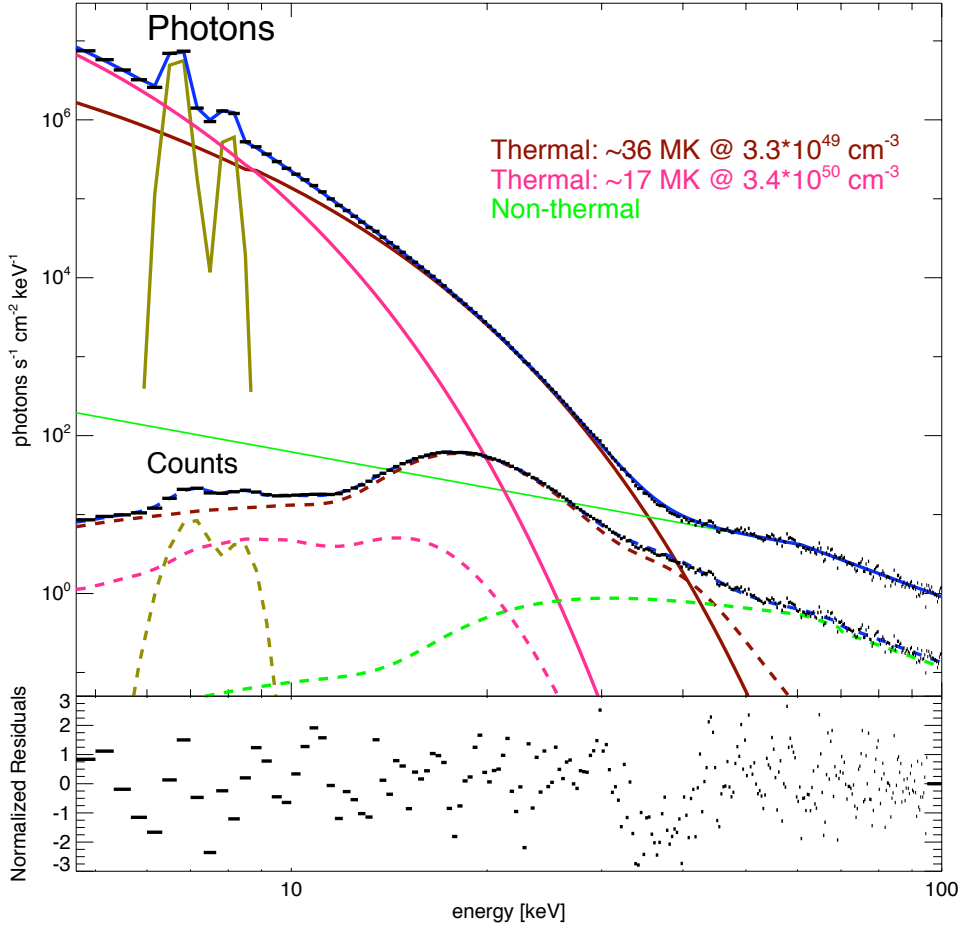

**Figure 4.6** – Example RHESSI count spectrum and corresponding inferred photon spectrum during the impulsive phase of Jul 23, with final model fits and normalized residuals (using 1/3-keV energy bins). The total model (blue) is the sum of super-hot (brick red) and cooler (magenta) isothermal continua, a non-thermal power-law continuum with low-energy electron cutoff (green), and the Fe and Fe/Ni line complexes (mustard). Residuals are normalized by the uncertainty for each energy bin.

The isothermal was modeled using the CHIANTI code with coronal (versus photospheric) values for the elemental abundances; the spectral shape follows equation (2.3), where the intensity at photon energy $\varepsilon$ is:

$$I(\varepsilon; T_e, Q) \propto Q g(\varepsilon, T_e) \frac{\exp(-\varepsilon/k_B T_e)}{\varepsilon \sqrt{T_e}} \quad \text{(photons / s / cm}^2 \text{ / keV)} \tag{4.1}$$

and the electron temperature $T_e$ and volume emission measure $Q$ are taken as free fit parameters. For further simplicity, we assume that the source has uniform electron number density $n_e$ over its volume $V$, thus $Q = n_e^2 V$. The Gaunt factor $g$ includes contributions from both free-free (bremsstrahlung) and free-bound (radiative recombination) interactions, including the dependence on the elemental abundances. The non-thermal continuum was modeled by a power-law, or a double power-law where needed, with a low-energy electron cutoff corresponding to a low-energy break in the photon spectrum (§2.1.1):

$$I(\varepsilon; A, \gamma_L, \varepsilon_C, \gamma_H, \varepsilon_B) = \begin{cases} A\varepsilon^{-1.5} & \varepsilon \le \varepsilon_C \\ A\varepsilon_C^{-1.5+\gamma_L}\varepsilon^{-\gamma_L} & \varepsilon_C \le \varepsilon \le \varepsilon_B \\ A\varepsilon_C^{-1.5+\gamma_L}\varepsilon_B^{-\gamma_L+\gamma_H}\varepsilon^{-\gamma_H} & \varepsilon_B \le \varepsilon \end{cases} \quad \text{(photons / s / cm}^2\text{ / keV)} \quad (4.2)$$

with the normalization $A$, low-energy cutoff $\varepsilon_C$, spectral index $\gamma_L$ and, where needed, upper break energy $\varepsilon_B$ and spectral index $\gamma_H$ taken as free fit parameters. The spectral indices are defined positive > 1.5; the low-energy cutoff is typically constrained only as an upper bound due to the dominant thermal emission at lower energies.

For most intervals, subtracting the best-fit model from the observations left a significant remaining continuum below ~15 keV, along with the Fe and Fe/Ni line complexes (Figure 4.7). The local (~5 to ~10 keV) residual continuum around the lines was well-fit by a power-law, and could therefore be accurately subtracted to isolate the line complexes, which were then modeled as Gaussian functions with intrinsic FWHMs of 0.15 keV centered at their mean energies of 6.680 and 8.015 keV, respectively (Phillips 2004), to obtain the integrated line fluxes.

The residual low-energy continuum decreased rapidly above ~10 keV, faster than at lower energies; it appeared to fit well to a cool isothermal. We therefore replaced the initial power-law fit with a second isothermal model. Holding the Gaussian line components fixed, as they were now well-determined, we refit the entire ~4.67-100 keV continuum with a revised total model, including both the initial super-hot and the new cool isothermal components plus the non-thermal power-law, throughout the flare. The fits obtained with this new model yielded reduced $\chi^2$ values of ~0.7 to ~2.4 (averaging ~1.4 over all intervals), leaving no significant residual continuum (Figure 4.6).

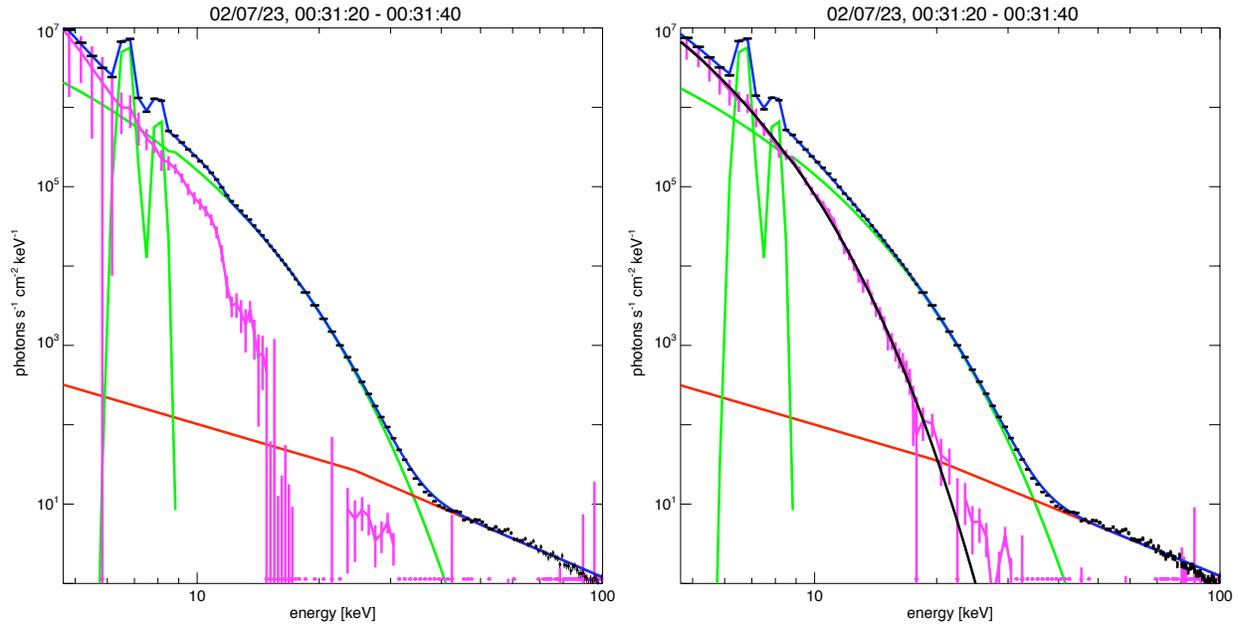

**Figure 4.7** – [left] Example RHESSI photon spectrum (black) with initial model components of super-hot isothermal (green) and non-thermal power-law (red) continua, plus Fe & Fe/Ni lines (also green). The residual continuum (fuchsia) drops off sharply with energy and suggests a second, cool isothermal. [right] The same photon spectrum fit with an expanded model that includes a cooler isothermal; the super-hot and non-thermal continua were also refit. The new residual continuum (data minus super-hot and non-thermal continua) matches the cool continuum well.

### 4.2.3 Imaging analysis

RHESSI images offer vital spatial information to complement the spatially-integrated spectra. Examining the images as a function of energy reveals that the two isothermal plasmas are also spatially distinct. For example, during the RHESSI SXR peak at ~00:31:30 UT, we accumulated 20-second images at 6.3-7.3 keV, 9-12 keV, and 17-18 keV for the thermal emission, as well as 60-100 keV for the non-thermal emission (Figure 4.8). These show that the non-thermal emission is from three compact footpoint sources, while the thermal emission is from a large source higher in the corona. The image contours show that the thermal emission in the three energy bands overlaps but with some displacement, with higher-energy emission farther from the footpoints and thus higher in the corona.

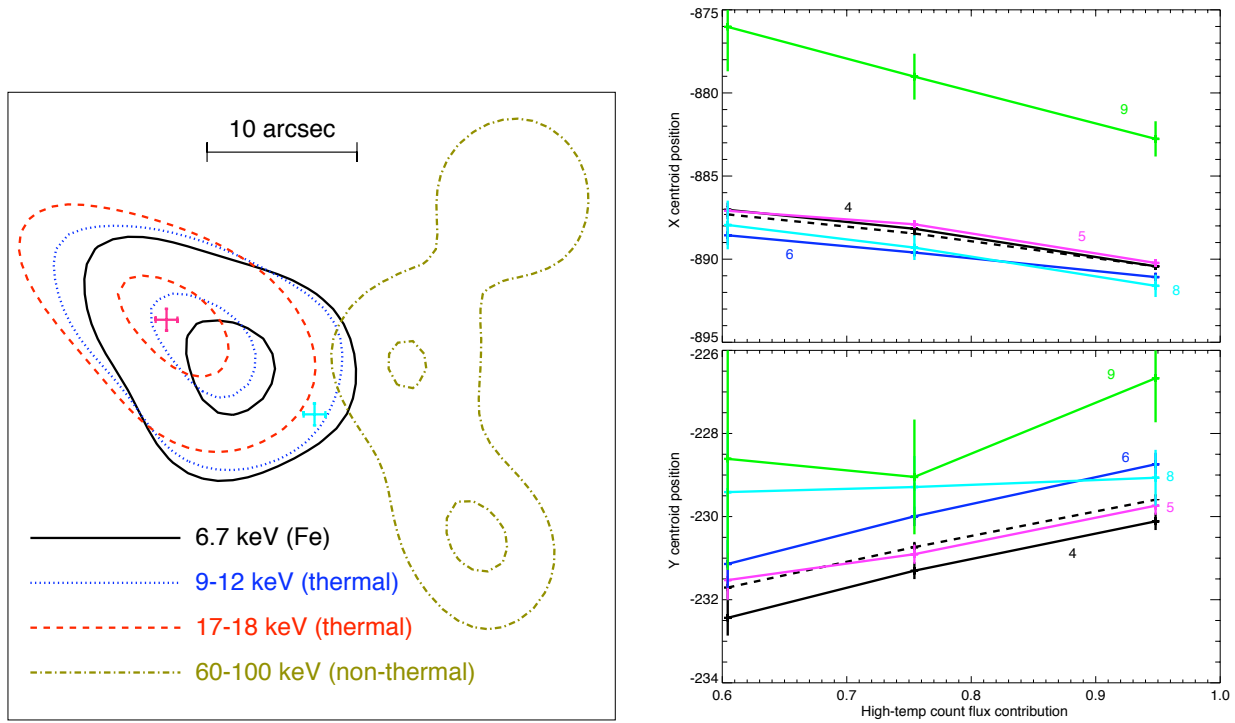

**Figure 4.8** – [left] RHESSI images (50% and 90% contours) for Jul 23 in various energy bands at ~00:31:30 UT, using grids 3-9 (excl. 7), CLEAN, and uniform weighting. The crosses indicate the derived centroid positions and uncertainties (see right) for the super-hot (magenta) and cool (cyan) thermal sources, respectively. [right] Measured centroid X & Y positions, with uncertainties, for grids 4-9 (excl. 7) for the three thermal energy bands, which each contain a different fractional contribution from the super-hot component (determined from spectral modeling); the dashed curve is the weighted average of the measurements from the 5 grids, and is linear with the super-hot fractional contribution. The intercepts at x = 1 and x = 0 therefore represent the extrapolated centroid positions for the super-hot and cool components, respectively.

From the spectral model fit for this interval, we can determine the fractional contribution of the super-hot and cool components to the total counts (for the Fe line, which was fit separately from the continuum, the fractional contribution was determined using the ratio of the predicted Fe line emissivity [Phillips 2004] for the two components, based on their fit temperatures and emission measures), suggesting that the super-hot component contributes ~63%, ~76%, and ~95% of the counts at 6.3-7.3, 9-12, and 17-18 keV, respectively. The centroid position (first moment of intensity) of the emission at these energies varies linearly (with $\chi^2 < 1$) with the fractional count contribution of the super-hot component, and is thus consistent with each centroid position being a weighted average of the centroids of two separate sources; the slope of this linear dependence corresponds to the centroid separation of the two sources and is found to be ~11.7 ± ~0.73 arcsec, showing that the two sources are significantly separated.

A more advanced analysis using visibilities (§3.3.4) allows us to decompose the two sources and actually image them separately, revealing not just the source separation but also their morphologies. Visibilities are complex numbers that are, in essence, the spatial Fourier components;

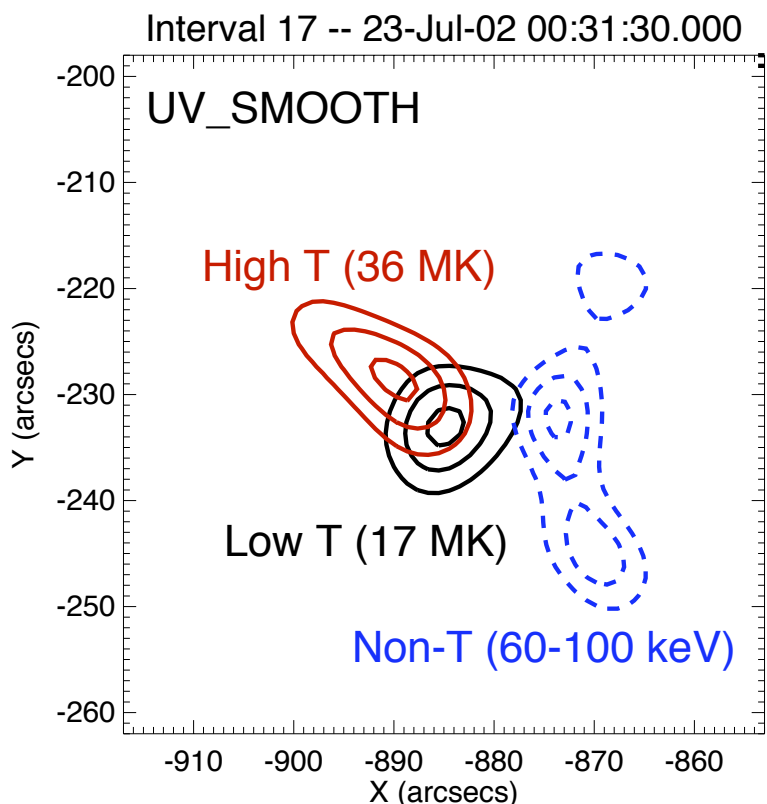

**Figure 4.9** – Images (50%, 75%, 90% contours) of the super-hot and cool isothermal sources derived from linear combinations of the 6.3-7.3 and 17-18 keV visibilities weighted by the fractional contribution from the respective thermal components (see Appendix C) for the same time period as Figure 4.8. The 60-100 keV non-thermal footpoints are shown for reference.

they are inherently linear, and the observed visibilities can therefore be treated as the sum of the visibilities from each thermal source. Using the visibilities at two different energies and assuming that the respective fractional contributions of the two thermal sources at each energy are as determined from spectroscopy (in the same manner as was done for the centroid analysis, above), we can invert this relationship (see §C.4) to derive the individual source visibilities, from which we can then directly create images of the super-hot and cool sources individually. Applying this analysis to the SXR peak time, per above, reveals that the super-hot source is elongated and indeed well-separated from the rounder cool source, which is closer to the footpoints (Figure 4.9).

Images can also provide other quantitative measurements throughout the flare, in particular the thermal source volume (§C.2), which can be combined with the spectral information to derive the source's electron density and thermal energy. The 6.3-7.3 keV images show no significant overlap with the 60-100 keV HXR footpoints, suggesting little or no thermal emission from the footpoints. Images at 6.2-8.5 keV (to improve statistics) were therefore accumulated using the CLEAN algorithm (Hurford *et al.* 2002) with uniform weighting (§3.3.3) and ~1-minute cadence for the entire flare; the sources were roughly elliptical throughout the flare, and so they were manually measured to determine the approximate length *2a* and width *2b* of the 50% contour, which represents the FWHM of the source emission. The width and length were corrected to compensate for broadening from the point-spread function, a consequence of the instrument. The source is assumed to have an ellipsoidal geometry, with the (unseen) depth equal to the short dimension *2b*; thus, the volume $V = (4/3)\,\pi ab^2$. When tested on simulated elliptical Gaussian sources, this measurement method yields a ~7% uncertainty in *a* and *b*, and thus a ~23% uncertainty in *V* (subject to the assumption of source geometry, which is likely the biggest source of uncertainty). The visibility source-decomposition analysis is not yet sufficiently robust to yield precise quantitative measurements for the individual sources, but shows that the sources have areas of the same general magnitude; thus, the same *V* is assumed for both the super-hot and cool plasmas, from which we then derive their electron densities $n_e = \sqrt{Q/V}$ and energies $E_{th} = (3/2)n_e V k_B T_e$. We note that these quantities vary as $V^{\pm 1/2}$, so they are not very sensitive to uncertainties in the volume estimate – a ~20% error in *V* yields only a ~10% error in $n_e$ and $E_{th}$.

Throughout the spectral and imaging analysis, we ignored the effect of Compton backscatter of coronal X-rays from the photosphere ("albedo," cf. §2.2 & §3.2.3) on the spectral shape and the image morphology. This is justified by the flare's ~73° heliocentric angle, which reduces the albedo contribution – already a second-order effect – by a factor of cos(73°) ≈ 0.3. Because albedo is a Compton scattering process, its effect can be incorporated into the instrument response matrix in the same way as Compton scattering within the spacecraft; the albedo contribution to the response depends primarily on the flare heliocentric angle and on the directionality of the flare HXR emission (Kontar *et al.* 2006). If we re-examine the spectrum during the SXR peak (Figure 4.6) and this time include the effect of albedo from an isotropically-emitting source in the response matrix, the subsequent model fit yields only small (≲10%) changes in both the super-hot and cool continuum temperatures and in the cool emission measure, although the super-hot emission measure, which is more sensitive to changes in the spectral shape at higher energies, does drop by 33%. However, the super-hot density and energy change by only ~18%, and the derived centroid separation of the super-hot and cool sources changes by only ~15% (~2 sigma) to ~9.9 ± ~0.63 arcsec, so our results are nevertheless not significantly affected by ignoring albedo.

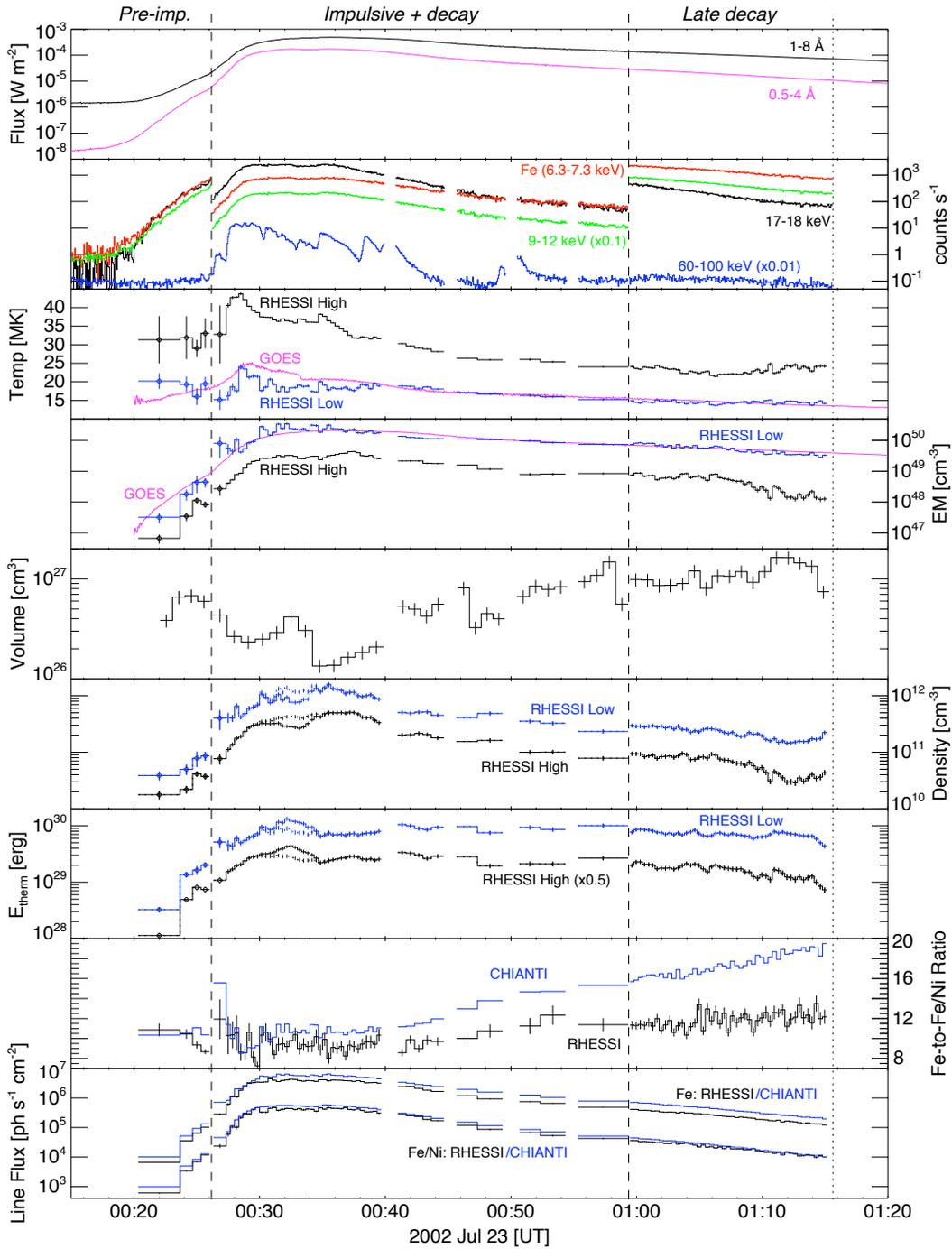

**Figure 4.10** – Observational results for Jul 23 determined from spectral modeling and image analysis. Error bars – for temperature, emission measure, & line flux: as reported by OSPEX (see Appendix B); for volume, a uniform 23% (see §C.2); for density, thermal energy, & line ratio: propagated from above. During the early times (diamonds), the temperature & emission measure could be determined only within upper & lower bounds (see §4.3); the diamonds show the mean value, while the error bars show the limits.

## 4.3 Observational results

Figure 4.10 shows the evolution of the temperatures, emission measures, and other derived quantities for the super-hot and cool sources, determined using the spectral and imaging analyses. The cool temperature and emission measure exhibit large fluctuations over short timescales (~20-60 sec), e.g. around the temperature peak at ~00:29-00:33 UT; these fluctuations are often anti-correlated and are thus likely artifacts of fitting. However, the underlying trend is consistent with the data – replacing the fit cool temperature and emission measure with 3- or 5-interval boxcar-smoothed values yields equally acceptable fits to the spectra. We note that the cool continuum contributes only ~10-20% of the total low-energy counts, yielding large uncertainties in its fit parameters.

The unusual variation in thermal volume during ~00:30-00:35 UT occurs when two small, spatially-distinct sources (one visible prior to ~00:32 UT, the other after ~00:34 UT) are simultaneously bright (Figure 4.11). The SXR flux and fit emission measures do not change signifi-cantly during this period, suggesting that one source dims as the other brightens, with equal brightness at ~00:33 UT (the volume maximum) when a significant co-spatial non-thermal coronal source also appears briefly; if this non-thermal source is the signature of intense particle acceleration from the onset of magnetic reconnection (cf. §1.3), this would suggest that a new region of the loop arcade may have "turned on," and the simultaneous thermal sources may thus be, respectively, the fading emission from the cooling plasma in the old loop and the brightening emission from the heating plasma in the newly-active loop. Because the density and thermal energy as we have derived them are only well-defined for a single source, we therefore define the "single-source volume" during this period as the linear interpolation between the two minimum individual source volumes (at ~00:32 and ~00:34 UT), consistent with the interpretation of one fading and one brightening source, and thus obtain corrected "single-source" densities and energies during this period (shown as the dotted lines in Figure 4.10).

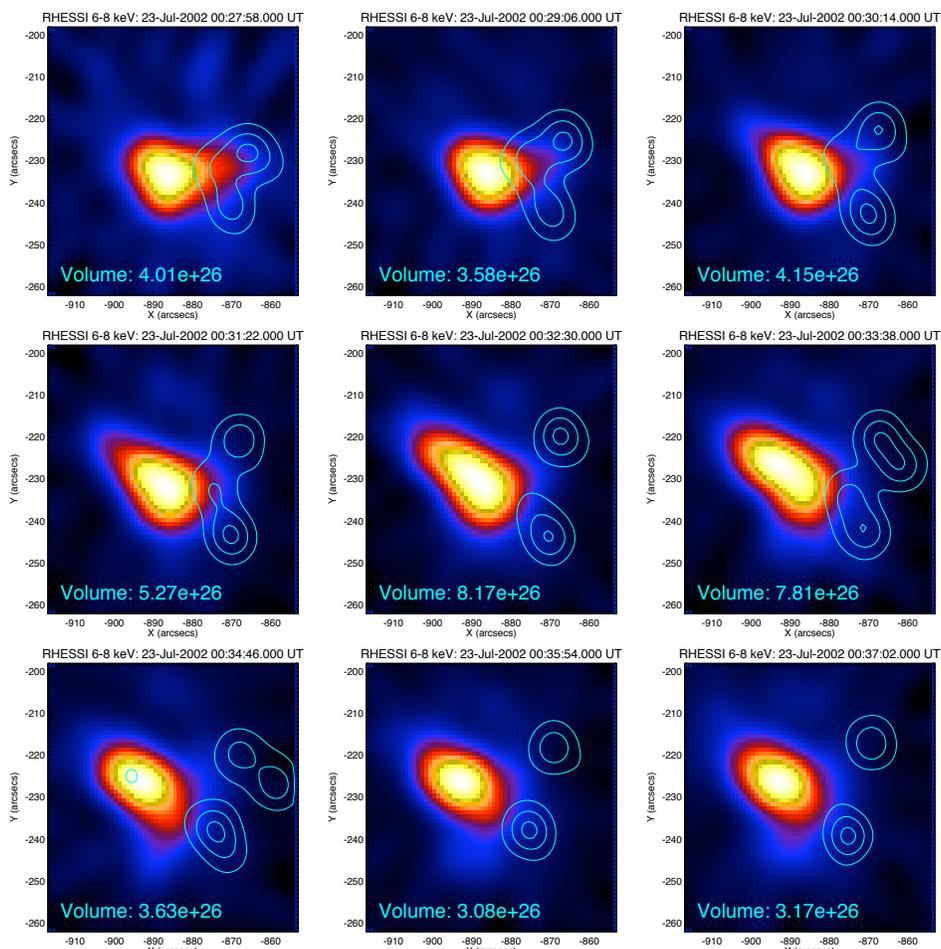

**Figure 4.11** – RHESSI 6.2-8.5 keV images (with 60-100 keV contours for reference) during ~00:30-00:35 UT. The initial compact source appears to elongate, then shrink to a different position; the intermediate images are consistent with two separate sources (the ones visible before and after) being simultaneously bright. A single volume measurement is not applicable, as the two sources are not separately resolved.

As the super-hot continuum temperature increases, the line fluxes generally also increase while the Fe-to-Fe/Ni line flux ratio decreases, steeply below ~25 MK and more gradually above (Figure 4.12). This agrees qualitatively with the CHIANTI-based predictions (Phillips 2004) from the two observed isothermal continua, but quantitatively, the Fe and Fe/Ni line fluxes and their ratio are significantly smaller – by, on average, ~55%, ~20%, and ~34%, respectively, with larger deviations at lower temperatures. The ionization timescales (cf. Jordan 1970; Phillips 2004) for Fe XXV and Ni XXVII – the primary line contributors – at the measured temperatures and densities are generally below 1 sec and never exceed ~13 sec; this is much shorter than the temperature change timescale, $T(\partial T/\partial t)^{-1}$, which always exceeds ~130 sec (often by an order of magnitude), suggesting that ionization equilibrium is always maintained. Including the effect of isotropic-source albedo into the response matrix, per §4.2.2, changes the CHIANTI-predicted fluxes by only ~2%, with negligible change in the line ratio, and so also cannot explain the overall disagreement. Phillips *et al.* (2006) suggested a potential inaccuracy in CHIANTI's Fe ionization fractions that may account for these systematic discrepancies, although the root cause remains undetermined.

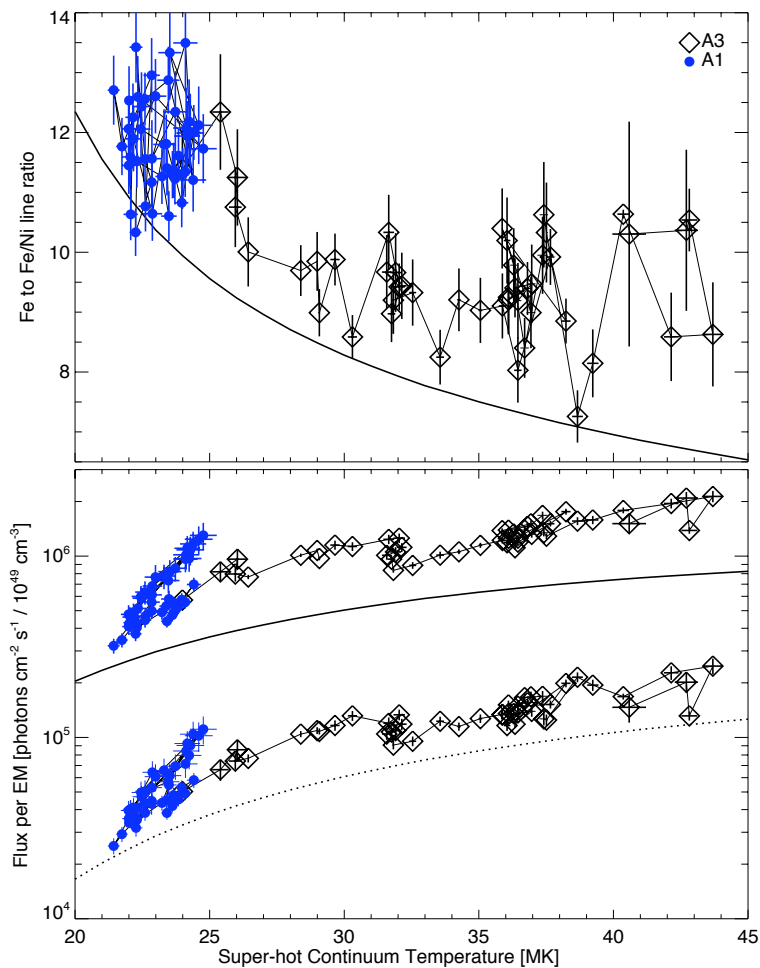

**Figure 4.12** – [top] Measured ratio of the Fe and Fe/Ni line integrated fluxes versus measured super-hot continuum temperature, along with predicted (isothermal) curve from Phillips (2004). [bottom] As above, for the absolute line fluxes, normalized per $10^{49}$ cm$^{-3}$ by dividing out the fit emission measure of the super-hot plasma.

During the pre-impulsive phase, when the HXR emission is dominated by an apparently-nonthermal coronal source, the continuum model fits are ambiguous – a wide range of thermal and non-thermal model parameters yield equally acceptable fits to the continuum spectrum (cf. Holman *et al.* 2003). The observation of the line complexes indicates that some thermal emission is present, although the continuum alone does not constrain the thermal model well. However, if we assume that the measured line fluxes and ratio share the same empirical relationship with the super-hot continuum temperature as observed during the rest of the flare (cf. Figure 4.12), the line observations during the pre-impulsive phase can thus constrain the thermal continuum model. At the pre-impulsive phase peak (~00:25:30 UT; see Figure 4.13), when faint footpoints are visible, we find using this method that the super-hot temperature is constrained to be between ~29 and ~37 MK; a cool component is also required by the observed spectrum, with a temperature between ~21 and ~18 MK, respectively. Earlier in the pre-impulsive phase, when no footpoints are discernible, the data still appear to require a cool component but are insufficient to indicate a large separation in temperatures, so the cooler component may not be well-defined.

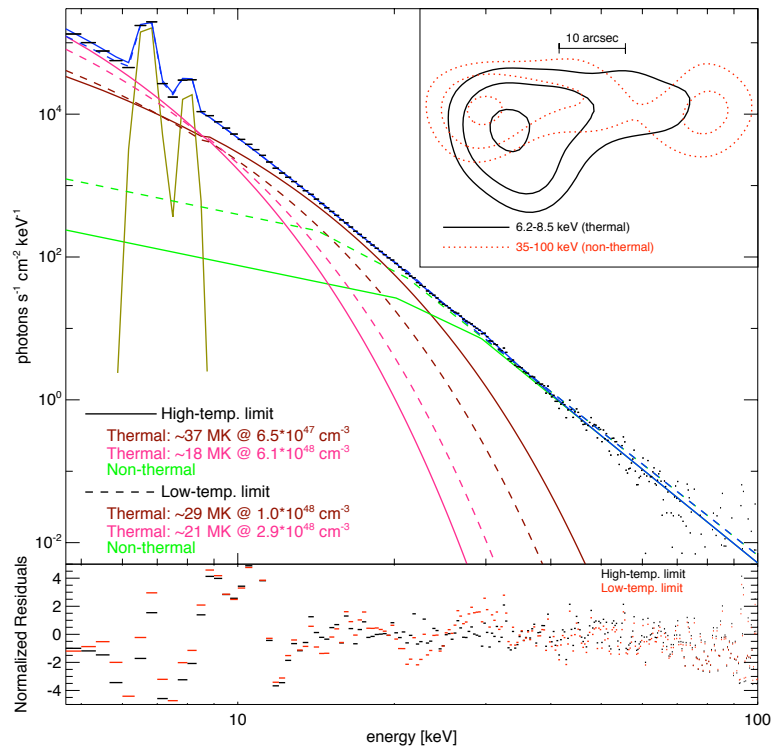

**Figure 4.13** – Photon spectrum during the late pre-impulsive phase, with two possible model fits; the super-hot temperature and emission measure is fixed at the upper (solid) or lower (dashed) limits determined by the line fits, while the non-thermal and cool thermal component are varied freely to fit the rest of the spectrum. [inset] Images in the thermal and non-thermal energy ranges (contours at 30%, 50%, 80%).

In this way, we obtain the temporal evolution of the super-hot and cool thermal plasmas throughout the Jul 23 flare. The super-hot component is observed at ≳25 MK from the very start of the flare at ~00:22 UT and reaches a peak temperature of ~44 MK at the non-thermal HXR (60-100 keV) peak (~00:28:30 UT). The super-hot emission measure at that time is only ~20% of the later peak value of ~4.4×10$^{49}$ cm$^{-3}$, reached at ~00:37:30 UT. The temperature decreases rapidly after ~00:35-00:37 UT, when the HXR emission drops by a factor of ~10; it drops below 30 MK at ~00:42 UT, when the HXR emission nears background levels, and reaches a minimum of ~21 MK during the flare decay at ~01:06:18 UT. The total super-hot thermal energy, however, decreases relatively slowly, dropping by only a factor of ~4.6 over the ~33 minutes following its maximum at ~00:41:20 UT. Images using the visibility decomposition method show that throughout the flare, the super-hot source is farther from the footpoints than the cool source and, at times (e.g. ~00:31-00:34 UT), it is elongated up to ~2× in that direction; its separation from the footpoints increases over time.

The cool plasma is present at ≲18 MK at least as early as ~00:25 UT, when footpoints begin to be visible (though the line observations suggest that the plasma may be present from the beginning of the flare). It achieves a peak temperature of ~24 MK ~1 min after the super-hot temperature peak, decaying relatively slowly thereafter. The cool isothermal is always significantly (~7.5-25 MK) colder than the super-hot component, and its temperature varies within a much narrower range (~13-24 MK). It is noteworthy that, without any *a priori* assumptions, the best-fit cool temperature and emission measure agree closely with those derived from GOES – to within ~5%, and ~20%, respectively – except before ~00:37:48 UT, when the super-hot and non-thermal emission are intense and thus likely contaminate the GOES measurements.

Both the super-hot and cool thermal sources show evidence of the Neupert effect, in that the flux of the thermal emission rises similarly to the time-integral of the power contained in the non-thermal electrons, although the thermal flux appears to rise more gradually, suggesting that not all of the non-thermal power goes into heating the plasma. Both thermal sources also appear to move progressively further from the footpoints over time.

### 4.4 Discussion

The spectra and images show that, for the Jul 23 flare, two thermal components exist simultaneously: the usual ~10-20 MK plasma normally detected by GOES, and a distinct super-hot component; both sources are well-approximated as isothermal. The imaging geometry suggests that the super-hot component is likely at higher altitude than the cool plasma. Sui & Holman (2003) also observed higher energy/temperature emission at higher altitudes, along with the reverse behavior (decreasing energy/temperature with increasing height) above, interpreting this as thermal plasma in layered reconnecting loops and hence evidence of magnetic reconnection.

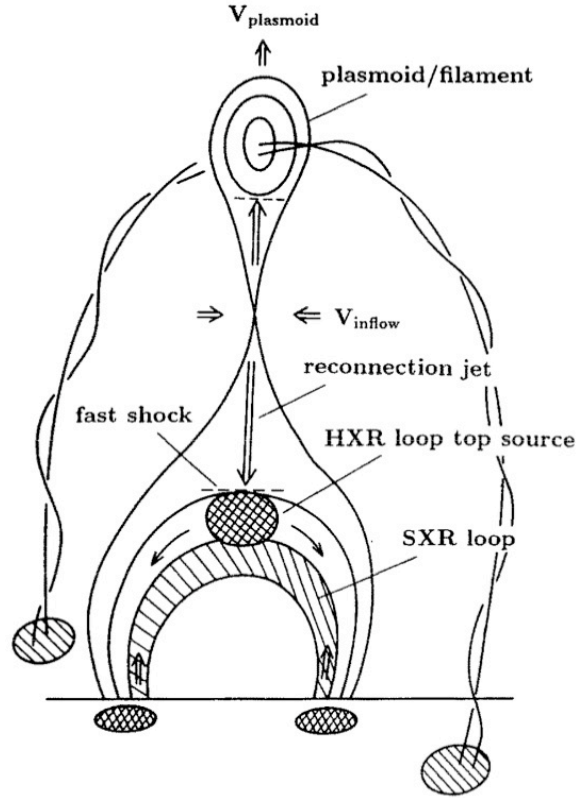

**Figure 4.14** – Cartoon model of flare energy release via magnetic reconnection, from Shibata (1996). Interpreting our results in this context, the super-hot component corresponds to the HXR looptop source, heated by reconnection outflows and subsequent loop compression, while the cool component corresponds to the SXR loops, full of evaporated chromospheric material.

While we do not observe the higher temperature-inverted sources, the relatively small (factor of ~10) dynamic range of RHESSI imaging limits our sensitivity, if these sources are much fainter than the observed ones. When the super-hot source is elongated, its elongation also appears to be towards higher altitudes, although it may alternatively be along the top of the loop arcade later observed in post-flare TRACE images; in either case, the super-hot source appears to progress further from the footpoints and suggests an association with newly-reconnecting loops.

Although we have no direct observations of the reconnection region, we may speculate that it lies above the super-hot plasma, in the direction of its elongation. Then, interpreting our observations in the context of the Shibata (1996) reconnection flare model (Figure 4.14), the super-hot plasma resides at the top of a cusp-shaped reconnecting loop and corresponds to the above-the-looptop HXR source, heated directly by the reconnection outflow, as first suggested by Lin *et al.* (2003). The cooler, denser, lower-altitude GOES plasma corresponds to the model SXR source and resides in a previously-reconnected loop, formed as the hot plasma, carried down by the reconnected magnetic field as it relaxes into a more dipolar configuration, mixes with cooler chromospheric material evaporated into the loop (as suggested by the rising densities) by the impacting accelerated electrons.

This scenario remains consistent even during the pre-impulsive phase, when the non-thermal HXR source substantially overlaps the coronal thermal source both spectrally and spatially. Dur-ing the peak pre-impulsive period (Figure 4.13), for the (line-constrained) ~29-MK limit of the super-hot temperature, the total (super-hot plus cool) thermal density is ~$1.1 \times 10^{11}$ cm$^{-3}$, yielding ~$6.6 \times 10^{37}$ thermal electrons with a total energy of ~$3.3 \times 10^{29}$ ergs. Non-thermal emission dominates the spectrum down to ~15 keV, and is well-fit by bremsstrahlung from a power-law electron spectrum with a sharp low-energy cutoff. Assuming that the ambient density in the non-thermal source is no greater than the thermal density, the instantaneous number of non-thermal electrons deduced from the spectrum is $\geq 1.0 \times 10^{35}$ with a total energy of $\geq 4.2 \times 10^{27}$ ergs. For the ~37-MK limit, these numbers are, respectively: ~$1.3 \times 10^{11}$ cm$^{-3}$, ~$7.9 \times 10^{37}$ thermal electrons, ~$3.7 \times 10^{29}$ ergs, ~23 keV, $\geq 1.7 \times 10^{34}$ non-thermal electrons, and $\geq 1.0 \times 10^{27}$ ergs. Images at 6.2-8.5 and 35-100 keV (Figure 4.13, inset) during this time show that the thermal and non-thermal sources are offset by ~$4.1 \pm 0.54$ arcsec; the geometry suggests that the peak non-thermal emission is above the thermal loop. The faint footpoint contains only ~16% of the total non-thermal flux within the 50% contour (~29% of the total flux within the 30% contour), suggesting that most of the non-thermal electrons never reach the chromosphere – they may be trapped in the corona, or may escape the flare site entirely. The instantaneous thermal energy is ~79 (~29-MK limit) to ~370 (~37-MK limit) times greater than the total instantaneous energy in non-thermal electrons; if the thermal loop is filled by chromospheric plasma evaporated by the few non-thermal electrons that reach the footpoint, this large energy difference suggests that the non-thermal electrons are replenished via continuous injection (although if the ambient density in the non-thermal source is much smaller than assumed, the number of non-thermal electrons – and hence non-thermal energy – would be proportionately larger). At the measured thermal densities, the collisional energy loss time (cf. equation [2.4]) for 20-100 keV non-thermal electrons is only ~0.03 to ~0.3 seconds. If we assume a thick-target model for the bremsstrahlung emission (cf. Brown 1971), the spectrum suggests that the total energy deposited by non-thermal electrons during the entire pre-impulsive period is $\geq 1.0 \times 10^{31}$ ergs for the ~29-MK super-hot limit, somewhat smaller than the lower limit of ~$1.7 \times 10^{31}$ ergs estimated by Holman *et al.* (2003); for the ~37-MK super-hot temperature, however, the energy deposition drops to only $\geq 2.4 \times 10^{30}$ ergs. In both cases, these numbers are significantly smaller than the upper limit of ~$4 \times 10^{32}$ ergs determined by assuming that the non-thermal electrons extend down to ~10 keV (cf. Lin *et al.* 2003).

# Chapter 5: Observations of other flares

The 2002 July 23 flare is an excellent case study for the temporal and spatial evolution of super-hot plasma. However, this single flare cannot provide insight into how common super-hot plasma is, in general, nor can it answer whether the existence of super-hot plasma is correlated with other flare attributes such as energy or X-ray intensity. To determine these more global parameters, a statistical survey of numerous flares is required.

The discovery of non-thermal coronal sources that dominate pre-impulsive flare emission is a triumph of RHESSI's design; the choice to use movable rather than fixed attenuators provides the dynamic range necessary to observe these faint sources from their very inception, while RHESSI's good energy and angular resolution allows their spectra and position to be precisely characterized. The pre-impulsive phase of Jul 23 is a particularly intriguing time, as the thermal source density is already well above ambient values but there is little to no discernible footpoint emission, signifying that chromospheric evaporation is unlikely to be significant; it is therefore an open question as to how the density can be so high, something we explore further in Chapter 6. With only one observed instance of pre-impulsive emission – in an X-class flare reaching super-hot temperatures – one is tempted to wonder whether pre-impulsive emission is present only for X-class flares and/or is a precursor to super-hot temperatures, or whether it exists in all flares but has not been observed due to sensitivity issues. Additional examples of pre-impulsive emission would help to answer these questions.

## 5.1 Statistical survey of super-hot plasma

Observations from both *Yohkoh* and *Hinotori* suggest that super-hot plasmas may be common among larger (M- and X-class) flares. The primary indications of super-hot temperatures were from Bragg crystal spectrometer (BCS) observations of Fe XXVI excitation lines, interpreted as evidence of temperatures above ~30 MK (see, e.g., Tanaka 1987; Pike *et al*. 1996). However, observations of such super-hot flares were rare in comparison to the total number of flares observed, and neither spacecraft included a high-resolution HXR spectrometer – *Yohkoh*'s HXT had only 4 energy channels from ~14-93 keV (Kosugi *et al*. 1991), while *Hinotori*'s HXM had only 7 channels from ~17-340 keV (Tanaka 1987) – that would allow them to resolve the steeply-dropping super-hot continuum emission to provide complementary temperature measurements. (*Hinotori* did include a gas scintillation detector, but its energy range was limited to no more than ~17 keV and it often saturated during large flares.)

Since its launch, RHESSI has observed over 500 M- and X-class flares – the larger flares that are presumably most likely to reach super-hot temperatures – with the high resolution needed for precise measurements of the thermal continuum. Its imaging capabilities allow determinations of source sizes, and hence densities, finally providing a single comprehensive data set for measurements of super-hot plasma.

To that end, we began a systematic statistical survey of M- and X-class flares observed by RHESSI, to determine the maximum temperature reached during those flares and relate that measurement to other flare attributes such as intensity or energy content. In particular, we sought to answer whether there is an intrinsic limit to the maximum achievable temperature, and if so, on what does it depend, as well as to determine whether super-hot flares are a unique class of flare or whether they are merely the high-temperature tail of a global distribution. The survey is not yet complete, but the latest results are presented below.

### 5.1.1 Flare selection criteria

For this survey, we wanted to focus on flares most likely to produce super-hot plasma. GOES class has been previously found to be roughly correlated with the temperatures derived from the GOES fluxes (e.g. Garcia & McIntosh 1992; Feldman *et al.* 1996). Thus, we restricted our analysis to only M- and X-class flares. Because the RHESSI detector performance began to degrade in 2006 due to accumulated radiation damage, we also chose only those flares that occurred between 2002 Feb 18 (the first observed flare after RHESSI's launch) and 2005 Dec 31; since the solar cycle was near minimum in 2006, this restriction excluded only a small fraction of RHESSI-observed flares.

To maximize the likelihood of observing the temperature peak, we required that the flare be well-observed, here defined as uninterrupted coverage of the GOES SXR peak and the entire preceding 10-minute interval. We further required that the RHESSI HXR (25-50 keV) and SXR (6-12 keV) peaks be contained within this 10-minute interval and that they both occur in order prior to the GOES SXR peak. (The 10-minute length was chosen to include most flares while reducing extraneous processing; the peak-order requirement eliminates extremely long flares for which the 10-minute interval is not sufficiently long.)

Because we needed a volume measurement to estimate the thermal electron density, we required that selected flares be imageable at the GOES SXR peak time (40-sec duration) in the 6 to

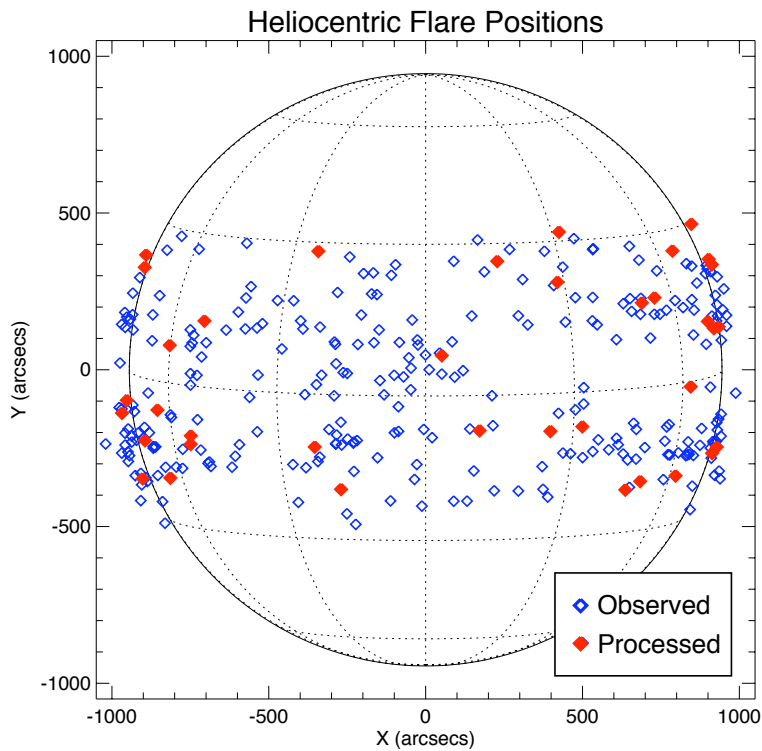

**Figure 5.1** – Synoptic map showing the heliocentric position of all flares chosen for the survey; the flares in red have been analyzed to date, and are presented here.

15 keV energy range, using grids 3 through 9 (excluding 7) and the CLEAN image reconstruction algorithm with uniform weighting (§3.3.3); here, "imageable" is defined simply as having the CLEAN algorithm work to completion without error, although 2 flares with clearly-identifiable imaging artifacts were manually culled. Finally, because we wanted to be able to compare all flares equally, we required that the time-series spectra over the 10-minute analysis interval be well-fit by the spectral model described below.

Given these criteria, 260 total flares – 234 M-class and 26 X-class – were selected for analysis; for this preliminary study, we completed spectral analysis for 37 (25 M-class, 12 X-class) of the selected flares, chosen simply in chronological order (all analyzed M-class flares occurred in 2002, while the X-class flares, being observed with lesser frequency, occurred from 2002 to 2004). The 260 selected flares and the analyzed subset were both distributed fairly randomly across the solar disk, though with some higher density nearer the limbs as expected from projection, so there was no overall position bias for this subset of flares (Figure 5.1).

### 5.1.2 Analysis methodology

Because of the large amount of data, the analysis was as largely automated as possible, though manually monitored at every step to ensure reliability of the results. Consequently, it was not possible to go through the detailed calibration process as performed for Jul 23 (see Appendix A); instead, the nominal calibration was used for the detector response and for pulse pileup, although observations in the A3 state were approximately corrected for the small calibration er-ror of the thick attenuator via the addition of a component to the photon model (rather than an adjustment to the response matrix, as was done for Jul 23).

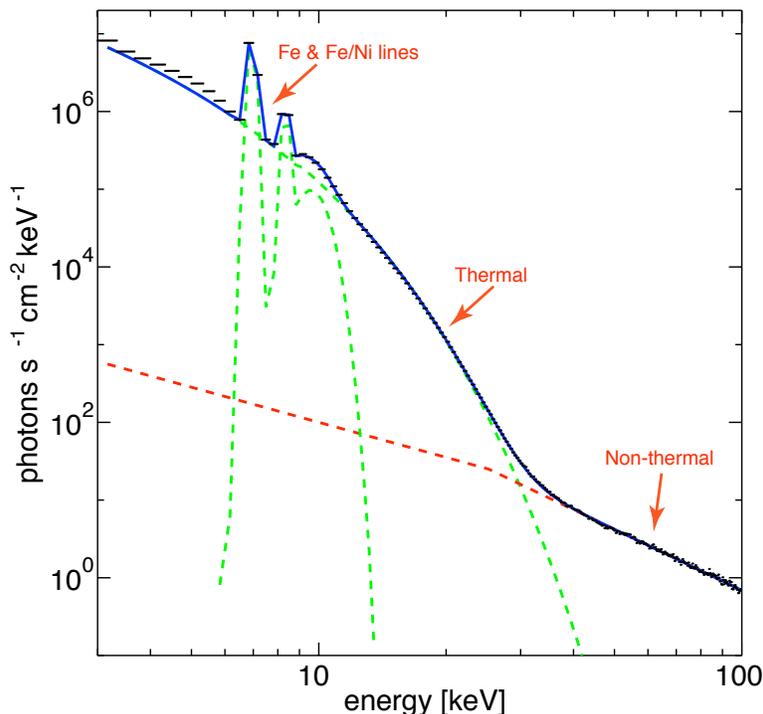

**Figure 5.2** – Example photon spectrum depicting the model fit used for the automated survey; the model included a single isothermal component, a non-thermal power-law, and the Fe & Fe/Ni line complexes. A third feature at ~10 keV was added to the model to compensate for the calibration error of the thick attenuator, since the many individual flares could not be manually calibrated beforehand.

For each selected flare, forward-modeling spectral analysis was performed with OSPEX (see Appendix B). The accumulated 10-minute observation period was partitioned into 20-second intervals; any intervals that spanned an attenuator transition were ignored to prevent mixed-state observations, where the detector response matrix is not well-defined. To maximize statistics, we used spectra averaged over all detectors except 2 and 7 (neither of which is useful for low-energy spectroscopy; see §3.2.1). The non-solar background was subtracted using the same procedure as for Jul 23 (§B.2.1). At each 20-second interval, we used OSPEX to forward-fit a photon model including a single isothermal continuum, a non-thermal power-law continuum, and two Gaussian functions for the Fe and Fe/Ni lines (cf. Figure 5.2). The starting parameter values for the first fit were set manually; each subsequent fit was initialized with the best-fit values from the previous interval. For any interval, if the model failed to converge or if the reduced chi-square for the best fit was greater than 4, the flare is not well-defined. This occurred for only 3 flares, not included in the analyzed sample of 37.

After achieving reasonable fits at each interval, the interval with the maximum isothermal continuum temperature was identified; focusing solely on this one interval per flare allows an equal comparison between flares, regardless of their duration or temporal variations. (This also reduces the overall data volume, to make the final results less unwieldy.) We then generated an image at this time in the 6-15 keV energy range, which was without exception dominated by the thermal component. Just as was done for flare selection, the image duration was 40 seconds, using grids 3 through 9 (excluding 7), CLEAN, and uniform weighting. The thermal source volume was then approximated by calculating the area within the 50% contour, corrected for broad-ening by the point-spread function, and extrapolated to a volume assuming spherical symmetry, $V = (4/3) \pi (A/\pi)^{3/2}$; this was done entirely automatically (see §C.2 for full details).

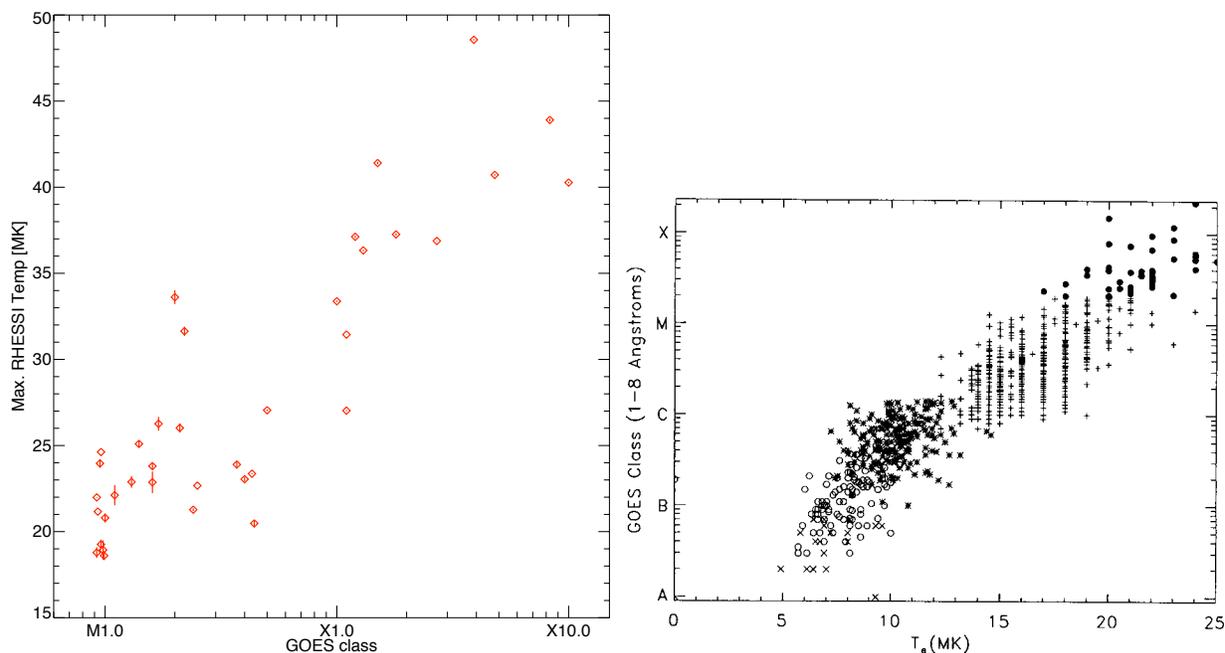

**Figure 5.3** – [left] Maximum measured isothermal continuum temperature versus GOES class for the 37 analyzed flares. [right] Maximum temperature versus GOES class (plotted in the inverse sense), measured using ratios of Si & Ca excitation lines (from Feldman *et al.* 1996). Our results are qualitatively, but not quantitatively, similar.

From the spectral model fit, we obtain the peak isothermal temperature and its emission measure. With the volume obtained from imaging, we can then calculate the source density, the thermal energy, and the thermal energy density at this time, following the same procedures as in §4.2.3. These measurements were compiled for all 37 analyzed flares.

### 5.1.3 Results

Figure 5.3 shows how the maximum RHESSI temperature varies with GOES class. Although the data points are somewhat sparse and the spread is large, there does appear to be a general power-law relationship between the two quantities. This is qualitatively similar to the results obtained for non-super-hot temperatures by Feldman *et al.* (1996), who used BCS observations of Si and Ca excitation lines to determine plasma temperature and found a power-law relationship between GOES class and the maximum temperature. Quantitatively, the results differ significantly; we observe the maximum temperature to rise much faster with GOES class than do Feldman *et al*. However, their method of temperature determination uses ions with low peak formation temperatures of ~10-20 MK, and thus is sensitive primarily to temperatures in this range – they are effectively measuring the maximum temperature of only these ions, which necessarily exist only within a specific temperature range. In contrast, RHESSI's broadband, high-resolution continuum observations observe the bremsstrahlung of free electrons, which can exist at any temperature, and are thus most sensitive to the hottest electron temperatures throughout the flare; the faster rise that we observe may therefore indicate that the hottest temperatures have a steeper dependence on the physical factors that influence flare intensity, whatever they may be.

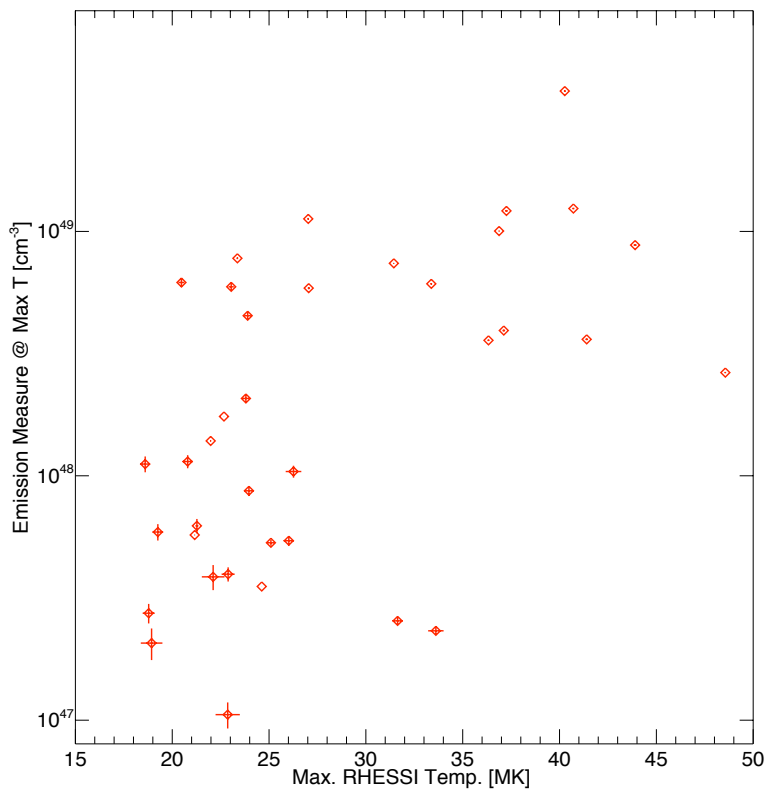

**Figure 5.4** – Emission measure at the time of maximum temperature, versus the maximum temperature. While there is no apparent functional dependence, all but two of the super-hot flares have an emission measure exceeding ~3×$10^{48}$ cm$^{-3}$, while the lower-temperature flares vary widely.

However, we note that our analysis utilizes the nominal calibration for the pulse pileup correction; since GOES class corresponds to integrated X-ray intensity, the average RHESSI live time decreases – and hence the pileup increases – with increasing GOES class. An inadequate pileup correction would yield increasing errors with intensity, which would contribute to an apparent increase in the maximum flare temperature and therefore to a faster-than-normal increase in temperature with GOES class; the impact of this potential effect has not yet been investigated.

As a sanity check, we examine the maximum temperature versus its emission measure (Figure 5.4). If our temperature variations were the result of fitting error, we would expect the temperature to be anti-correlated with emission measure; this is not the case, and suggests that our temperature variations are likely real, barring other sources of systematic error. Indeed, while there is no specific correlation between the temperature and emission measure, it is intriguing that, with two outliers, all of the super-hot (T > 30 MK) flares have an emission measure exceeding ~3×10$^{48}$ cm$^{-3}$, while the cooler flares appear more equally distributed over the entire range of emission measures. The super-hot emission measure behavior is in direct contrast to the results of Garcia & McIntosh (1992), who found that flares with GOES-derived temperatures exceeding ~25 MK have a generally *low* emission measure compared to the cooler flares that vary widely. However, given that GOES observes only in two very broad channels, which thus have a broad temperature response, the anti-correlation of temperature and emission measure they report appears likely attributable to the crudeness of the GOES temperature determination.

To test this relationship, it is instructive to examine how the derived volume, and hence density, vary with maximum temperature and with GOES class (Figure 5.5). The source volume shows no correlation with either quantity, being relatively evenly distributed (though this distribution is not very well-defined given the sparseness of data points at higher temperature and GOES class). While density shows no specific correlation, neither does it appear evenly distributed; rather, all of the X-class flares and 10 of 12 super-hot flares have a density exceeding ~6×10¹⁰ cm⁻³, while the weaker/cooler flares have densities that vary across the entire range. This clustering may simply be a consequence of small-number statistics, but if real, it may suggest a link between super-hot temperatures and a rough minimum density threshold, e.g. super-hot temperatures can only be achieved when the source density is beyond a minimum value. If we consider super-hot plasma being formed by the reconnection outflow as in the Shibata model (§4.4), this density threshold is not inconsistent, as the looptop density dictates how quickly the accelerated particles lose energy to collisions – and hence how quickly they heat the ambient plasma – as they traverse the looptop region; however, other explanations are also possible.

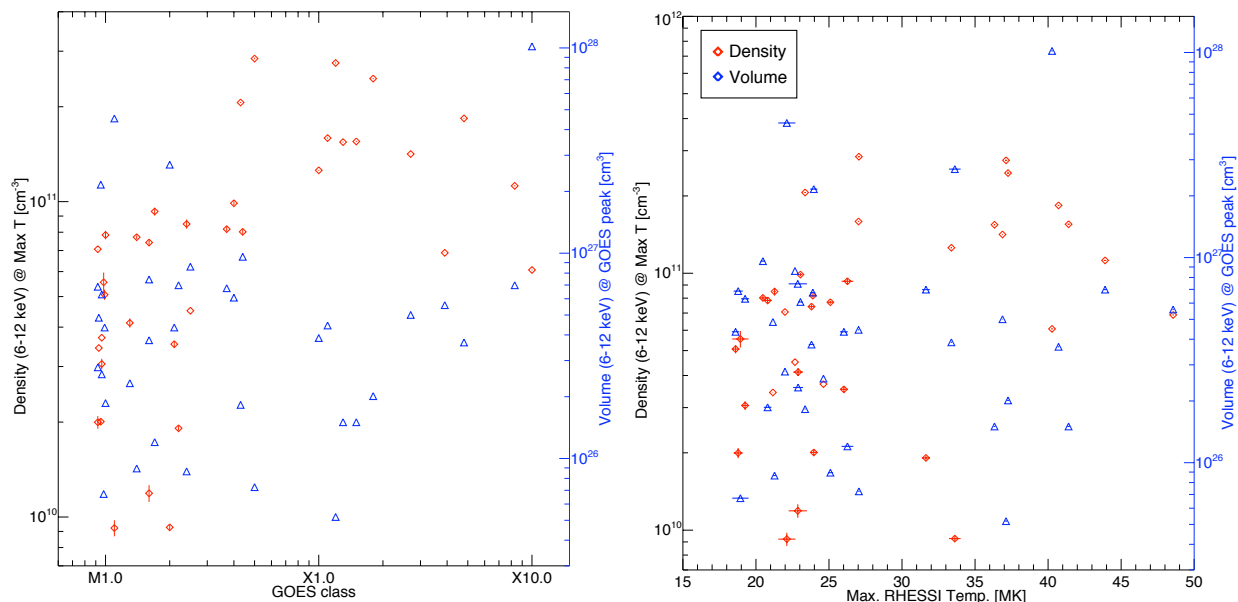

**Figure 5.5** – Density (red) and volume (blue) versus GOES class [left] and maximum RHESSI temperature [right]. The volume appears evenly distributed for both quantities, while density does not – X-class and/or super-hot flares appear to have a lower bound on density of ~6×10$^{10}$ cm$^{-3}$, while cooler/less intense flares vary widely.

We also examine how the thermal energy varies with both temperature and GOES class (Figure 5.6). Given the behavior of the density, it is perhaps not surprising that the total thermal energy shows a rough, but shallow, correlation with both temperature and GOES class – higher temperatures, and larger GOES class, appear to suggest a larger total thermal energy. The thermal energy density yields a more interesting result, however. If we consider the plasma β for an isothermal source contained by magnetic fields, then per equation (1.1) we require that β never exceed 1 – if it did, the plasma kinetic pressure would dominate the magnetic pressure and could push the fields apart, allowing the plasma to expand and cool adiabatically; if β is always smaller than 1, the field pressure dominates and it can keep the plasma confined, preventing it from cooling by expansion. From the measured thermal energy density (plasma pressure), then, we can determine the minimum field strength that is required to contain the plasma – this is represented by the horizontal dotted lines in Figure 5.6. Interestingly, all of the X-class flares, and 10 of 12 super-hot flares, appear to require a coronal field strength of *at least* ~170 Gauss (or higher by $\sqrt{2}$, if we also include the ion kinetic pressure, which we omitted from equation [1.1]). As with the density, if this clustering is real, it could indicate a minimum threshold for field strength, under which a super-hot plasma cannot form.

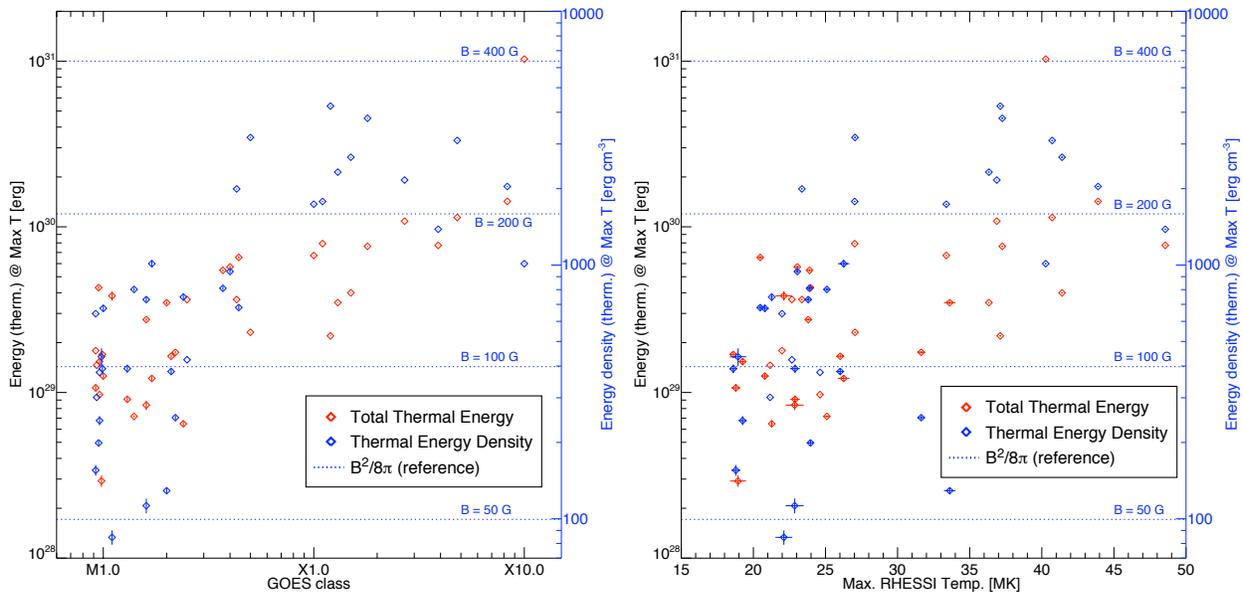

**Figure 5.6** – Total thermal energy (red) and thermal energy *density* (blue) versus GOES class [left] and maximum RHESSI temperature [right]. Although the total energy appears generally widely distributed (with slight trending to high values with larger GOES class), the thermal energy density appears to have a lower bound for X-class and/or super-hot flares. If we require that β < 1 then we can determine the minimum B-field strength required to contain the plasma (some reference values are shown, denoted by dotted lines).

We acknowledge that our flare selection criteria may introduce some selection bias. The requirement that a flare be imageable with grids 3-9, for example, effectively places upper limits on the size of the thermal source features – grid 3 has an angular resolution of ~7 arcsec, so sources that are significantly larger in all dimensions will be oversampled, leading to a noisy image; with uniform weighting, that noise will dominate the real signal from the higher grids, making the flare un-imageable according to our criteria. In actuality, this only affected 2 out of 260 flares, so this is not a significant effect, but it may exclude sources with volumes larger than a certain threshold. Additionally, our choice of only a 10-minute interval prior to the GOES SXR peak with the requirement that the interval also contain identifiable RHESSI HXR and SXR peaks may exclude certain long-duration events that take a long time to cool from the initial HXR burst, or ones that do not exhibit a traditional Neupert effect (§1.3). However, such flares may be governed by different physical processes; the establishment of these criteria help ensure that we are comparing flares of similar behavior. Finally, the requirement that the GOES SXR peak must succeed the RHESSI SXR peak presupposes that the maximum temperature occurs early in the flare and that the GOES peak manifests as the flare cools, which thus makes certain assumptions about the flare heating and cooling processes. However, none of the initial flare candidates were excluded due to peaks being "out of order," so these assumptions appear justified for this study.

Finally, we note that while the flares in our survey were well-fit by a single isothermal component, the carefully-calibrated analysis of Jul 23 showed that, for that flare, two isothermal components existed simultaneously throughout the flare. Our survey offers no specific insight as to the prevalence of such dual-isothermal configurations in other flares, and a thorough analysis would require careful calibration for each flare. As a quick test, we chose a single flare – the 2003 Nov 02 X8.3 event – and analyzed the spectrum during its primary HXR peak. Rather than obtaining a specific calibration for this flare, we instead applied the calibration corrections determined for Jul 23. The K-escape and attenuator calibration corrections appeared to work well for this flare, but the calibrated pileup parameters did not – for as-yet unknown reasons, the parameters determined for Jul 23 yielded a predicted pileup response that was significantly larger than observed. As such, we instead used the nominal pileup calibration, which provided good results, while retaining the improved attenuator and K-escape calibrations from Jul 23. Then, we followed the same iterative fitting procedure as we did for Jul 23 (§4.2.2), fitting and fixing the line fluxes before fitting the continuum with two isothermals and a non-thermal power-law. This quick test revealed that during the Nov 02 HXR peak, with the improved calibration, the continuum emission was *also* well-fit by two distinct isothermals with temperatures of ~43 MK and ~15 MK, respectively. While not conclusive, this result suggests that double-isothermal distributions may be common in super-hot flares, and may thus help identify the origins of the super-hot plasma. A survey using improved calibration over all flares would help investigate this.

## 5.2 Pre-impulsive phase observations of 2002 August 24

Gradual rise phases that precede the impulsive phase, which we term "pre-impulsive," are observed in many flares, but almost always with a primarily thermal spectrum (e.g. Fárník & Savý 1998; Veronig *et al.* 2002b; Battaglia *et al.* 2009). Those which exhibit significant non-thermal coronal sources appear quite rare; to date, only 3 flares have been identified as such – 2002 July 23 (Chapter 4), 2002 August 24 (discussed below), and 2003 November 3 – and all three were X-class events (Krucker *et al.* 2008). This may indicate a connection between pre-impulsive coronal particle acceleration and intense, powerful eruptive events, but it is also plausible that our measurements of less-intense flares are simply sensitivity-limited during this early phase; there is unfortunately no clear distinction between these scenarios at present.

Nevertheless, the few existing observations provide a wealth of information that shed light not only on this unique and intriguing period of the flare, but may also yield insight into particle acceleration processes and the physical conditions that possibly trigger the explosive energy release during the flare impulsive phase.

### 5.2.1 Observational overview

The X3.1 event on 2002 Aug 24 is perfectly situated to allow a detailed study of the coronal HXR source. It was a limb flare with occulted footpoints whose loop is oriented broadside to the line of sight, eliminating any projection effects both from loop orientation and heliocentric angle and ensuring that any HXR emission sources are located unambiguously in the corona; the occultation of the footpoints further allows identification of the spatially-integrated spectrum entirely with the coronal source.

The GOES SXR time profiles for Aug 24 behave similarly as for Jul 23 (Figure 5.7), rising gradually over the ~6-8 minutes of the pre-impulsive phase; the fluxes in the higher-energy (0.5-4 Å) channel rise faster than in the lower-energy one, indicating a hardening of the spectrum, either from heating or from non-thermal emission (see below). Although RHESSI went into eclipse just prior to the impulsive phase onset, the gradual rise of the SXR and HXR emission during the Aug 24 pre-impulsive phase also behaves similarly to Jul 23, with only a mildly-bursty behavior at high energies. The correspondence of the flare intensity and X-ray time profiles between the two flares suggests that they likely evolved under similar conditions, but the limb position of Aug 24 allows for imaging analysis of the source that was not possible for Jul 23 (see below). Additionally, the flare was well-observed both by TRACE and by the *Nobeyama Radio Heliograph* (NoRH) in both of its frequency passbands, providing a wealth of supporting observations.

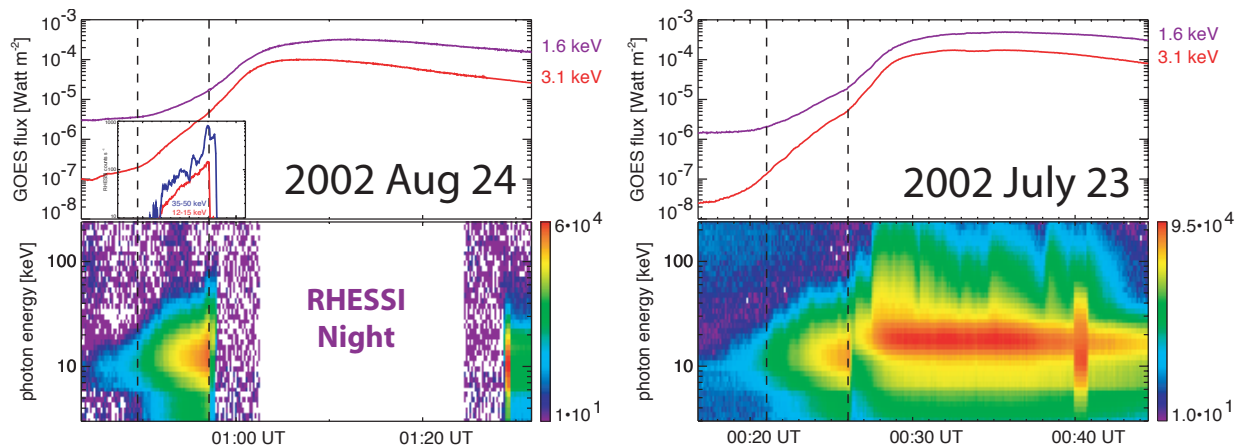

**Figure 5.7** – GOES lightcurve [top] and RHESSI spectrogram [bottom] showing the pre-impulsive rise of the 2002 Aug 24 [left] and, for comparison, 2002 Jul 23 [right] flares. For Aug 24, the 12-15 and 35-50 keV RHESSI lightcurves are also super-posed, for reference. The lightcurves and spectra behave very similarly during the pre-impulsive phases of both flares.

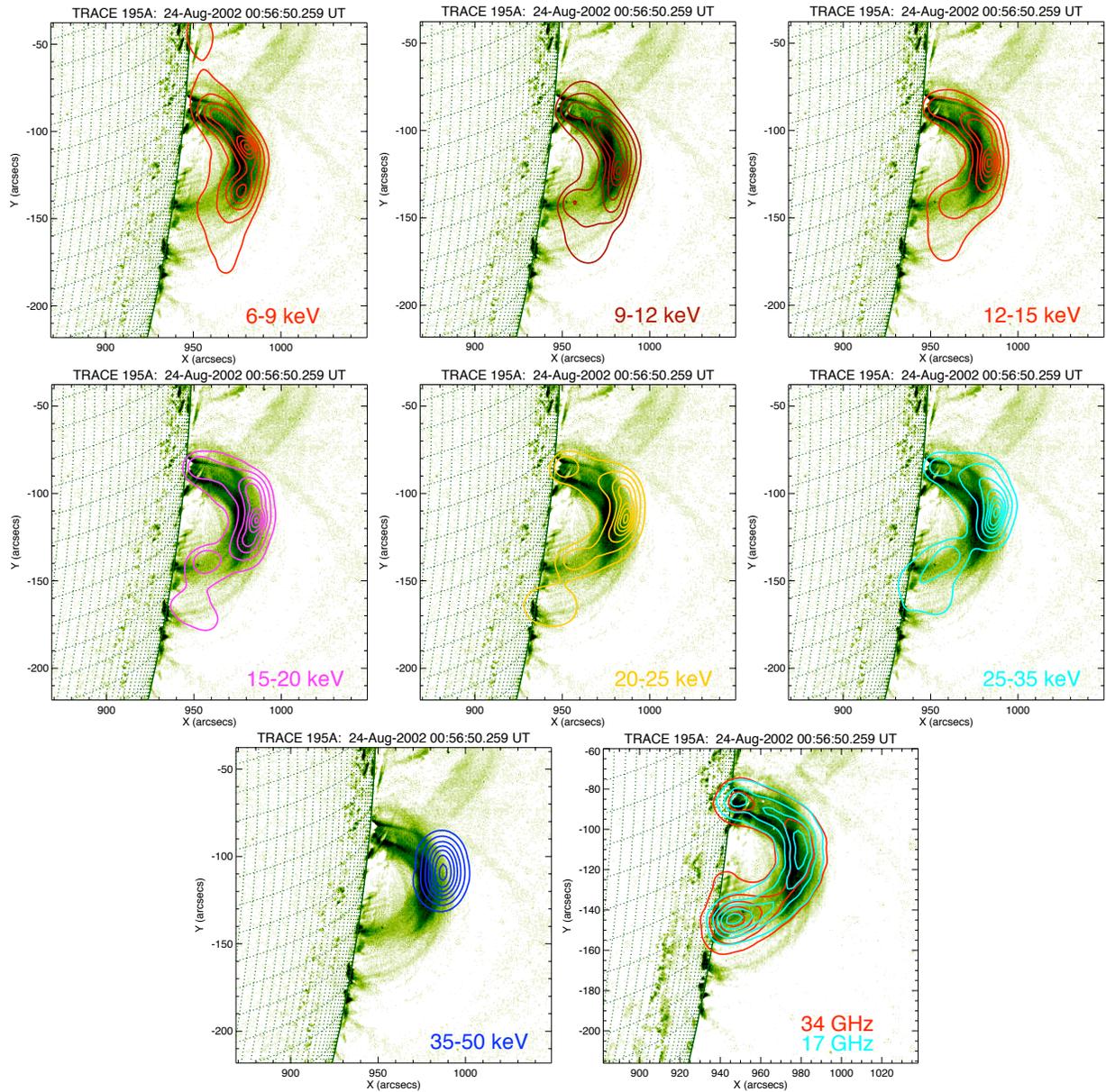

**Figure 5.8** – RHESSI SXR/HXR contours (evenly spaced from 10% to 95%) and NoRH radio contours overlaid on TRACE 195 Å emission for the peak of the pre-impulsive phase of Aug 24. The X-ray emission is increasingly concentrated towards the loop-top with increasing energy; the highest-energy emission is *entirely* from the looptop.

Figure 5.8 shows RHESSI SXR and HXR contours during the peak of the pre-impulsive phase, just prior to the spacecraft eclipse, overlaid on a TRACE 195 Å EUV image from the same time period; NoRH radio contours at 17 and 34 GHz are also shown. The SXR emission from 6 to 15 keV is cospatial with the TRACE loop and traces the loop outline quite well, consistent with emission from hot thermal plasma. At higher energies, the emission contours begin to contract and break up into regions; this is most noticeable for the 25-35 keV contours, where the emission is still cospatial with the loop but slightly offset and no longer smoothly varying. The 35-50 keV contours are radically different; they are localized to a looptop source with a FWHM of only ~20% the total loop length, completely unlike the lower-energy contours. This localized emission may be from a super-hot plasma or it may be from non-thermal bremsstrahlung from accelerated electrons; the compactness of the source suggests a potential density enhancement at the looptop or, if the emission is non-thermal, possible trapping of the >35 keV electrons. The NoRH radio contours trace the loop well, similar to the SXR contours. However, the spectral index of the emission between the two frequencies indicates that the emission is primarily non-thermal, suggesting that despite the lack of footpoint emission due to occultation, non-thermal electrons are still being accelerated down the legs of the loop; however, rather than hitting the chromosphere, they may be mirroring back into the corona. This is supported by the apparent gap between the southern loop leg and the limb, which suggests a stronger magnetic field and therefore likely magnetic mirroring (Reznikova *et al.* 2009).

### 5.2.2 Spectral analysis

We can test the possible emission models suggested by the images via the usual forward-modeling analysis of the spectra (Figure 5.9). Since the footpoints are occulted, one might expect that the looptop HXR emission is thermal; the compactness of the source would then indicate that the hottest temperature plasma is localized to the looptop. A single isothermal model cannot simultaneously fit the SXR and 35-50 keV emission; to fit both regions with thermal emission requires at least a double-thermal model. Then, under a two-thermal model, the spectrum would suggest that the hot temperature is ~55 MK, if one assumes that the 50-100 keV emission is non-thermal; a thermal interpretation for *all* of the emission up to ~100 keV requires temperatures up to ~100 MK.

However, we can use the Fe and Fe/Ni lines to restrict the allowable temperature range, as we did for the pre-impulsive phase of Jul 23 (§4.3). Since we do not have a full time-series of observations for Aug 24, we can instead use the line-to-continuum relationship derived from Jul 23 (e.g. Figure 4.12); fitting the lines with Gaussian functions and assuming that the line ratio depends on the continuum temperature as measured during Jul 23 yields predicted continuum temperatures of only ~25 MK, inconsistent with the double- and triple-thermal models that include much higher temperatures. The line analysis suggests a single-isothermal model with a temperature of ~26 MK, indicating that all of the emission down to ~20 keV is likely non-thermal; with a broken power-law for the non-thermal emission, this yields equally good fits, compared to the multi-thermal models, while retaining consistency with the line observations.

We applied this procedure for multiple intervals throughout the pre-impulsive phase, fitting a photon model consisting of an isothermal continuum, a broken power-law, and two Gaussians to the spectrum at each interval, to examine the time evolution of the model parameters. As above, the temperature of the isothermal was constrained by the measured line ratio using the relationship obtained from the analysis of July 23. The non-solar background was subtracted using the procedure outlined in §B.2.1. As all spectra were observed in the A1 state, no correction for the thick attenuator calibration was needed; however, because there was no A3→A1 transition observed, we used the nominal calibration for the pileup parameters.

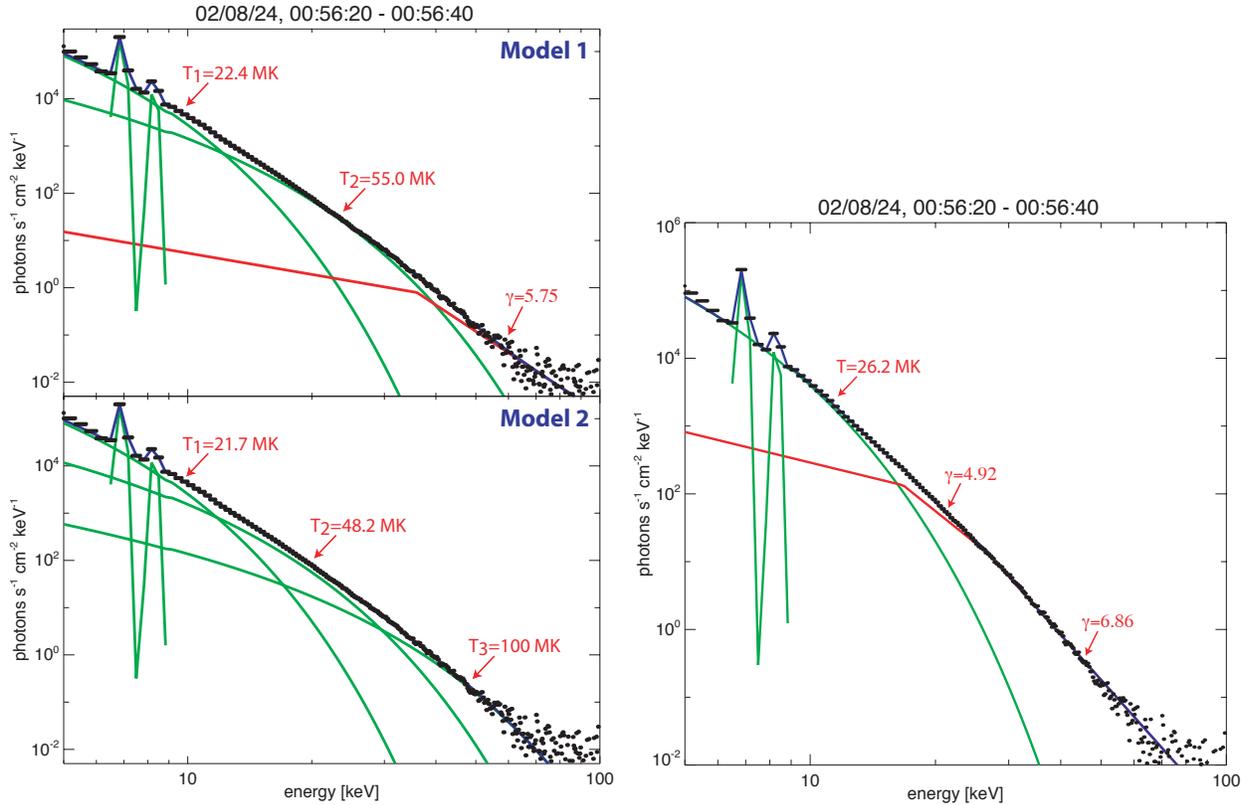

**Figure 5.9** – Potential model fits to the spectrum for the peak of the pre-impulsive phase of Aug 24. Explaining the HXR continuum with thermal emission [left] requires multiple thermal components and very high temperatures (~50 MK or greater), but the Fe & Fe/Ni line emission indicates temperatures ≲26 MK, suggesting that the HXR emission is predominantly non-thermal [right], as in Jul 23. All three continuum models fit equally well, but only the last is consistent with the line observations.

Figure 5.10 shows the time evolution of the model parameters obtained via fitting. As dictated by the Fe and Fe/Ni lines, the temperature of the isothermal continuum rises early and levels off, while the emission measure rises relatively slowly until the latter half of the pre-impulsive phase. From the 6-15 keV image contours, we estimate the loop half-length to be ~3.5×10^9 cm; approximating the loop as a curved cylinder therefore suggests a thermal source volume of ~8×10^26 cm^3, with both of these values remaining roughly constant throughout the phase. Assuming this constant volume, the rising emission measure indicates that the thermal density rises from ~10^10 to ~10^11 cm^-3 over the observed time period.

The column density inferred from this, combined with the collisional energy loss equation (2.5), suggests that electrons with kinetic energy >35 keV will not be stopped by the ambient plasma and will see it as a thin target, while the lower-energy electrons will be stopped as if in a thick target. Electrons >35 keV trapped at the looptop would thus emit thin-target bremsstrahlung as they repeatedly mirror in the trap, while escaping electrons would quickly reach the oc-culed footpoints without interacting strongly in the loop, and thus would not be observed. Lower-energy electrons would lose all of their energy via collisions; those trapped at the looptop would thus enhance the looptop SXR emission, while those escaping from the looptop would stop somewhere along the legs of the loop, yielding an extended SXR source. Both of these scenarios are consistent with the imaging observations.

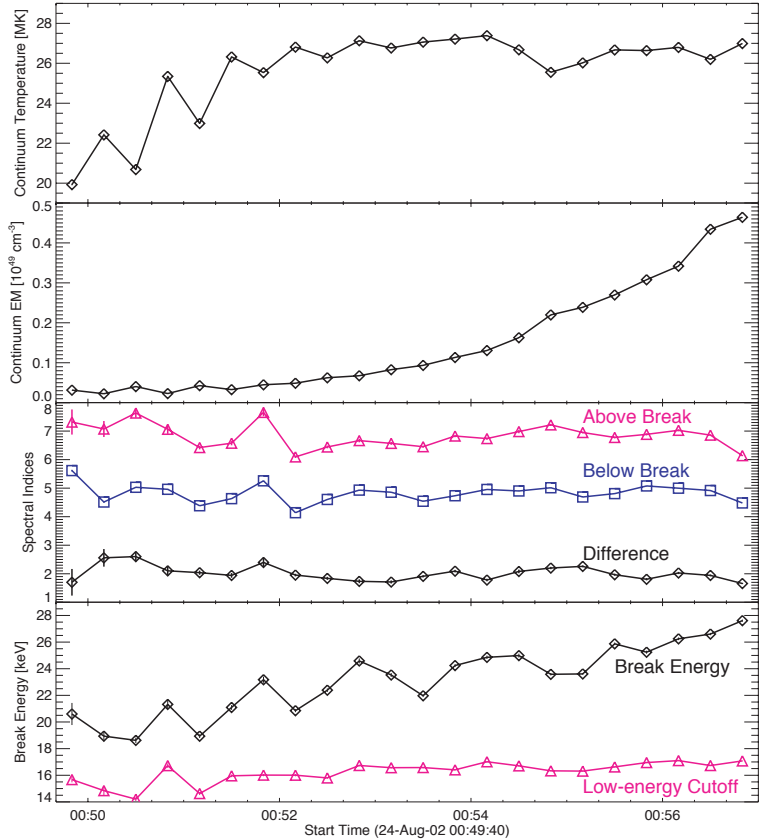

**Figure 5.10** – Time evolution of the best-fit model parameters for Aug 24, including the isothermal continuum temperature and emission measure, the non-thermal spectral indices above and below the break, and the break energy. Assuming a roughly constant volume, the rising emission measure indicates a rising density.

It is intriguing that throughout the analysis period, the difference between the spectral indices above and below the break is ~2, the same as for the difference between thin- and thick-target bremsstrahlung emission; this could suggest that the break energy is the transition between the two regimes, as implied above. Indeed, the break energy is near the thin/thick transition energy inferred from equation (2.5), and rises with time consistent with the rising density. However, such a sharp break between the thin- and thick-target regimes is likely unphysical, suggesting that the photon spectral break also, or perhaps instead, indicates a true break in the underlying electron spectrum. We note that the collisional energy loss time (cf. Lin 1974) for >35 keV electrons at the inferred densities is below ~10 sec, while we observe steady or increasing non-thermal emission at all times, implying that there is continuous particle injection replenishing (and adding to) the non-thermal electron population.

Having determined that the HXR continuum is primarily non-thermal, we may calculate the energy in non-thermal electrons as we did for July 23, to compare the two flares. For a single power-law photon spectrum and assuming a thick-target model, the total power in non-thermal electrons may be determined using equation (2.6). For a broken power-law, as observed for Aug 24, this formula is no longer entirely correct, but we may nevertheless approximate the total power by summing the contribution from the two separate pieces of the power-law, above and below the break energy. In this manner, using the spectral indices, break energies, and cutoff energies determined during fitting (Figure 5.10), we calculate that the total power deposited during the pre-impulsive phase by non-thermal electrons above the measured cutoff energy (assuming thick-target energy deposition) is $\sim 3 \times 10^{31}$ ergs, good to order-of-magnitude accuracy. This is the same order of magnitude as the energy determined in the low-temperature limit for the Jul 23 pre-impulsive phase (§4.4). Similarly, the maximum thermal energy during the pre-impulsive phase was calculated at $\sim 3.4 \times 10^{29}$ ergs, nearly identical to the values derived for Jul 23; in both cases, under a thick-target assumption, the non-thermal particles deposit $\sim 100$ times more energy than is seen in the thermal plasma. The remarkable similarities between the two pre-impulsive phases suggest that identical physical processes are occurring in both flares, and offers the intriguing possibility that these mechanisms operate in all large flares – more observations of pre-impulsive phases are needed to answer that question.

**Chapter 6: Discussion of the Origins of Super-Hot Plasma, and Future Directions**

RHESSI's rich data set and flexible analysis tools enable unprecedented precision for measurements of X-ray emission from solar flares, and especially of hot thermal plasma, providing valuable insights into the physical mechanisms that drive particle acceleration and heating of super-hot plasma. Our statistical study shows that super-hot plasmas are common in X-class flares, and that the maximum temperature appears strongly correlated with the logarithmic flare class. For the 2002 July 23 X4.8 flare, we have shown for the first time that the super-hot plasma exists separately from and simultaneously with the traditional GOES-temperature plasma throughout the flare. A cursory examination of the 2003 Nov 02 X8.3 flare also reveals a two-temperature structure, suggesting that such a distribution is common to most super-hot flares. These observations provide the first clues as to the ultimate origins of super-hot thermal plasma.

*6.1 Origins of super-hot plasma*

The statistical survey of thermal plasmas in large flares shows that super-hot temperatures are routinely achieved by X-class flares, but only rarely so by weaker, M-class flares. Although there is no clear functional dependence of the thermal energy density on the maximum flare temperature, those flares reaching super-hot temperatures do exhibit an apparent lower bound on the energy density, while cooler flares show no such lower bound and are more uniformly distributed. Assuming that $\beta \leq 1$, this lower bound indicates that the magnetic field strength in the corona must be at least ~150-400 Gauss for super-hot flares. There is no significant dependence of the *total* thermal energy on flare temperature, however, suggesting that the existence of super-hot plasma is related to the field configuration and instantaneous energy, but not necessarily to the amount of energy released during the flare.

For the specific case of July 23, we can compare the super-hot and GOES-temperature plasmas directly. The cooler plasma's temperature time profile is slightly (~1 min) delayed from that of the non-thermal HXR emission, and it is at lower altitude and higher density compared to the super-hot plasma, suggesting that it originates from the traditional picture of chromospheric evaporation, wherein chromospheric material is heated by collisions with the downward-moving accelerated particles and subsequently evaporates into the loop. In contrast, the position and morphology of the super-hot source – at higher altitudes than the cooler plasma, and elongated away from the footpoints – along with its temperature time profile, which peaks at the same time as the non-thermal HXR emission, suggest that the super-hot plasma is more directly related to the accelerated non-thermal electrons and hence to the reconnection process. Indeed, the most compelling evidence for this is the existence of the super-hot plasma even during the non-thermally-dominated pre-impulsive phase, when there is little to no footpoint emission that would signify energy deposition in the chromosphere, and hence likely negligible chromospheric evaporation.

At the start of the pre-impulsive phase, the super-hot plasma density is still ~1-5×10$^{10}$ cm$^{-3}$, ~10-50 times higher than the ambient density. Without chromospheric evaporation that would add new particles, one means of achieving this increased density is through compression; absent any significant cooling through radiation or conduction, compression would also heat the plasma, and may explain the early high temperatures as well as densities. If we consider the open-field reconnection models of Sturrock (Figure 1.9, left) or Shibata (Figure 4.14), an available energy source for compressing the plasma is the coronal magnetic field, or more specifically, the magnetic tension that pulls the just-reconnected field lines – which are in a highly non-potential configuration – into a more dipolar loop configuration. After reconnection, the field lines are reduced in length and, by conservation of magnetic flux, the loops are also reduced in cross-section; with $\beta < 1$, the plasma within the loop is carried downward and compressed, thus increasing its density and temperature.

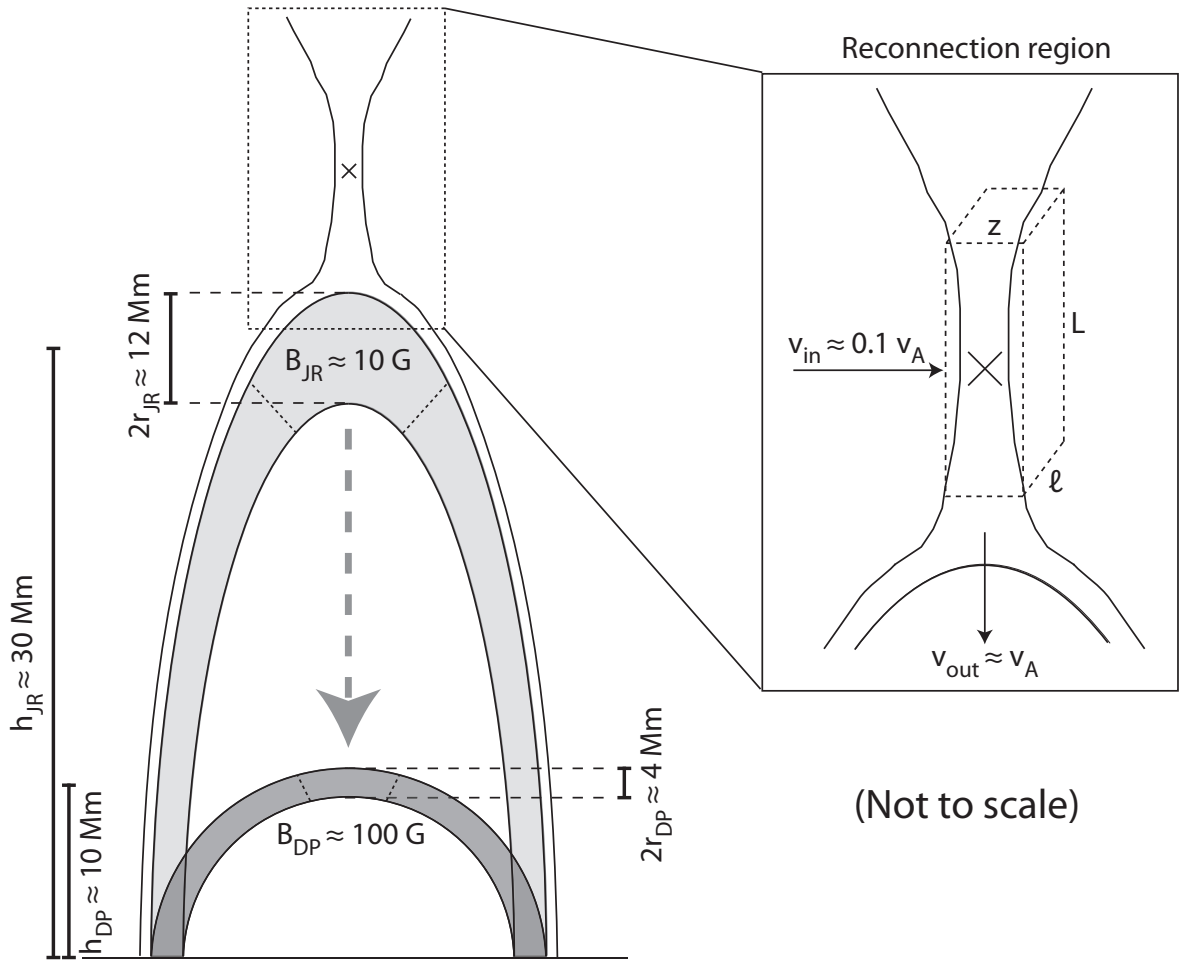

**Figure 6.1** – An order-of-magnitude cartoon model to explain the origins of super-hot plasma. Magnetic field lines reconnect at the × -point, forming the upper "just-reconnected" loop (light gray); magnetic tension compresses the loop into the lower quasi-dipolar configuration (dark gray). Ambient particles initially energized by reconnection will be further heated by compression of the loop.

During the impulsive phase, when strong non-thermal footpoints are observed, the super-hot temperature may be explained by collisions of high-energy reconnection-accelerated particles with the ambient plasma in the reconnected loop – assuming an atmospheric heating model such as that of Fisher *et al.* (1985), the accelerated electrons inferred from the non-thermal bremsstrahlung spectrum (assuming a thick-target model with low-energy cutoff) can deposit sufficient energy in the corona as they traverse down to the footpoints. Chromospheric evaporation may partly account for the rise of the super-hot density during the impulsive phase, as the most ener-getic (and thus fastest) particles may evaporate some material into the loop even while the lower-energy particles are still heating the plasma. However, such a scenario cannot apply early in the pre-impulsive phase, where the thermal density is still high without evidence for evaporation. This picture, then, suggests that the origin of the super-hot plasma is indeed the compression and heating of ambient material within just-reconnected loops as they relax into more dipolar configurations, along with subsequent thermalization of accelerated particles further energized by the field relaxation (e.g. Somov *et al.* 2005).

We can make rough order-of-magnitude calculations to test whether this hypothesis yields parameters for the plasma that are reasonably consistent with the observations; all values quoted below are to be considered only to order of magnitude. Consider the Sturrock-like reconnection schematic in Figure 6.1: magnetic field lines (dark curves) reconnect at the × point, forming the "just-reconnected" (hereafter "JR") loop at the top (light gray shading); magnetic tension then pulls the loop down, eventually leading to a more compact, quasi-dipolar (hereafter "DP") loop below (dark gray shading). The DP loop is the one generally observed; for simplicity, we treat it as a "solid" half-torus (a cylinder of constant radius bent into a semicircle). We assume the DP looptop altitude to be $h_{DP} \approx 10$ Mm (thus the loop length is $\sim \pi h_{DP} \approx \sim 30$ Mm and the footpoint separation is $\sim 2h_{DP} \approx \sim 20$ Mm) and its cross-sectional radius to be $r_{DP} \approx 2$ Mm; we take the height of the reconnection region and JR looptop, which we assume to be equal to order of magnitude, to be $h_{JR} \approx 30$ Mm. These values are in order-of-magnitude agreement with the respective observed quantities for large flares (e.g. our observations of Jul 23 and Aug 24, and the observations by Sui & Holman [2003]). We assume the magnetic field strength at the DP looptop to be $B_{DP} \approx 100$ Gauss, in order-of-magnitude agreement with our observations of the minimum field strength required to contain a hot plasma in a large flare (cf. §5.1.3).

The JR loop is generally not observed, but we can infer its properties by comparison with the dipolar loop and by analogy with the Earth's magnetotail (cf. Lin *et al.* 1977). The Earth's magnetic field is essentially dipolar within $\sim 6$ $R_E$ (where $R_E \approx 6$ Mm), but has a long tail where reconnection occurs in an ×-point configuration down the tail, qualitatively analogous to the model considered here. The field strength at Earth's surface is $\sim 0.5$ Gauss; at a radius of $r = 6$ $R_E$, the equatorial dipolar field (in analogy with the top of a flare loop) would thus be $\sim 2 \times 10^{-3}$ G. In comparison, the field strength in the magnetotail at $\sim 20$ $R_E$, where reconnection has been commonly observed to occur (Nagai *et al.* 1998), is $\sim 20$ nT $= \sim 2 \times 10^{-4}$ G (e.g. Slavin *et al.* 1985), thus a factor of $\sim 3$ increase in distance yields a factor of $\sim 10$ decrease in field strength. (We note that this is a far smaller decrease than would be expected from a dipolar field, immediately showing that the tail is significantly non-potential and therefore stores a lot of "free" energy.) Applying the same scaling factors to our flare model yields that the reconnection region and JR looptop height are $h_{JR} \approx 3h_{DP} \approx 30$ Mm (as we had previously assumed) and that the magnetic field strength there is $B_{JR} \approx 0.1$ $B_{DP} \approx 10$ G. Then, the cross-sectional radius $r_{JR}$ can be determined by conservation of magnetic flux:

$$\int_S \mathbf{B} \cdot \hat{\mathbf{n}} \, dS = \text{constant} \approx B\pi r^2 \implies r_{\mathrm{JR}} = \left(\sqrt{B_{DP}/B_{JR}}\right) r_{DP} \approx \left(\sqrt{10}\right) r_{DP} \approx 6 \text{ Mm} \tag{6.1}$$

Having calculated our scaling parameters, we can now consider what happens to the initial energy of particles within the JR loop (discussed in more detail below) as it contracts into the DP loop; we'll consider only electrons, as these are what generate the bremsstrahlung emission that we observe. The Alfvén velocity $v_A \equiv B/\sqrt{4\pi\rho} \approx B/\sqrt{4\pi\bar{Z}m_p n}$ describes the speed at which perturbations of the magnetic field can propagate in an ion-electron plasma of number density $n$ and average ion mass $\bar{Z}m_p$ ($\approx 1.2\ m_p$ for the corona); the loop will contract at approximately this rate. Assuming an ambient plasma density within the JR loop of $n_{JR} \approx 10^9$ cm$^{-3}$ (such a low density would not be detectable with existing instruments), then with $B_{JR}$ as calculated previously, we get $v_{A\text{-}JR} \approx 600$ km/s. Assuming equipartition of energy between the electrons and ions, the average electron velocity is ~43 times greater than the average ion velocity (which is ~$v_A$), therefore the electron motion is fast compared to the rate of change of the loop parameters and the adiabatic invariants $\mu$ and $J$ should be approximately conserved. Then, for a particle in the JR loop with initial energy $E_{JR} = E_{JR\perp} + E_{JR\parallel}$, we compute its final energy $E_{DP}$ in the DP loop as:

$$\mu \equiv \frac{mv_\perp^2}{2B} = \frac{E_\perp}{B} = \text{constant} \implies E_{DP\perp} = \frac{B_{DP}}{B_{JR}} E_{JR\perp} \approx 10 E_{JR\perp} \tag{6.2}$$

$$J \equiv \int v_\parallel ds \approx \sqrt{\frac{2E_\parallel}{m}} \pi h = \text{constant} \implies E_{DP\parallel} = \left(\frac{h_{JR}}{h_{DP}}\right)^2 E_{JR\parallel} \approx 9 E_{JR\parallel} \tag{6.3}$$

To order of magnitude, the perpendicular and parallel increase by the same factor, and we can approximate $E_{DP} \approx 10\ E_{JR}$ – the compression of the loop increases the particle energy tenfold (on average, since we're ignoring the particle's initial pitch angle). If the plasma is originally thermal and we equate the average energy $E_{av} = k_B T$, then by the same token, the temperature of a thermal plasma will increase tenfold (although there's no particular reason to pre-assume a thermal distribution). Additionally, the plasma density will increase by a factor of ~30, since the total number of particles $nV$ is assumed constant and the loop volume $V$ decreases by ~30 (since $V_{JR} \approx \pi^2 h_{JR}(r_{JR})^2 \approx \pi^2 (3h_{DP})(10[r_{DP}]^2) \approx 30 V_{DP}$), thus the density in the DP loop is $n_{DP} \approx (V_{JR}/V_{DP}) n_{JR} \approx 30 n_{JR} \approx 3 \times 10^{10}$ cm$^{-3}$, a 30-fold increase over the JR loop density.

We therefore need only approximate the initial particle energy $E_{JR}$ to assess the validity of our order-of-magnitude model, which we do following Lin *et al.* (1977). Consider a box of dimensions $L \times \ell \times z$ inside the reconnection region (see Figure 6.1, inset) within which the open field lines approach the ×-point with inflow velocity $v_{in}$; this is the current sheet required by such a magnetic configuration. Assuming for simplicity that **B** is parallel to the box sides and of constant magnitude, we can immediately derive the current within the box from Ampère's Law as:

$$\oint \mathbf{B} \cdot d\mathbf{l} \approx 2BL = \frac{4\pi}{c} I \implies I = \frac{cBL}{2\pi} \tag{6.4}$$

In a plasma of finite resistivity, a current necessitates an electric field, by Ohm's Law; if the resistivity is small and can be neglected, the generalized Ohm's Law dictates the electric field to be $\mathbf{E} = -\mathbf{v} \times \mathbf{B}/c$ (out of the page for the geometry given in Figure 6.1), where **v** is the inflow velocity into the reconnection region (and also corresponds exactly to the **E**×**B** drift of the particles). The potential drop along the current sheet is thus $V = E\ell = v_{in}B\ell/c$. With a finite resistivity, the particles within the reconnection region must experience Joule heating, corresponding to the energy lost by the magnetic field as it diffuses through the plasma; the total power dissipated is therefore $P = IV = v_{in}B^2L\ell/2\pi$. Given the geometry, a particle will transit through the current sheet in an *average* time $t = L/(4v_{out})$; by conservation of mass flow, $Lv_{in} = zv_{out}$, therefore $t = z/(4v_{in})$, whence the average energy gain *per particle* is:

$$\Delta E = \frac{Pt}{nV} = \frac{\left(v_{in}B^2L\ell/2\pi\right)(z/4v_{in})}{nL\ell z} = \frac{B^2}{8\pi n} \tag{6.5}$$

Using $B_{JR}$ and $n_{JR}$, we therefore get $\Delta E \approx 2.5$ keV per particle gained from reconnection.

The JR loop is in the state immediately after reconnection, and thus the looptop is filled with plasma that has just transited through, and been energized in, the reconnection region. The average particle energy $E_{JR}$ will be the sum of the ambient thermal energy and the energy gained from reconnection, per above. For an ambient coronal temperature of ~1 MK, the thermal energy is ~0.1 keV and the energy gain from reconnection dominates, hence $E_{JR} \approx 2.5$ keV for those particles that have transited the reconnection region. Then, from the above analysis, our order-of-magnitude cartoon model predicts that the *average* particle energy in the DP loop is increased tenfold, thus $E_{DP} \approx 25$ keV (although we have neglected the fact that the reconnection-energized particles "injected" into the looptop with initial energy of ~2.5 keV will necessarily mix with the ambient plasma [with initial thermal energy of ~0.1 keV] already present in the legs of the loop, thereby reducing the average energy by a factor of a few).

With $n_{JR} \approx 10^9$ cm$^{-3}$, the predicted density in the DP loop is $n_{DP} \approx 3 \times 10^{10}$ cm$^{-3}$, and since $V_{DP} \approx \pi^2 h_{DP}(r_{DP})^2 \approx 4 \times 10^{26}$ cm$^3$, the total number of particles $N_{DP} = n_{DP}V_{DP} \approx 10^{37}$ with a resulting emission measure $Q_{DP} = (n_{DP})^2 V_{DP} \approx 4 \times 10^{47}$ cm$^{-3}$. The energized particles (with average energy $E_{DP} \approx 25$ keV) will be initially primarily non-thermal, but will thermalize through collisions. In comparison, during the peak of the pre-impulsive phase for Jul 23 (cf. §4.4), the total (instantaneous) number of particles was ~$10^{38}$ thermal electrons and ~$10^{34}$-$10^{35}$ non-thermal electrons, with average energies of ~3 keV and ~26-31 keV, respectively. Our simplistic model yields average energies that are in relative agreement with the observed average non-thermal energies, especially accounting for losses during thermalization. While the predicted number of particles is an order of magnitude too low compared to the observed *total* (thermal plus non-thermal) number of particles (which is the number to which we must compare if considering that the thermal population results primarily from thermalization of the non-thermal population), we note that the number of particles in the model was constant and determined entirely through our (somewhat arbitrary) choice of initial loop length, loop cross-section, and ambient density; we could adjust any or all of those parameters to obtain reasonable agreement for the particle numbers while also maintaining agreement of the average energies.

Hence, accounting for the above and to within an order-of-magnitude precision, the predictions from our simple model *are* entirely consistent with the densities and average energies (and temperatures) actually observed in flares – *such a model may plausibly explain the origins of super-hot plasma.*

### 6.2 Future observations

While our model appears to provide a plausible mechanism for the formation of super-hot plasma, existing observations seem to support this model only indirectly. Imaging and spectral observations of the hot plasma in the already-compressed DP loop are plentiful, such as the ones we describe in great detail for July 23. High-altitude HXR emission from energetic particles, potentially from the reconnection region itself, has also been observed during the impulsive and decay phases (Sui & Holman 2003; Krucker *et al.* 2008), though not at earlier times (the HXR coronal sources observed in pre-impulsive phases appear to be lower-altitude, e.g. at the top of the DP loops). There have also been observations by RHESSI and TRACE of apparent compression of existing DP loops prior to the flare impulsive phase, e.g. during July 23 (Lin *et al.* 2003). However, there have been no published measurements, whether in images or spectra, of a ~2.5-keV-average-energy plasma in a JR loop, either before or during compression.

In large part, this may be due to the limited sensitivity of existing measurements. Prior to compression, the emission measure of the JR loop is only $Q_{JR} = (n_{JR})^2 V_{JR} \approx 10^{46}$ cm$^{-3}$; already-compressed DP loops have emission measures ~40 times greater, but their total intensity is increased by at least ~2-3 orders of magnitude compared to the JR loop because of the higher average energy (or temperature) in addition to the higher emission measure. Distinguishing the emission from the JR loop against the much-brighter DP loops thus requires a dynamic range of ~$10^3$-$10^4$, something very few (if any) existing instruments can achieve. This suggests that the optimal time to observe the JR loop would be just at the flare onset, when there are no existing DP loops to overwhelm its emission. However, if the loop contracts with a speed of ~$v_A$ then, using the numerical parameters above and accounting for the approximately linear increase of the magnetic field strength with decreasing altitude (per our model), the entire compression phase takes only a few tens of seconds; to isolate the JR loop before or just during compression and observe its dynamic evolution thus requires an observational cadence of no more than a few to ~10 seconds.

Assuming observations just at the flare onset (to mitigate the issue of dynamic range), direct-imaging instruments such as TRACE or EIT are sufficiently sensitive to observe emission measures of ~$10^{46}$ cm$^{-3}$, but they are based on excitation line emission and have very narrow bandwidths that limit their temperature response, so that the emission from the JR loop may not fall within their bandpasses. (Moreover, since reconnection and the loop compression are highly dynamic processes, there is also no guarantee of thermal or ionization equilibrium, so the specific ionization states to which the instrument bandpasses are tuned may not be highly populated.) The GOES Soft X-ray Imager (SXI) uses broadband filters and has a better temperature response, but it has decreased imaging fidelity and sensitivity due to instrumental issues (Hill *et al.* 2005). All three instruments also have relatively low cadence, and none are capable of high-resolution spectroscopy.

A broadband spectroscopic imager with high temporal and spectral resolution, such as RHESSI, thus affords the best chances of a positive observation. At an emission measure of ~$10^{46}$ cm$^{-3}$ and if observing at the beginning of the flare when other sources are not present to dominate the emission, RHESSI should be able to easily detect emission from a ~2.5-keV-average-energy plasma above the non-solar background even through the thin attenuator (§3.2.1). The fact that such sources do not appear to be commonly observed (with only one or two potential exceptions, e.g. the observations of Sui & Holman [2003]) strongly suggests that either or both of the emission measure or the average particle energy in the reconnection region and/or JR loop are smaller than predicted by our model; the average energy would only have to decrease by a factor of a few to make the plasma nearly, or entirely, unobservable. If, for example, the average particle energy were only ~1 keV, then only A0-state observations, when no attenuators are engaged, would be sufficiently sensitive: when the attenuators are engaged, RHESSI's sensitivity at low energies is significantly reduced; to distinguish the spatially-integrated spectrum of a ~1-keV plasma from the non-solar background requires an emission measure exceeding ~$3 \times 10^{48}$ and ~$7 \times 10^{49}$ cm$^{-3}$ in the A1 and A3 attenuator states, respectively, ~300-7,000 times greater than the predicted $Q_{JR}$. As the loop compresses, the emission measure and temperature both rise quickly and the loop should become observable, consistent with the super-hot observations presented in §4.3, but RHESSI is not sufficiently sensitive in the A1 and A3 states to observe the emission from the uncompressed plasma if its average energy is only ~1-keV.

In the A0 state, the required emission measure is only ~$10^{45}$ cm$^{-3}$ for spatially-integrated spectra, so A0 observations should afford sufficient sensitivity to distinguish the JR loop emission from the non-solar background in this assumed limit. Imaging requires a somewhat (factor of ~few) higher emission measure, since the intensity is spread over multiple pixels and since image reconstruction enhances noise, and thus also requires increased integration times (resulting in lower cadence), but A0 imaging should nevertheless be at least marginally sensitive to ~1-keV JR loop emission. RHESSI has observed over 30 X-class flares since launch; a comprehensive survey of those flares which included A0 observations during their onset would hopefully yield an observation of the JR loop emission both before and during compression.

Higher sensitivity and improved dynamic range would improve the chances of a positive detection; a better dynamic range would also allow observations of the JR loop even in the presence of brighter sources. Higher sensitivity can be achieved through a larger effective area, while direct imaging can improve the dynamic range since the image is read out directly from individual and independent pixels rather than reconstructed using Fourier algorithms. A pixelated spectrometer coupled with focusing optics, such as that in the upcoming *Focusing Optics X-ray Solar Imager* (FOXSI) instrument, would be optimal for our purposes. FOXSI is expected to be ~50 times more sensitive than RHESSI at ~10 keV and to have a dynamic range up to ~100 times larger, with comparable (~1 keV) energy resolution (Krucker *et al.* 2009). Such an instrument would maximize the chances of detecting the emission from the JR loop and of studying its time evolution even as other sources (such as previously-compressed loops) brighten within the field of view.

If no positive detections can be made with either RHESSI or FOXSI, we can still place upper limits on the parameters of the JR loop, such as its maximum possible emission measure and/or average energy, just as we did for the simple model above. A positive observation, however, especially if it allows us to investigate the time evolution of the compressing loop, would allow verification of our model, including quantification of the essential parameters, and would provide the first measurements of the earliest physical processes that lead to the creation of super-hot plasmas. Discovering the origins of the super-hot plasma would also, ultimately, provide insight into the physical mechanisms that drive the explosive energy release and efficient energy transport in solar flares.

## Appendix A: In-flight spectral calibration

RHESSI's spectral response was studied extensively prior to launch, both through mass modeling (using the GEANT modeling software[1]) and ground calibration. However, mass modeling is idealized and does not necessarily capture the true response of a real detector; ground calibration is done with line sources and therefore cannot measure the response across all energies. With attenuators engaged (see §3.2.1), the response at low energies (below ~10 keV in the A1 state, and below ~18 keV in A3) becomes quickly dominated by off-diagonal effects (see §3.2.3); because of the decreasing significance of the diagonal response, small changes in the count spectrum yield enormous changes in the inferred incident photon spectrum at these energies. At high counting rates, pulse pileup is also a concern; the spectrum falls so steeply with energy that even a ~1% pileup rate can contribute 10% or more of the counts at higher energies (above ~20 keV in A1, ~36 keV in A3), significantly affecting the spectral shape. Both effects can severely alter the physical parameters (e.g. plasma temperature) inferred from the spectra.

Thus, for precise and accurate SXR/HXR spectroscopy, it is important to achieve the best possible spectral calibration across the continuum of energies from ~3-100 keV. Although solar flares do not offer any "standard candles" from which to obtain an absolute calibration, the responses in different attenuator states and live time fractions can be cross-calibrated with one another, yielding a relative calibration that can then be normalized (if necessary) by comparison with other instruments.

### A.1 General principles

The measured count rate is determined from the incident photon spectrum and the instrument response; if we represent the observed count and incident photon spectra as vectors $\mathbf{c}$ and $\mathbf{p}$, respectively, where each vector component represents an energy bin, then we can write this relationship as:

$$\mathbf{c} = \mathbf{R} \cdot \mathbf{p} \qquad (A.1)$$

where $\mathbf{R}$ is the "detector response matrix" (see §3.2.3) that maps photon energies to count energies based on the instrument response, including the attenuator state and any dynamic effects such as pulse pileup (hence $\mathbf{R}$ is not necessarily fixed in time and is applicable only to the one measurement, but it varies in a known fashion). Thus, any given count energy bin (represented as a given component of $\mathbf{c}$) contains contributions from all incident photon energies (all components of $\mathbf{p}$) weighted by the instrument response for those incident energies (the corresponding row of $\mathbf{R}$).

For general spectroscopy, the goal is to determine the unknown $\mathbf{p}$ from the measured $\mathbf{c}$ and calibrated $\mathbf{R}$. This can be done via inversion or forward-modeling. For in-flight calibration, the goal is to determine $\mathbf{R}$; however, because $\mathbf{p}$ remains unknown, the equation is not solvable on its own. Instead, what can be done is to make multiple observations with "fixed" $\mathbf{p}$ but varying $\mathbf{R}$ (e.g. in different attenuator states), allowing a calibration of the variable components of $\mathbf{R}$ with respect to one reference state; this is cross-calibration.

---

[1] http://wwwasd.web.cern.ch/wwwasd/geant/

Assume for the moment that $\mathbf{p}$ is fixed, as can be approximately true for short timescales during the decay phase of a large flare, and that it is known (e.g. through forward-modeling). Then, defining vector multiplication as the component-by-component product vector:

$$\mathbf{v} \equiv (v_1, v_2, ...); \; \mathbf{w} \equiv (w_1, w_2, ...) \; \Rightarrow \; \mathbf{vw} = (v_1 w_1, v_2 w_2, ...); \; \mathbf{v/w} = (v_1/w_1, v_2/w_2, ...) \quad (A.2)$$

we can take the ratio of the count spectra observed in attenuator states A3 and A1 to yield:

$$\frac{\mathbf{c}_{A3}}{\mathbf{c}_{A1}} = \frac{\mathbf{R}_{A3} \cdot \mathbf{p}}{\mathbf{R}_{A1} \cdot \mathbf{p}} \quad (A.3)$$

For the energy range of ~3-100 keV, neglecting the live time-dependent offset and the detector resolution, the dominant factors in the detector response are the attenuation, K-escape, and pileup (the grid transmission is effectively uniform in this energy range, and other effects such as Compton scattering within the spacecraft are small in comparison). Then the response matrix can be written as the matrix product of these effects:

$$\mathbf{R} = \mathbf{P} \cdot \mathbf{K} \cdot \mathbf{A} \quad (A.4)$$

where the attenuation $\mathbf{A}$ is diagonal, and the K-escape $\mathbf{K}$ and pulse pileup $\mathbf{P}$ include off-diagonal terms; $\mathbf{A}$ and $\mathbf{K}$ are fixed by the detector geometry, while $\mathbf{P}$ varies in a determinable fashion based on the spectral shape, and thus is dynamic – a given value of $\mathbf{P}$ applies only to one specific measurement. For an A3→A1 transition during the flare decay, pileup is effectively negligible in the A3 state, since the live time must be near 100% to allow a transition to A1, thus $\mathbf{P}_{A3}$ can be taken as the identity matrix for this scenario. Additionally, because the only difference between A1 and A3 is the presence of the thick attenuator, we note that we may write $\mathbf{A}_3 = \mathbf{A}_{thick} \cdot \mathbf{A}_1$. With this, equation (A.3) becomes:

$$\frac{\mathbf{c}_{A3}}{\mathbf{c}_{A1}} = \frac{\mathbf{K} \cdot \mathbf{A}_{thick} \cdot \mathbf{A}_1 \cdot \mathbf{p}}{\mathbf{P} \cdot \mathbf{K} \cdot \mathbf{A}_1 \cdot \mathbf{p}} \quad (A.5)$$

Thus, the two observations differ only by the application of the thick attenuator in one, or pileup in the other. The count spectra $\mathbf{c}$ are measured; if $\mathbf{p}$ is known, then the ratio of the left-hand side (LHS) to the right-hand side (RHS) should be identically unity (a vector with all components equal to 1); any deviations from unity therefore represent a calibration error of either $\mathbf{A}_{thick}$ or $\mathbf{P}$.

The above requires knowledge of $\mathbf{p}$, which is generally obtained via forward-modeling (see Appendix B). Any error in the model could therefore be mistakenly construed as an error in the calibration, although it would only be a second-order effect since the error should be canceled to first order by taking the ratio. However, we can instead rewrite the above in an inverse rather than forward sense, as follows:

$$\mathbf{p} = \left(\mathbf{R}_{A3}\right)^{-1} \cdot \mathbf{c}_{A3} \; \Rightarrow \; \mathbf{c}_{A1} = \mathbf{R}_{A1} \cdot \left(\mathbf{R}_{A3}\right)^{-1} \cdot \mathbf{c}_{A3} \quad (A.6)$$

In principle, this is independent of any modeling of $\mathbf{p}$ and therefore depends only on the observations and calibrations. In practice, however, $\mathbf{R}_{A3}$ is often not nicely invertible; when using the fine (1/3-keV) energy bins required for precise spectroscopy, the inversion is "noisy," in that small variations in $\mathbf{c}_{A3}$ can lead to unrealistically-large variations in $\mathbf{p}$, which should vary smoothly, and often yields unphysically-small (i.e. zero) values. This noise is due primarily to the off-diagonal elements of $\mathbf{R}$; we can mitigate the effect by considering only the diagonal com-ponents. If we decompose $\mathbf{R}$ and $\mathbf{c}_{A3}$ into diagonal and off-diagonal components as $\mathbf{c}_{A3} = (\mathbf{R}_{A3})_{diag} \cdot \mathbf{p} + (\mathbf{R}_{A3})_{off} \cdot \mathbf{p} = (\mathbf{c}_{A3})_{diag} + (\mathbf{c}_{A3})_{off}$, then we have:

$$(\mathbf{c}_{A3})_{diag} = \left( \frac{(\mathbf{R}_{A3})_{diag} \cdot \mathbf{p}}{\mathbf{R}_{A3} \cdot \mathbf{p}} \right) \mathbf{c}_{A3} \quad \Rightarrow \quad \mathbf{c}_{A1} = \mathbf{R}_{A1} \cdot (\mathbf{R}_{A3})_{diag}^{-1} \cdot (\mathbf{c}_{A3})_{diag} \tag{A.7}$$

While this again introduces some dependence on modeling of $\mathbf{p}$, any uncertainty in $\mathbf{p}$ is now folded only into an uncertainty in $\mathbf{R}_{A3}$ rather than amongst both $\mathbf{R}_{A3}$ and $\mathbf{R}_{A1}$ as before. We note that while equation (A.7) is analytically identical to equation (A.3) [and hence to equation (A.5)], in practice, the actual numerical result (and model dependence) is somewhat different because of the ordering of operations. As above, taking the ratio of the LHS and RHS should yield a unity vector, and deviations from unity indicate calibration errors in the pileup $\mathbf{P}$ or the thick attenuation $\mathbf{A}_{thick}$.

The live time-dependent energy offset complicates matters somewhat. It is not incorporated into the general gain solution, which maps digital measurement channels to analog energy values, but is instead measured using the general gain as a reference. The offset thus appears to shift the entire spectrum towards higher energies and hence acts as a translation matrix; with this, equation (A.3) becomes:

$$\mathbf{R} = \mathbf{G} \cdot \mathbf{P} \cdot \mathbf{K} \cdot \mathbf{A} \tag{A.8}$$

Since $\mathbf{G}$ is dynamic, determined from the counting rate, it is applicable only to a given observation (much like $\mathbf{P}$). The offset is never negative because of how it is thought to arise (see §3.2.3), and therefore even at a time-averaged live time of near 100%, the average offset will be positive, though small. $\mathbf{G}$ can be determined by observations of sharp features with known positions (e.g. the Fe line complex; see below).

Finally, we consider the detector resolution. As with $\mathbf{A}$ and $\mathbf{K}$, this should be a static component determined solely from the properties of the detector; for the front segments, it is affected primarily by noise in the electronics when measuring the count energy, and thus is the final step in the instrument response chain. Incorporating the resolution into the general response matrix, equation (A.8) becomes:

$$\mathbf{R} = \mathbf{D} \cdot \mathbf{G} \cdot \mathbf{P} \cdot \mathbf{K} \cdot \mathbf{A} \tag{A.9}$$

When the energy bins are wider than the nominal resolution of ~1 keV, $\mathbf{D}$ is effectively the identity matrix; for finer bining, $\mathbf{D}$ is a quasi-diagonal matrix where the number of filled sub- and super-diagonal chords represent how much the resolution is oversampled (e.g. for 1/3-keV binning, the first two sub- and super-diagonal chords will be significant). As with $\mathbf{G}$, $\mathbf{D}$ can be determined by observations of sharp features, e.g. the Fe line complex.

These principles are equally applicable for transitions between A1 and A0, although such cases were not treated for this dissertation.

*A.2 Methodology*

As described in §4.2.1, comparison of RHESSI observations (using only detector G4) of flares in the A1 state with simultaneous observations from SOXS showed close agreement of the spectra from the two instruments in the 6-12 keV range, the usable range of overlap between them. The RHESSI observations were made with relatively high live time, so that pulse pileup was not significant. The agreement between RHESSI and SOXS (Figure A.1) confirmed the calibration of the A1 response for detector G4, which was taken as the reference observation state for cross-calibration.

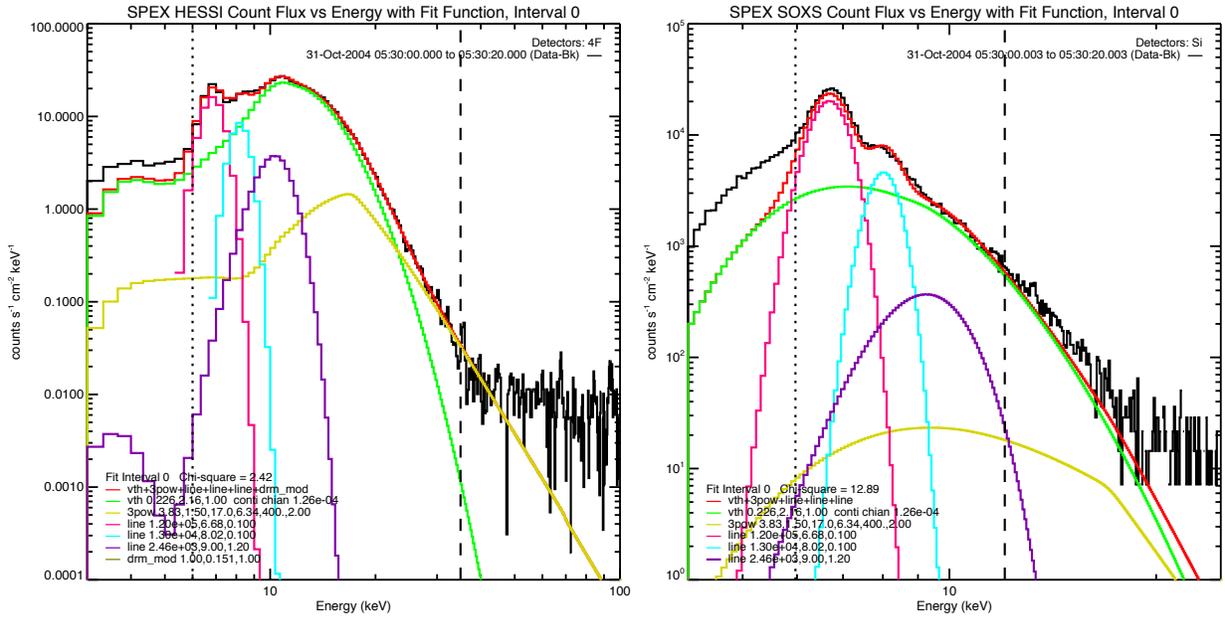

**Figure A.1** – [left] RHESSI spectrum in the A1 attenuator state, with model fit between 6 and 32 keV. [right] SOXS spectrum for the same time period with the model fit from the RHESSI spectrum; note the good agreement (within ~5-10%) in the 6-12 keV range, thus validating the A1 response above 6 keV. (Below 6 keV, the RHESSI response is dominated by off-diagonal contributions; above 12 keV, SOXS suffers from uncorrected pulse pileup.)

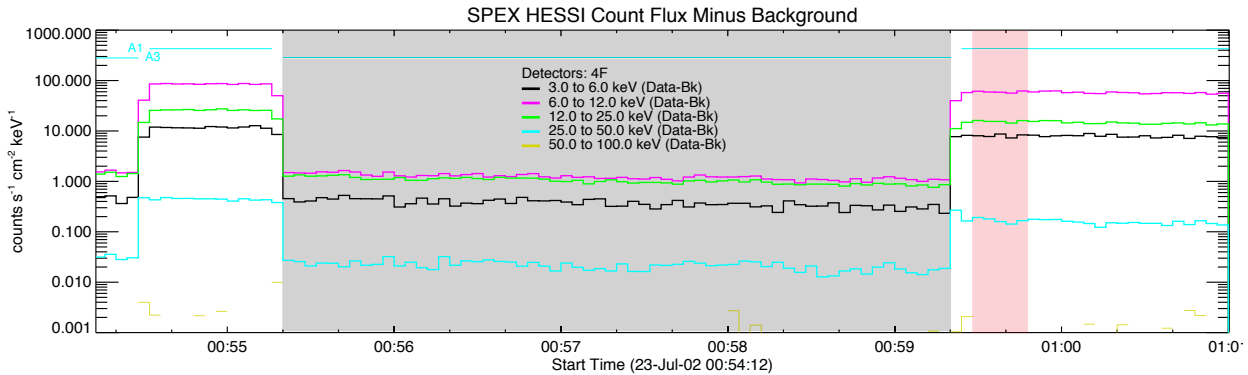

**Figure A.2** – RHESSI lightcurves in various energy bands during the flare decay, around the final A3→A1 attenuator transition. The A3 spectrum was integrated over the entire 4 minutes prior to the transition (gray shading) to improve statistics; the A1 spectrum was a 20-second integration just after the transition (red shading).

For cross-calibration, the decay phase of a flare provides the best possible conditions, as the incident photon spectrum is (presumably) changing very slowly and, given how the attenuators are controlled, the live time is near 100% just before a transition from A3 to A1, thereby allow-ing some separation of the effects of calibration error of the thick attenuator and pileup, respectively. For this research, we chose the final A3→A1 transition during the 2002 Jul 23 flare decay (Figure A.2), around ~01:00 UT – all subsequent discussion refers only to this flare – and used only detector G4, which had the best nominal ground-calibrated resolution of 0.98 keV (Smith *et al.* 2002). Since the A3 count rates were quite low, to maximize statistics, the A3 spectrum was integrated for 4 minutes (the time period between the final and next-to-final attenuator transitions); the A1 spectrum was integrated for 20 seconds. The 8 seconds surrounding the attenuator transition (two 4-second time bins) were omitted, to avoid cross-transition observations that would contaminate the results.

For both the A3 and A1 spectra, the live time-dependent offset was determined by fits to the Fe line complex. The complex's intrinsic width is only ~0.15 keV, finer than the instrument resolution; thus, a Gaussian photon model with this intrinsic FWHM was assumed, with a centroid fixed at 6.68 keV (the average centroid of the Fe line complex at flare temperatures of ~20-50 MK). The photon model was then convolved with the nominally-calibrated instrument response, and the resulting count-model Gaussian was forward-fit (see Appendix B) against the observations to determine the best-fit centroid offsets, which were taken as the offset values for the respective spectra. (Although the model fitting to determine the energy offset is dependent upon the calibration of the response matrix, any error in the fit offset due to an error in the response calibration is therefore second-order; it is further reduced by iterating the calibration procedure – see §A.3.) Because of how the offset is incorporated into the forward modeling soft-ware, it is not part of the static response matrix and therefore must be compensated for prior to calibration; as such, the A3 and A1 count spectra were each translated by their respective measured offset, to co-align their energy axes.

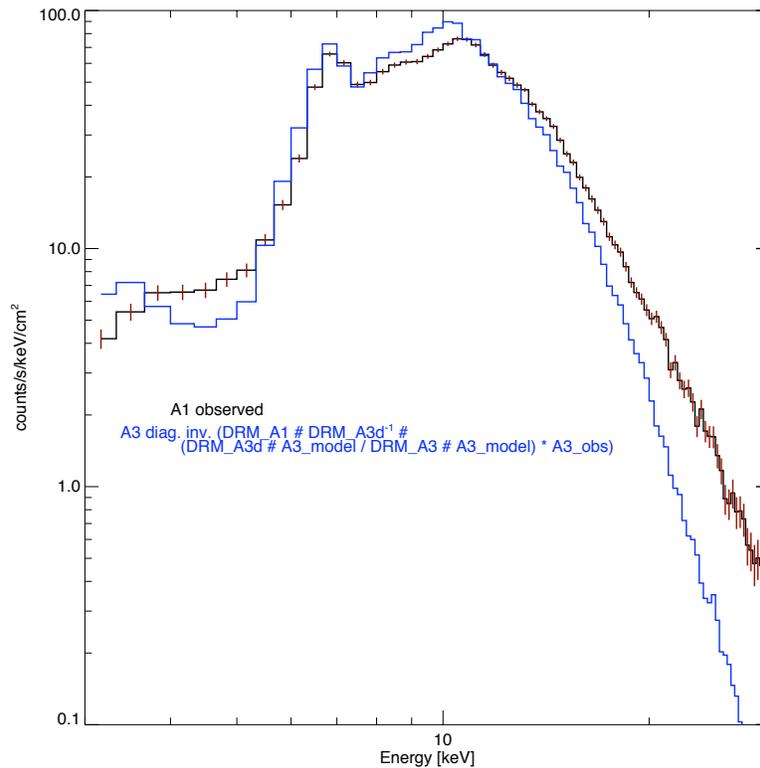

**Figure A.3** – Pre-calibration count spectra during the decay of Jul 23: the observed A1 spectrum (black) and the A3 spectrum "cast" into the A1 state following equation [A.7] (blue). The disagreements at low and high energies were used to calibrate the A3 attenuator response and pulse pileup, respectively.

Because the A3 spectrum is integrated over a fairly long period, the flare does decay noticeably, albeit slowly, over this time period. The A3 spectrum was multiplied by a constant factor (~87%) to account for the decay of the X-ray intensity from the midpoint time of the A3 spectrum to that of the A1 spectrum, normalizing the two spectra relative to one another. (In reality, the higher energies, e.g. 12-25 keV, decay slightly more rapidly than the lower energies, e.g. 6-12 keV, indicating that the flare also cools somewhat during this period; the energy dependence of the decay rate was ignored for this analysis, as it is a second-order effect, but may affect the pileup calibration and may thus be worthy of more thorough consideration in future analyses.)

Then, following the principle of equation (A.7), a photon model was forward-fit to the A3 spectrum; the diagonal contribution to the count spectrum was isolated and was then "cast" into an A1 state by multiplying by the inverted diagonal portion of the A3 response matrix and then by the full A1 response matrix, including the contributions from (nominally-calibrated) K-escape and pulse pileup (cf. Figure A.3). This "cast" spectrum (the RHS of equation [A.7]) was then compared to the actual observed A1 spectrum (the LHS of equation [A.7]); by comparing the ratio of the two with a unity vector, corrections to the thick attenuation and the pileup parameters were calculated as follows.

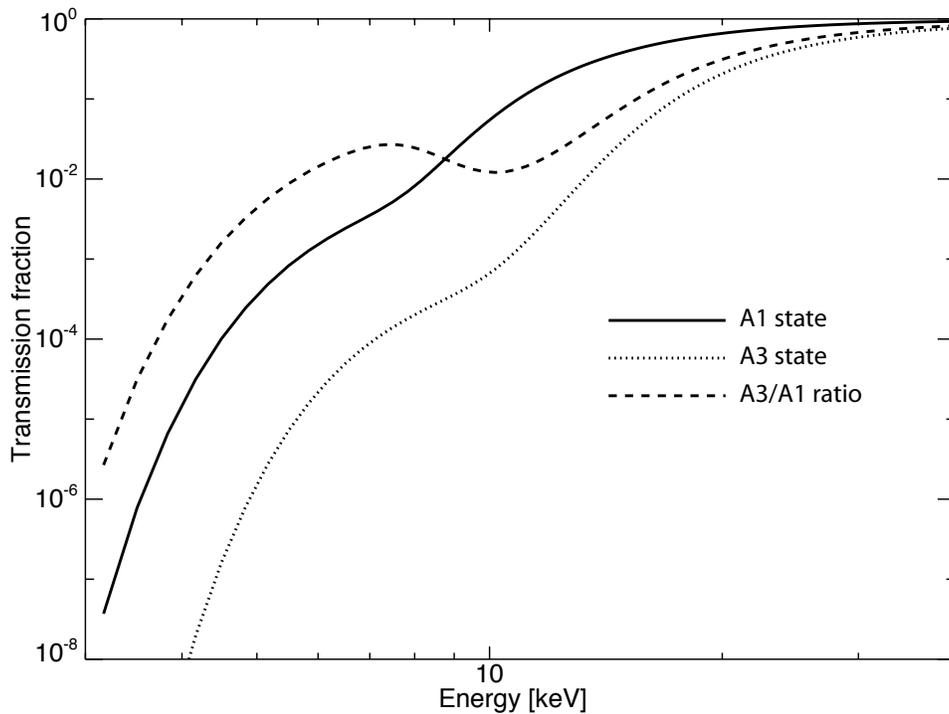

**Figure A.4** – Transmission fraction through the thin (A1) and thick+thin (A3) attenuators; unity represents the transmission without attenuators (A0). The ratio of the transmission fractions has a dip at ~10 keV.

### A.2.1 Thick attenuator calibration

In the A1 state, the count-rate peak is at ~10 keV, and pileup therefore becomes significant only above ~15-17 keV. Thus, any discrepancies between the A3 "cast" spectrum and the A1 observed spectrum below ~15-17 keV can be attributed primarily to calibration of the thick attenuator, per equation (A.3). The ratio of the two spectra was observed to deviate from unity between ~6 and ~17 keV (cf. Figure A.3), peaking at ~10 keV and trending smoothly to unity on either side. Examining the nominal transmission fraction for the A1 and A3 states (area-averaged over the 3 or 5 annular regions, for the respective states, of differing thickness [cf. Figure 3.3]), the ratio of the A3 and A1 transmission fractions (Figure A.4) shows a wide dip also centered at ~10 keV; the observed discrepancy between the two spectra is exactly what one might expect if this dip were somewhat too deep, consistent with a small error in the radius or thickness of one or more of the annular regions from which the A3 transmission fraction is calculated. The discrepancy from 6 to 17 keV – and unity elsewhere – was therefore taken as a multiplier for the A3 transmission and applied as a diagonal correction matrix, $\left(\mathbf{A}_3\right)_{calib} = \mathbf{A}_{corr} \cdot \mathbf{A}_{thick} \cdot \mathbf{A}_1$ (Figure A.5).

### A.2.2 Pulse pileup (software correction) calibration

Because pileup is a dynamic effect, it is not incorporated into the static response matrix directly but is instead applied separately; the specific values of the effective pileup matrix $\mathbf{P}$ depend on the spectral shape and counting rate, and differ for each time interval. However, they are calculated formulaically from a few time-independent parameters (see §B.2.5), and it is these parameters that must be calibrated. Therefore, no fixed diagonal-matrix correction can be determined, as above. Instead, we can use the forward-modeling framework to determine the optimal parameter values from which **P** is dynamically calculated. We have already determined an approximate **p**, needed to "diagonalize" the A3 observations (per equation [A.7]); we now need to determine **P**, and thus **R**$_{A1}$. Hence, rather than varying **p** with a fixed **R** to obtain a best fit to **c** (per equation [A.1]), we can instead vary **R** (or, more specifically, **P**) with a fixed **p** to do the same.

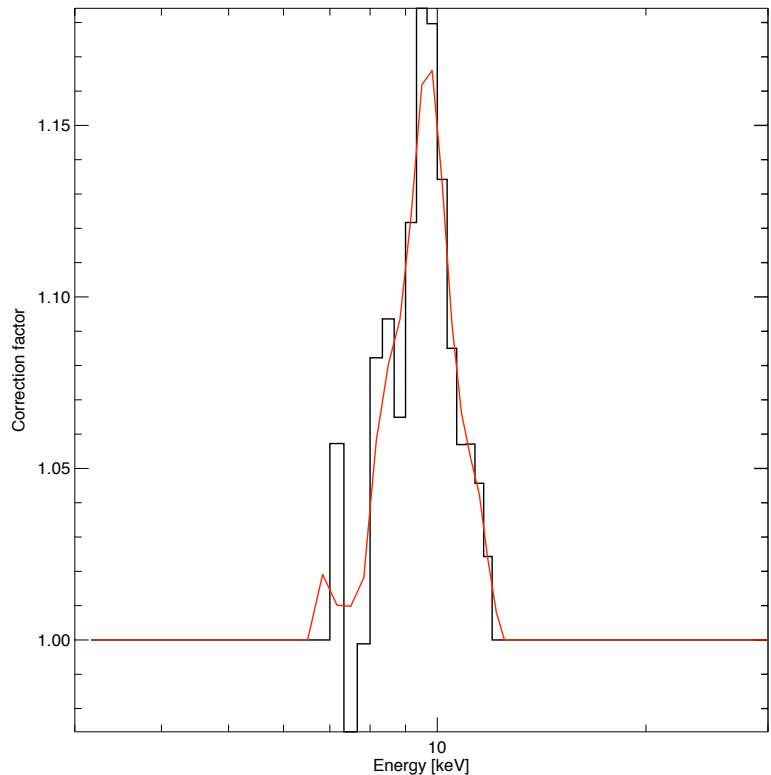

**Figure A.5** – Multiplicative correction factor for the A3 attenuator, derived from calibration; smoothing with a 3-interval boxcar algorithm yields the red curve, to reduce statistical noise from the narrow energy bins.

Thus, after determining the correction for the thick attenuator, a new photon model was determined by forward-fitting to the A3 count spectrum using the now-corrected A3 response matrix. This photon model was fixed and convolved with the nominal A1 static response matrix, not including pileup, to obtain an A1 count model. Pileup was then applied to the count model and the optimal pileup parameters determined by forward-fitting to the observed A1 spectrum.

### A.2.3 K-escape calibration

The K-escape matrix **K** factors equally into both **R**$_{A1}$ and **R**$_{A3}$, thus the comparison of A1 and A3, especially via the ratio method used above, is insensitive to calibration errors in **K**; the K-escape contribution cannot be verified or calibrated via the cross-calibration procedure. Using the nominal K-escape calibration, a photon model that yields a good fit to the observations at

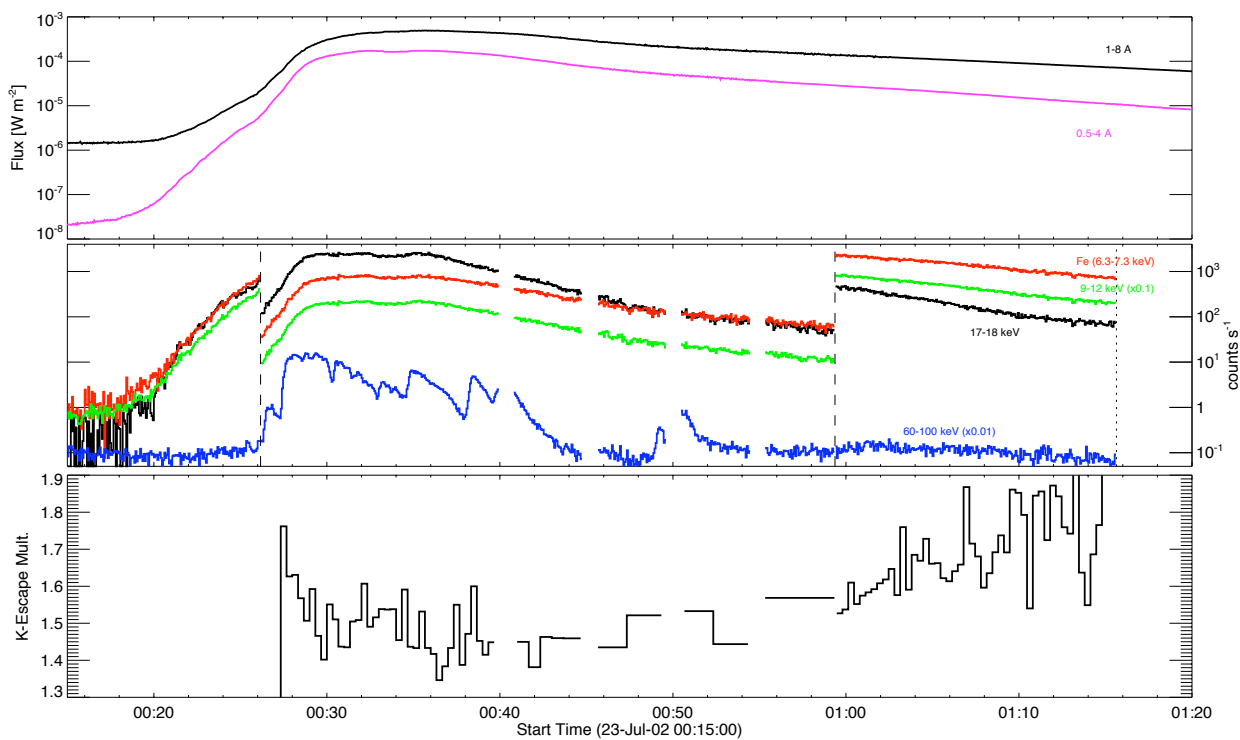

**Figure A.6** – Time profile of K-escape multiplier that yields a best fit to the ~4-6 keV data. The multiplier trends to a minimum value of ~1.45 during the A3 state, which must therefore be the optimal correction. (In the A1 state, the direct ~4-6 keV flux is not entirely negligible, and the best-fit multiplier is thus not independent of the incident photon flux.)

~14-15 keV nevertheless underestimates the observed ~4-5 keV counts – the energy range where a K-escaped ~14-15 keV photon would be measured – by ~31%. The "excess" observed counts must therefore reflect either attenuated ~4-5 keV solar photons not included in the model, or an error in the K-escape calibration; the former possibility can be largely excluded by comparing to observed GOES fluxes – GOES is more sensitive to <5 keV photons than is RHESSI, whose A3 attenuation of direct ~4-5 keV photons is greater than $10^7$. The ~4-5 keV photon fluxes inferred from the GOES observations, after attenuation, account for less than ~10% of the observed excess; the K-escape calibration is thus the dominant factor in explaining the excess.

Knowing this, the K-escape contribution can be calibrated in a manner similar to pulse pileup, by keeping fixed the photon model that fits the observations well at ~14-15 keV and varying the response to achieve a good fit at ~4-5 keV. We are limited to this very narrow energy range since we must ensure that K-escape is the dominant contribution to the response; below ~4 keV, the low-level discriminator calibration may be a factor, while above ~5 keV, the attenuator transmission is sufficiently high that direct photon flux is no longer small compared to the K-escape contribution. With such a narrow range, we can only obtain a zeroth-order correction to **K**, i.e. a scalar multiplier. However, since we are no longer cross-calibrating, we are not restricted to observing only the final A3→A1 transition; we can therefore fit across as many A3-state time intervals as possible to determine the multiplier more robustly. The best-fit multiplier value varies with time (Figure A.6), but does so primarily during the impulsive phase, when the HXR fluxes are high and hence when other sources of background (e.g. scattering of HXRs or bremsstrahlung generated within the spacecraft) may be significant. At other times, the multiplier reaches a constant minimum value; since the other contributing factors (direct ~4-5 keV flux and/or other background) must vary with time, this minimum value is therefore the most likely correction factor to the K-escape contribution.

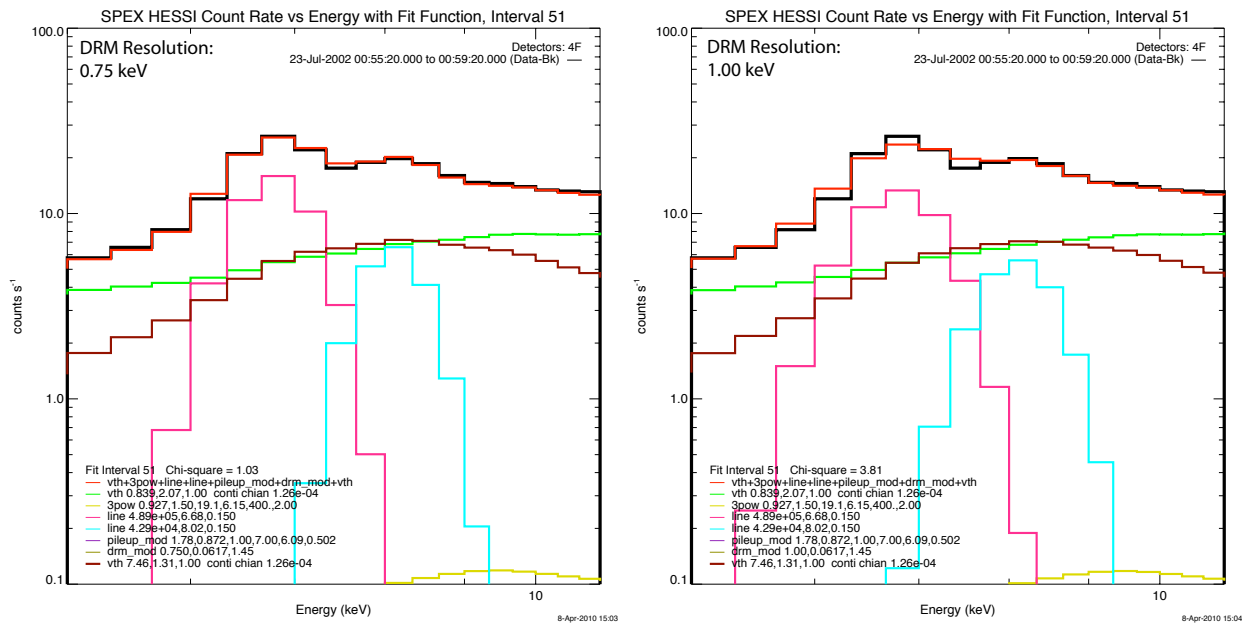

**Figure A.7** – Zoom of the Fe line complex in the A3 count spectrum; the detector resolution was calibrated by fitting to the observed width of the line. [left] Model fit for a resolution of ~0.75 keV. [right] The same model (with identical parameters) with the nominal resolution of ~1 keV. The smaller resolution is a significantly better fit.

*A.2.4 Detector resolution*

The analysis of the Fe line to determine the energy offset also revealed that the detector resolution was somewhat finer than expected for the A1 and A3 states; using the nominal resolution of ~1 keV, the count-model Gaussian was too wide compared to the observed width of the Fe line complex (Figure A.7). As with K-escape and pulse pileup, the detector resolution could be determined by fixing the photon model (in this case, the Gaussian model of the Fe line) and varying the response to achieve the best fit between the model Gaussian and observed line widths. Doing so revealed that the nominal detector resolution was overestimated by ~25% for the A1 and A3 states, as might be expected since all low-energy photons (e.g. the ~7-keV photons in the Fe line) are incident only in the very center of the front segment – a consequence of the attenuator geometry – where the electric field is strongest and charge collection is fastest – a consequence of the detector geometry.

*A.3 Results*

In "diagonalizing" the A3 count spectrum per equation (A.7), we effectively removed the (off-diagonal) K-escape contribution to the spectrum; this was required since it is precisely the off-diagonal terms of the response matrix that make its inversion so noisy. In doing so, however, we reduced the overall count rate at low energies (specifically below ~18 keV, where the K-escape contribution is non-negligible); the thick attenuator correction derived from this "reduced" spectrum is, consequently, somewhat too small. Additionally, because pileup does have a small effect even below the count-rate peak, fitting the pileup parameters changes the model fit somewhat at low energies, affecting the correction for the thick attenuator.

For both reasons, the calibration procedures detailed above were iterated; a convergent solution for all correction factors was achieved after 3 iterations. The detector resolution was determined to be ~0.75 keV in both the A1 and A3 states. The thick attenuation profile correction peaked at ~15% at ~11 keV, trailing smoothly to 0 at the energy boundaries of 5 and 17 keV. The probability of pulse pileup (see §B.2.5) was found to be nearly double the nominal value, with efficiency somewhat reduced from nominal, though we note that the pileup correction algorithm as currently implemented is still only an approximate model of pileup in the electronics – the best-fit parameters are likely to change as the algorithm is improved. The K-escape multiplier was determined from the A3 spectra to be ~145% of nominal; this multiplier also yields good agreement for A1 spectra, further validating the calibration.

These calibration improvements were implemented for the spectral analysis described in Chapter 4. The calibration procedure was performed only for one flare, so the results are not necessarily extendible, although preliminary analyses of another flare yield satisfactory results with these improvements. For robustness, however, the calibration procedure should be performed on multiple flares and the results averaged. If spectral analysis with detectors other than G4 is desired, this calibration should be performed for those detectors, as well. However, as currently implemented, the improvements derived from this procedure can be applied only to single-detector spectra; the response for detector-averaged spectra can not yet be corrected in this fashion.

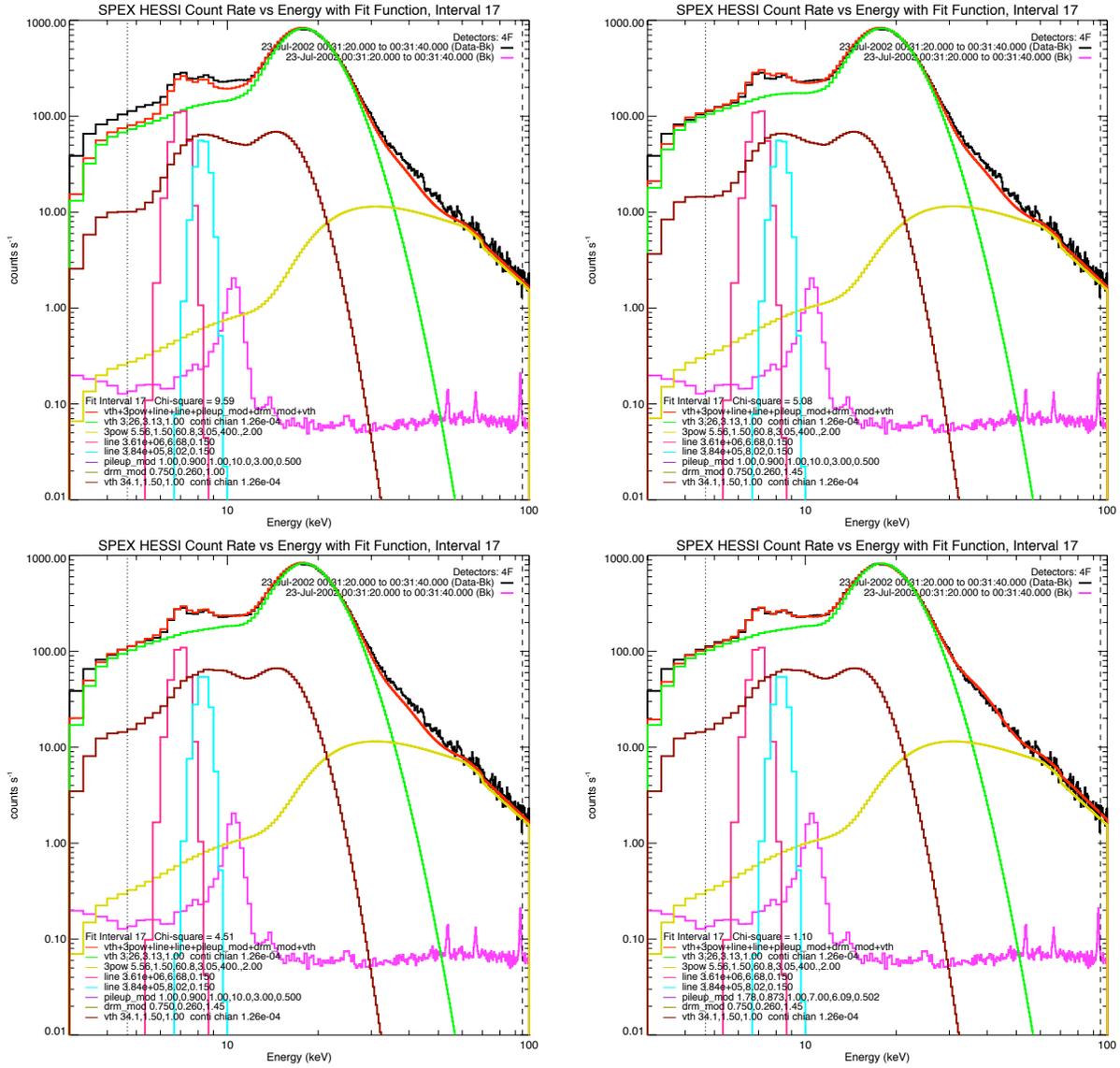

**Figure A.8** – Example A3 count spectrum with an identical model passed through the response matrix after different stages of calibration; the model parameters are determined from the *fully-calibrated* response. [top left] Nominal calibration; [top right] K-escape calibrated, other parameters nominal; [bottom left] K-escape and A3 response calibration, pileup nominal; [bottom right] all parameters calibrated.

## Appendix B: Forward-modeling with OSPEX

Forward-modeling analysis of RHESSI spectra is a straightforward, if involved, process. After accumulating an observed count spectrum for some time, with a given energy binning, the spectrum is corrected for decimation, dropouts, and live time to obtain a count rate (or flux, if the detector area is also divided out) spectrum. A parametric photon model is then assumed and folded through the response matrix, with dynamic effects such as pulse pileup included, to obtain a model count spectrum. By varying the model parameters and comparing with observations, the optimum parameters can be found through standard chi-squared minimization. Doing this manually for multiple spectra – which can be in different attenuator and decimation states with varying degrees of pileup and/or energy offset – can be impractical, however. The *Object Spectral Executive* (OSPEX) software package streamlines this process and affords us the flexibility to analyze spectra with arbitrary time and energy binning, attenuator state, decimation, etc.

### B.1 Operational overview

The detailed usage of OSPEX can be found in its online documentation[1]; we shall provide only a brief overview here. In general, OSPEX is independent of any particular instrument, and requires only the spectral data and instrument response matrix to be provided in a compatible format, with arbitrary time and energy binning. There is built-in compatibility for the spectrum (and response matrix) files produced by the RHESSI *SolarSoftWare* (SSW) IDL routines. Spectral analysis intervals can be selected to span an arbitrary number of time bins; the interval times and durations can be chosen on-the-fly so that multiple analyses with different series of intervals can be performed without having to reprocess the underlying spectral data. One or more intervals can also be designated as background times, and OSPEX will then fit a polynomial (from $0^{th}$ to $4^{th}$ order) to the fluxes over those intervals to approximate the background spectrum over the entire flare.

For a given analysis interval, OSPEX automatically corrects the count data for decimation, dropouts, and live time, determines the appropriate static response matrix (excluding pulse pileup and offset, as these are dynamic factors), and subtracts the calculated background. The user then defines a parametric photon model (see below), selects which parameters are fixed or free, and sets their initial values and allowable ranges; OSPEX convolves the photon model with the response matrix to obtain a count model and determines the optimum free parameter values via iterative chi-squared minimization, varying the free parameters to achieve the best fit (smallest chi-squared) between the model and observed spectra. The user can opt to fit over only a selected energy range (or set of ranges), such that only the selected energy bins contribute to the chi-squared calculation, allowing flexibility in avoiding irrelevant data (e.g. background) or fitting different model parameters over different parts of the spectrum. When analyzing multiple time intervals, OSPEX can automatically (if desired) set the initial parameter values of the current interval equal to the best-fit values of the previous interval, either forwards or backwards in time; the energy ranges used for fitting can differ between intervals.

Although the user may define any custom parametric function or set of functions for use in the photon model, OSPEX provides a built-in library of some commonly-used functions applicable for flare analysis. For the analysis presented in §4.2.2, the photon model included isothermal bremsstrahlung, broken power-law, and Gaussian components, corresponding to the OSPEX functional components *vth*, *3pow*, and *line*, respectively. Multiple instances of the same component can be included simultaneously, as was done for the *vth* and *line* components.

---

[1] http://rhessidatacenter.ssl.berkeley.edu/spectroscopy.html

## B.2 Advanced operation

Rather than using a polynomial fit to the background, it is possible to manually specify the background spectrum at each interval, for more precise background subtraction. Additionally, rather than fitting only the photon model parameters, OSPEX also allows fitting of limited aspects of the static and dynamic response, including the energy offset and pulse pileup. This allows for improved accuracy of the spectral fits.

### B.2.1 Background data replacement

The standard polynomial fit can only approximate the true background, which can vary differently at different energy ranges. For example, the non-solar X-ray background below ~10 keV is fairly constant, while the background above ~25 keV (and especially above ~50 keV) varies sinusoidally with the spacecraft orbit as it passes through areas of low and high magnetic latitude. Using the *replacedata* command within OSPEX, it is possible to feed in a custom time-varying background spectrum (appropriately binned in time and energy), which is then automatically subtracted from the data, to improve the accuracy of spectral analysis.

The non-solar X-ray background is due primarily to bremsstrahlung generated by cosmic rays or charged particles in the Earth's radiation belts as they interact in the atmosphere or within the spacecraft; the spectrum varies primarily with the spacecraft's position relative to Earth's magnetic field, i.e. its magnetic latitude and longitude. When RHESSI is in eclipse (in Earth's shadow), all measured counts must be non-solar; it is therefore possible to empirically determine the orbital dependence of the non-solar background by recording the spectra measuring during eclipse as a function of magnetic latitude and longitude. Statistics can be improved by averaging many spectra measured at the same magnetic point, e.g. over a few months or years (J. McTiernan 2007, private communication). From this, the non-solar background can be estimated even when RHESSI is in sunlight, given only its orbital position (Figure B.1). The background also depends on the fluxes of cosmic rays and energetic particles, which is not fixed in time – especially after a flare, which can release high fluxes of solar energetic particles towards the Earth – but these variations are assumed to average out over the months/years used for the background determination.

For the spectral analyses presented in this dissertation, we used this method to estimate the non-solar background as a function of time, since RHESSI's orbital parameters (from which its magnetic position can be calculated) are known to high precision. We then fed this time-varying background into OSPEX using its *replacedata* command, so that the spectrum in each time interval was as accurately representative of the true solar spectrum as possible. In practice, this has negligible effect for the data below ~20-25 keV, where the flare emission is 10-1000 times greater than the non-solar background flux. Because the spectrum is so steeply falling, however, the background can become important above ~25 keV, especially during the flare decay when the solar HXR emission is weak or negligible; this background estimation technique improves the accuracy of measurements that depend strongly on the spectrum above ~25 keV, e.g. the super-hot temperature.

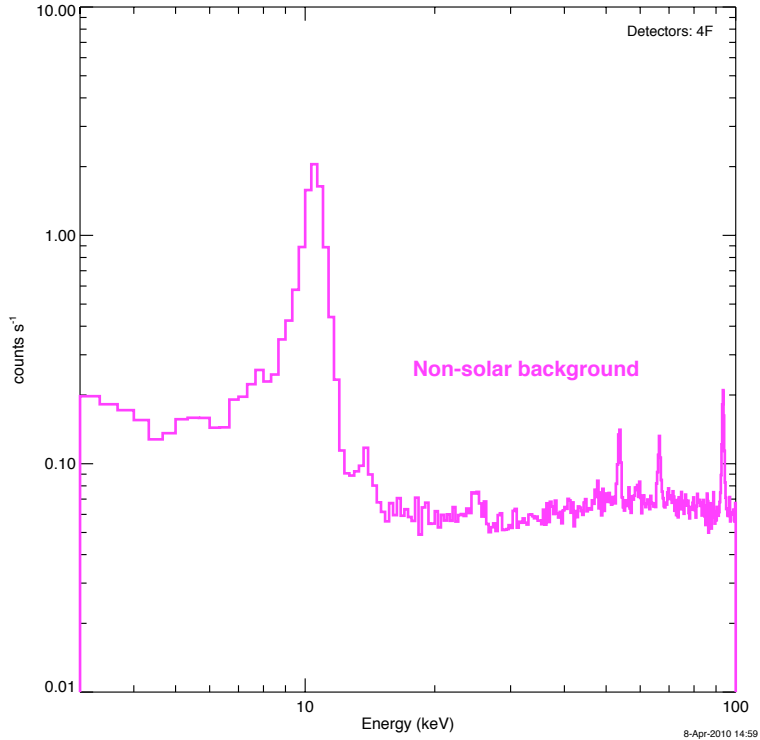

**Figure B.1** – Sample count rate spectrum for non-solar background, determined by averaging nighttime spectra at similar magnetic points over many orbits.

### B.2.2 Energy offset and detector resolution

When analyzing single-detector spectra, OSPEX allows modification of certain components of the instrument response (§3.2.3). The detector resolution and energy offset can both be modified using the *hsi_drm_mod* functional component; although presented as part of the photon model, this component influences the resulting model count spectrum by modifying the response matrix rather than the photon model spectrum directly. The detector resolution is implemented as a multiplier to the nominal value of ~1 keV; in the A1 and A3 states, analysis of the Fe line complex during calibration suggests that the resolution is ~0.75 keV (per §A.2.4), thus we fix this multiplier at 0.75 for the analysis of §4.2.2.

The offset is implemented as an absolute (positive) shift in keV, e.g. a value of 0.1 shifts the model spectrum upwards in energy by 0.1 keV. When used as a fit parameter, this value indicates how much higher in energy the observed count spectrum is, compared to the reference baseline. Although the offset depends on live time, a specific functional dependence has not yet been determined, thus we leave this parameter free during fitting of the Fe line complex, whose photon-model centroid is assumed fixed at 6.68 keV – the line is the sharpest feature in the flare spectrum and thus its centroid is currently the best measure of the offset.

### B.2.3 K-escape

Although not normally allowed by OSPEX, we wrote a custom modification to the software to implement the scalar multiplier for K-escape efficiency discussed in §A.2.3. We added a third parameter to the *hsi_drm_mod* component to multiply the nominal K-escape matrix component of the response matrix. During calibration, this was a free fit parameter to determine its optimum value, found to be ~1.45; it was subsequently fixed at this value for the analysis of §4.2.2.

### B.2.4 Correction to the thick attenuator

OSPEX also has no built-in capability to implement the calibration correction determined for the thick attenuator (§A.2.1). We wrote another custom modification to the software, implemented via a non-user-configurable (i.e. no changeable parameter) addition to *hsi_drm_mod*, that applies the thick attenuator calibration correction to the attenuation component of the response matrix when the matrix is generated and with the proper order of operations.

### B.2.5 Pulse pileup

Although not enabled by default, OSPEX includes a software correction to compensate for the small fraction of piled-up events that are missed by the on-board hardware rejection circuit, either because the photons were separated by less than ~800 ns (the shaping time constant of the fast pulse-shaper amplifier) or because one or more of the photons' energies fell below the fast shaper's low-level discriminator (LLD) threshold of ~7 keV (see §3.2.1). As is normal in OSPEX, the pileup correction is implemented in the forward sense, adding the effects of pileup to the model for comparison with the (unadulterated) observations.

The pileup behavior is controlled via the *pileup_mod* functional component that, like *hsi_drm_mod*, is presented as part of the photon model but actually operates by modifying the response rather than the model. However, the behavior of pileup is strongly dependent upon the amplitude and shape of the count spectrum; its matrix representation therefore depends not only upon its specific functional parameters (described below) but also upon the spectrum itself, unlike the other response components whose matrix representations are either static (e.g. the attenuation) or determined entirely from their functional parameters (e.g. the offset). The exact workings of the correction are beyond the scope of this appendix, but in brief, *pileup_mod* takes the model count spectrum – i.e. the spectrum after the application of all other response components (attenuation, K-escape, energy offset, and detector resolution) – and convolves it with itself to $n^{th}$ order, then sums the $n$ convolutions with weighting coefficients based on the total count rate and the specific functional parameters, which are as follows.

The probability of pileup is implemented as a multiplier for the nominal fast shaper time constant, thus a value of 1.5 represents a time constant of ~1200 ns. The efficiency of pileup, i.e. the ratio of the measured energy of the piled-up event compared to the sum of the energies of its constituent counts, is also a multiplier, implemented as part of the convolution coefficient. Two other parameters represent the effect of the fast LLD: an energy threshold, given in keV, and the "enhanced probability" of pileup for counts below this threshold (as they can pile up with counts of any energy, regardless of their timing), implemented as an effective multiplier for the probability parameter applied only below the threshold energy. (Two other parameters are also present in the software, but were not relevant for this study.)

During calibration of pileup (§A.2.2), the pileup probability, efficiency, and LLD "enhanced probability" were left as free parameters and fit to determine their optimum values, then fixed at these values for the spectral analysis of §4.2.2. The LLD threshold energy was always fixed at ~7 keV, the nominal value, as the calibration procedure was largely insensitive to reasonable (i.e. few keV) changes in the value, so no optimum value could be determined.

*B.3 Caveats*

Especially considering the advanced operations described above, OSPEX is fairly compli-cated, and it is important to have a solid understanding of how it and its components work before interpreting any spectral analysis results. In particular, one must understand how each functional component affects the model and/or response, and thus which parameters should be fixed or re-main free to fit. One must also understand over which energy ranges certain model parameters may be relevant and/or sensitive, so as to restrict the energy bins used for fitting to only those that are needed; additional energy bins outside the range of interest/relevance can still affect the chi-squared value and therefore the final best-fit parameter values. As the complexity of the model increases, so too does the complexity of the chi-squared space; local chi-squared minima may exist, and the fit parameter values obtained may thus not necessarily be the absolute best fit overall. As with all forward-modeling analyses, it is imperative to interpret the results within the context of the model and of physics – a good fit means the model is consistent with the data, but neither excludes other models nor implies that the fit model is actually physically plausible. All of these must be considered when interpreting model fit results.

The parameter uncertainties reported by OSPEX are also important to understand. OSPEX calculates the free parameter uncertainties from their covariance matrix, which assumes that the chi-squared space is "well-behaved" (smoothly-varying and non-degenerate, among other things). While these numbers can give some idea of the stability of the fit results, they do not account for interdependence between fit parameters (e.g. between the temperature and emission measure values of an isothermal emission component) and can often underestimate the true un-certainties when the chi-squared space is not well-behaved (e.g. when it is degenerate along one or more dimensions). In principle, it would be better to map the chi-squared space around the best-fit values to determine the parameter uncertainties more robustly, especially for interde-pendent parameters; in reality, this is often impractical to implement, but nevertheless these con-cerns must be kept in mind when interpreting OSPEX-reported uncertainties.

OSPEX also includes a global parameter to quantify systematic uncertainty, e.g. uncertainty regarding the detector calibration. The *uncert* parameter is user-selectable and is added to the statistical (Poisson) data error, thereby affecting the overall chi-squared value and the uncertain-ties for the fit parameters. This treats the systematic uncertainty as a stochastic one, which is principally incorrect; it is therefore advisable to leave this parameter at 0%, with the understand-ing that the reported chi-square value may thus be an upper limit as it does not account for the additional uncertainties associated with the instrument.

Finally, we note that the pileup correction in *pileup_mod* does not fully, or fully correctly, model the pileup behavior in the electronics, and this may introduce some error into the results. Specifically, the fast LLD threshold is not a sharp boundary in energy but is actually a sigmoid of some width; this is not yet modeled in the software. Additionally, *pileup_mod* as currently implemented applies the pileup correction after the detector resolution and energy offset have been applied; per equation (A.9), this order of operations is slightly incorrect, as it means that the effects of the energy offset and resolution broadening are themselves piled up when they should not be. Although the actual effect is likely negligible since the detector resolution is quasi-diagonal and the offset is small, in principle these effects should be applied in the correct order, although the software does not currently include that capability.

**Appendix C: Imaging analysis techniques**

RHESSI's imaging capabilities, and the flexibility with which one can manipulate the imaging data, offers a powerful complement to its spectroscopy. Images are useful both qualitatively, allowing spectroscopic results to be placed in a spatial context, and quantitatively, providing independent measurements that can be combined with spectral analysis to yield richer results than could be obtained from either method alone.

*C.1 Source morphology*

Determining the morphology of a flare source, especially as a function of energy, can be critical to placing spectral observations in context. The spatially-integrated spectrum may be fit with any arbitrary model, whether thermal or non-thermal; however, a good fit does not necessarily imply that the model is the correct one, and it may be possible to obtain equally good fits with quite disparate models (e.g. during the pre-impulsive phase of 2002 Jul 23). Source morphology as a function of energy can provide a reality check for the spectral modeling, with some assumptions and caveats. For example, multiple small compact sources, especially at high energies, may likely be non-thermal footpoint emission, while a large diffuse source, especially at low energies and especially in the shape of a loop, may likely be thermal emission. Placing RHESSI images in context with images from other instruments, e.g. TRACE, can further help distinguish between plausible models.

Of course, there are exceptions to the rules of thumb above, such as for extended non-thermal coronal sources or compact thermal loops whose size may be below the imaging resolution. Additionally, it is important to understand the limitations of the various imaging techniques (§3.3.3) being used, as some may introduce spurious sources (imaging artifacts) while others may smear multiple compact sources into a large extended source. It is also important to note that images are not currently corrected for pulse pileup or K-escape (or, indeed, any off-diagonal contribution to the response), so images at a given energy may contain contributions from photons of a different energy, either higher or lower depending on the effect being considered; the morphology of the observed source may therefore not reflect that of the true source entirely accurately. Nevertheless, with these limitations in mind, the general morphology of a source can serve as an invaluable sanity check on the plausibility of a model derived from spectral analysis.

*C.2 Source size*

Source size is a quantitative measurement that provides complementary information to spectral fits, under certain assumptions. For example, for a thermal source model, the emission measure combines with the volume estimated from imaging to yield the thermal electron density; for a non-thermal power-law model, the spectral index and cutoff combine with the source area to yield the non-thermal energy flux deposition. Sources sizes can also yield length scales for calculations of particle/wave travel times or temperature gradients.

For the thermal sources analyzed in this dissertation, we developed a semi-automated method of calculating source sizes, particularly applicable to round or elliptical sources. Images were generated in the ~6.2-8.5 keV energy band, which spans both the Fe and Fe/Ni lines as well as the underlying continuum. From spectral modeling, this energy range is generally entirely dominated by thermal emission. (The ~16.2-18.5 keV band is also most often thermal, so any K-escape contribution should not affect the source size or morphology significantly under the assumption that the same thermal source dominates the emission in both energy bands.) The images were reconstructed using CLEAN with uniform weighting (§3.3.3), using grids 3, 4, 5, 6, 8, and 9 (grid 7 was excluded due to detector G7's high LLD threshold, which exceeds the energy range of the image).

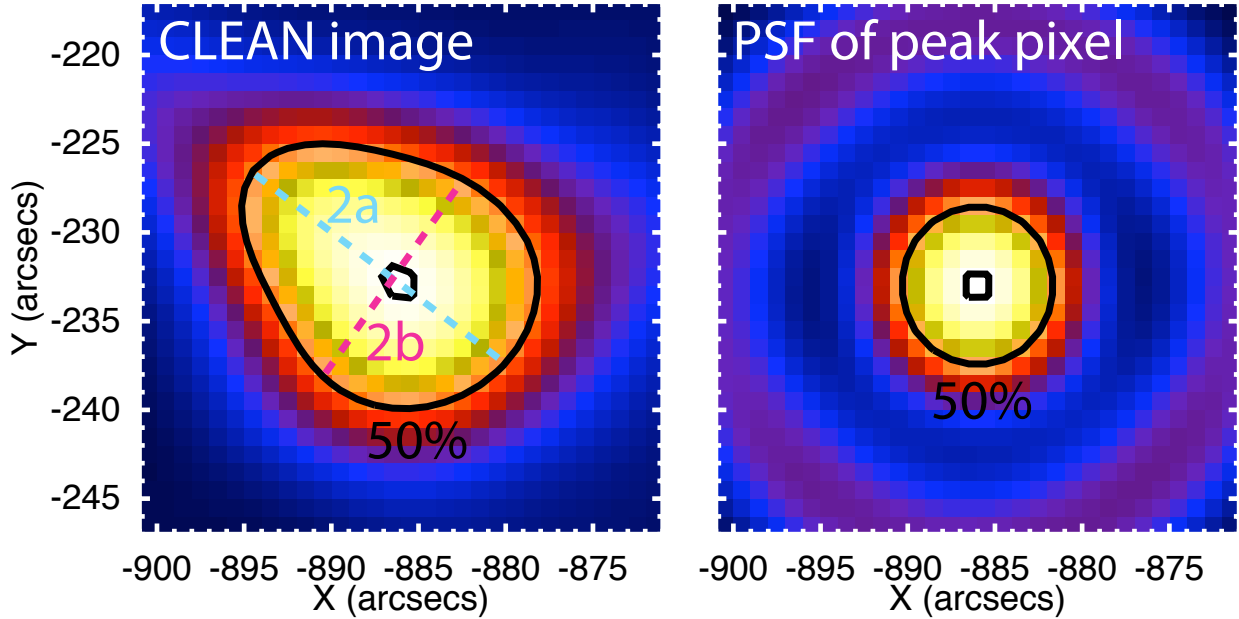

**Figure C.1** – [left] RHESSI 6.2-8.5 keV image used to estimate the thermal source volume. The "FWHM area" was measured by manually measuring the approximate axes of the 50% contour, and by automatically totaling the area within the contour. [right] The point-spread function for the peak pixel in the image at left. For the manual method, the PSF radius is subtracted in quadrature from the CLEAN axis measurements; for the automatic method, the area of the 50% PSF contour is subtracted from the CLEAN source area.

For each image, the contour of 50% of maximum intensity is effectively the source's FWHM; it was manually measured, via mouse cursor, to determine its length *2a* and width *2b*, or major and minor axes, respectively (cf. Figure C.1, left). Since CLEAN does not deconvolve the point-spread function (PSF) – the broadening of the image due to the finite angular resolution of the instrument – the measured axes were larger than the true values; the PSF adds in quadrature with the true source size to yield the CLEAN image, so the true source size can thus be recovered by subtracting, in quadrature, the PSF from the measured axes ($a = \sqrt{a_{meas}^2 - r_{PSF}^2}$, and similarly for *b*). Then, we define the true source area $A = \pi ab$, which represents the area of the source with >50% of the maximum intensity, i.e. the "FWHM area." Since the images are two-dimensional, there is no information about the source depth; if we assume that an elliptical area translates to an ellipsoidal geometry, however, with the (unseen) source depth equal to the smaller of the two measured dimensions *b*, then we can estimate the source volume $V_{ell} = (4/3)\pi ab^2$.

This procedure can potentially introduce some human bias into the measurement, since the length and width are determined manually. We can instead use a fully automated process whereby the "FWHM area" is calculated by summing the number of imaging pixels with intensity $\geq 50\%$ of the peak-intensity pixel; we can then define the effective source radius $r_{eff} = \sqrt{A_{meas}/\pi}$. However, we still need to subtract in quadrature the PSF; we do this by calculating the "FWHM area" of the PSF for the peak pixel (Figure C.1, right), which can be obtained from the CLEAN algorithm, and subtracting it from the measured area, yielding $r_{eff} = \sqrt{(A_{meas} - A_{PSF})/\pi}$. Since we now have only a single measurement, we must assume a spherical source geometry, thus $V_{sph} = (4/3)\pi r^3$. This method introduces no human error since it is entirely automated; however, because it assumes spherical geometry, it necessarily overestimates the volumes of elliptical sources. If we equate $r_{eff} = \sqrt{ab}$, then $V_{sph} = \left(\sqrt{a/b}\right)V_{ell}$, and since $a > b$, the spherically-calculated volume must be greater. This is compounded by the PSF subtraction method, which tends to overestimate $r_{eff}$, since

$$r_{eff}^4 = \left(\left(A_{meas} - A_{PSF}\right)/\pi\right)^2 = \left(a_m b_m - r_P^2\right)^2 = a_m^2 b_m^2 + r_P^4 - 2a_m b_m r_P^2$$

$$a^2 b^2 = \left(a_{meas}^2 - r_P^2\right)\left(b_{meas}^2 - r_P^2\right) = a_m^2 b_m^2 + r_P^4 - \left(a_m^2 + b_m^2\right)r_P^2$$

$$\Rightarrow r_{eff}^4 - a^2 b^2 = -2a_m b_m r_P^2 + (a_m^2 + b_m^2)r_P^2 = \left(a_m - b_m\right)^2 r_P^2 > 0$$

$$\Rightarrow r_{eff} > \sqrt{ab}$$

(C.1)

For both reasons, $V_{sph} > V_{ell}$, so the automated method yields larger calculated volumes, as can be seen in Figure C.2.

Both methods were evaluated to determine their robustness and relative uncertainties. Sources were simulated, using Monte Carlo code built into the RHESSI *SSW* package, for 13 different 2D Gaussian source configurations, all of equal intensity. The source areas were then measured using both of the above methods, and the tests repeated for 5 trials per source configuration; for each configuration and measurement technique, the 5 measurements were averaged and their standard deviation taken as the uncertainty. In all cases, the two techniques yielded measurements accurate to within ~15% of the true source area, with the larger uncertainties generally correlated with larger source sizes where the statistics of the 50% contour are poorer due to the diffusion of equal intensity over a greater area compared to the smaller sources. Both techniques yielded roughly equal uncertainty, indicating that the randomness from the simulation and manual measurement exceeds the overestimation effect of the spherical approximation. The ~15% area uncertainty translates to a ~23% uncertainty in the volume, which is likely smaller than the error introduced by assuming an *ad hoc* value for the (unseen) third dimension.

Source sizes may also be determined using visibilities (§3.3.4). The *vis_forward_fit* algorithm fits a model source to the observed visibilities; this is technically not an imaging technique since it does not require any actual image reconstruction, but is nevertheless included here because it provides the same information. The forward-fitting algorithm can fit a 2D elliptical Gaussian, including one with curvature (e.g. a loop) to yield the source axes; since visibilities are linear with definable uncertainties, this procedure also yields well-defined errors for the measurements. However, the goodness of fit is strongly dependent upon the actual source morphology and preliminary testing on simulated data revealed inconsistent results; therefore, the "FWHM" methods described above were used for our analyses here.

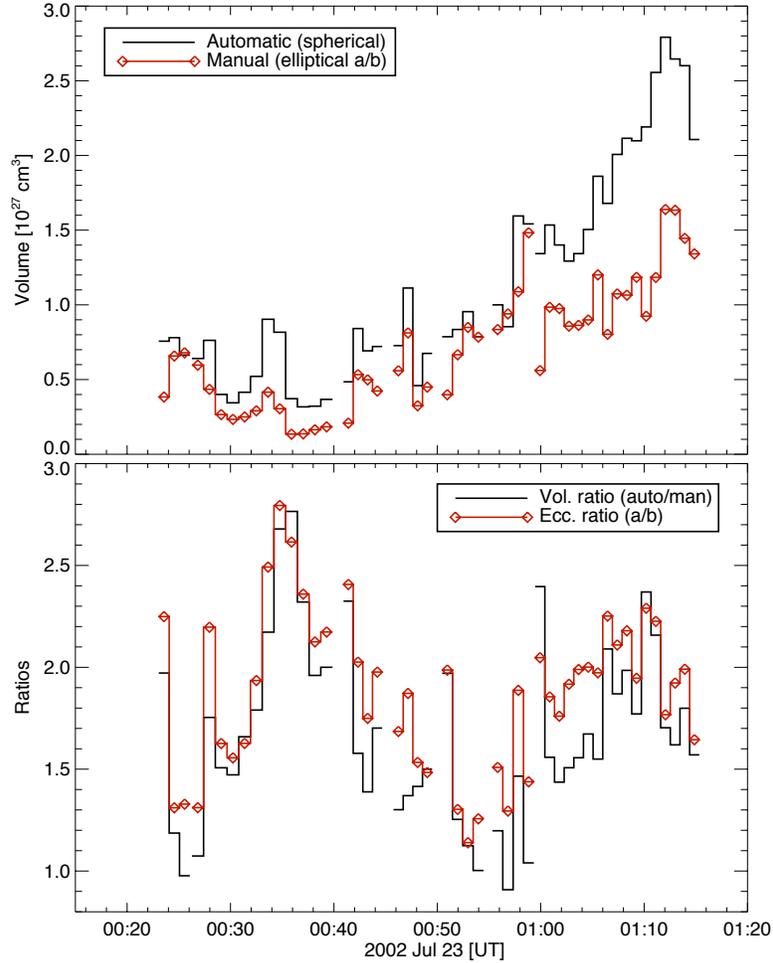

**Figure C.2** – [top] Thermal source volume as determined via manual axis measurement (red) and the fully-automated spherical approximation (black). The spherical approximation always yields a larger value. [bottom] Ratio of the automatically- and manually-determined volumes (black), compared with the aspect ratio (ratio of major and minor axes) of the manual measurements; the variations are strongly correlated.

## C.3 Imaging spectroscopy

Because images can be created at arbitrary energies, it is possible to perform spectroscopy of a specific spatial region by examining the image intensity within that region as a function of energy; this is limited by the statistics of the imaging, of course. This must be done carefully, however, because of how the images are reconstructed. Back-projection images are unsuitable for imaging spectroscopy as they still contain sidelobes, the artifacts of the reconstruction, thus spectroscopy should be performed on images created with higher-order algorithms such as CLEAN, UV_SMOOTH, Pixon, or others. However, as noted in §C.1, these images are not corrected for the off-diagonal contributions; therefore, once the intensity as a function of energy is determined, imaging spectroscopy follows much the same procedure as spatially-integrated spectroscopy, where a photon model is folded through a response matrix and fit to the observed count spectrum.

Despite poorer statistics from the image reconstruction, imaging spectroscopy does offer two distinct advantages over spatially-integrated spectroscopy. First and foremost, it is possible to analyze the spectra of multiple spatially-distinct regions, obtaining fit parameters for each region simultaneously which may reveal more complex information than is found in the spatially-averaged spectrum. Second, the image reconstruction by design excludes signal that is not modulated by the grids, which automatically removes essentially all non-solar background, obviating the need for advanced background estimation and subtraction techniques (§B.2.1).

*C.4 Source deconvolution with visibilities*

Regular images cannot be combined linearly, because their reconstruction algorithms are inherently non-linear (except for back-projection, which is linear but of limited quantitative use for measurements). Visibilities (§3.3.4) are linear, however, and therefore offer a means of essentially linearly combining images, allowing more accurate quantitative measurements to be made. For example, images can be corrected for K-escape by linearly combining visibilities in the desired energy range with those from energies ~10 keV higher, appropriately weighted based on the K-escape contribution for that energy range as determined from the (calibrated) instrument response (§3.2.3 and §A.2.3). Although more complicated, images could potentially be corrected for pulse pileup in a similar manner.

These methods combine visibilities with the spectral response, but it is also possible to combine the visibility information with the results of spectral modeling to yield even more powerful information. Specifically, by fitting a model to the spatially-integrated spectrum, one can determine the relative contribution of each model component to the count spectrum at each energy. Using this information, one can then linearly combine visibilities in order to isolate the visibilities from each specific model component, essentially decomposing a composite image into its constituent sources.

One application of this technique is to use "relative" visibilities to create a temperature map – imaging as a function of temperature rather than as a function of energy – as briefly described in §4.2.3. Relative visibilities are visibilities that have been normalized to unit amplitude; a regular visibility can thus be written as an amplitude times a relative visibility:

$$\Phi = A e^{iv} \equiv A\varphi \qquad (\text{C.2})$$

Say that we have obtained a good fit to an observed count spectrum using a model containing $N$ thermal components. Then, at each of the $N$ energies (or energy bands), we know both the total count rate and the count rate from each thermal component; we can thus write the total count rate as the sum of the fractional contributions. Since fractional contributions must sum to 1, then the "normalized" count rate for $N$ energies yields the matrix pseudo-equation:

$$\mathbf{c} = \mathbf{F} \cdot \mathbf{t} \implies \begin{bmatrix} 1 \\ 1 \\ \vdots \\ 1 \end{bmatrix}_c = \begin{bmatrix} f_{T_1 E_1} & f_{T_2 E_1} & \cdots & 1 - \sum_{i=1}^{N-1} f_{T_i E_1} \\ f_{T_1 E_2} & f_{T_2 E_2} & \cdots & 1 - \sum_{i=1}^{N-1} f_{T_i E_2} \\ \vdots & \vdots & \ddots & \vdots \\ f_{T_1 E_N} & f_{T_2 E_N} & \cdots & 1 - \sum_{i=1}^{N-1} f_{T_i E_N} \end{bmatrix} \cdot \begin{bmatrix} 1 \\ 1 \\ \vdots \\ 1 \end{bmatrix}_t \qquad (\text{C.3})$$

where the $N$-element vector $\mathbf{c}$ represents the total count rate at each of $N$ energies, normalized to 1 (thus it is an all-unity vector); the $N$-element vector $\mathbf{t}$ represents the thermal emission at each of $N$ temperatures (i.e. each of the $N$ thermal components); and the matrix $\mathbf{F}$ is the fractional contribution matrix (where the element $f_{T_i E_j}$ is the fractional contribution of the $i^{th}$ thermal component to the $j^{th}$ energy bin). Then, by definition, this equation also represents the scaling of the relative visibilities: the total observed relative visibilities at each energy (represented as the vector $\mathbf{c}$) are the sum of the relative visibilities from each thermal component (represented as the vector $\mathbf{t}$) weighted by their respective fractional contributions.

Thus, by inverting the equation, we obtain the relative intensity of each thermal component as a weighted sum of the relative count rate at each energy: $\mathbf{t} = \mathbf{F}^{-1} \mathbf{c}$. The inverted matrix $\mathbf{F}^{-1}$ is exactly the weighting matrix for the relative visibilities at each of the $N$ energies, which we can combine to obtain $N$ new sets of visibilities, each corresponding to a single thermal component. We can then use a visibility image reconstruction algorithm, such as UV_SMOOTH (Massone *et al.* 2009), to plot the images of these components, yielding images as a function of temperature. For the specific 2-temperature case treated in §4.2.3, equation (C.3) becomes:

$$\begin{bmatrix} v_{sh} \\ v_c \end{bmatrix}_t = \frac{1}{f_{sh2} - f_{sh1}} \begin{bmatrix} f_{sh2} - 1 & 1 - f_{sh1} \\ f_{sh2} & -f_{sh1} \end{bmatrix} \cdot \begin{bmatrix} E_1 \\ E_2 \end{bmatrix}_c \tag{C.4}$$

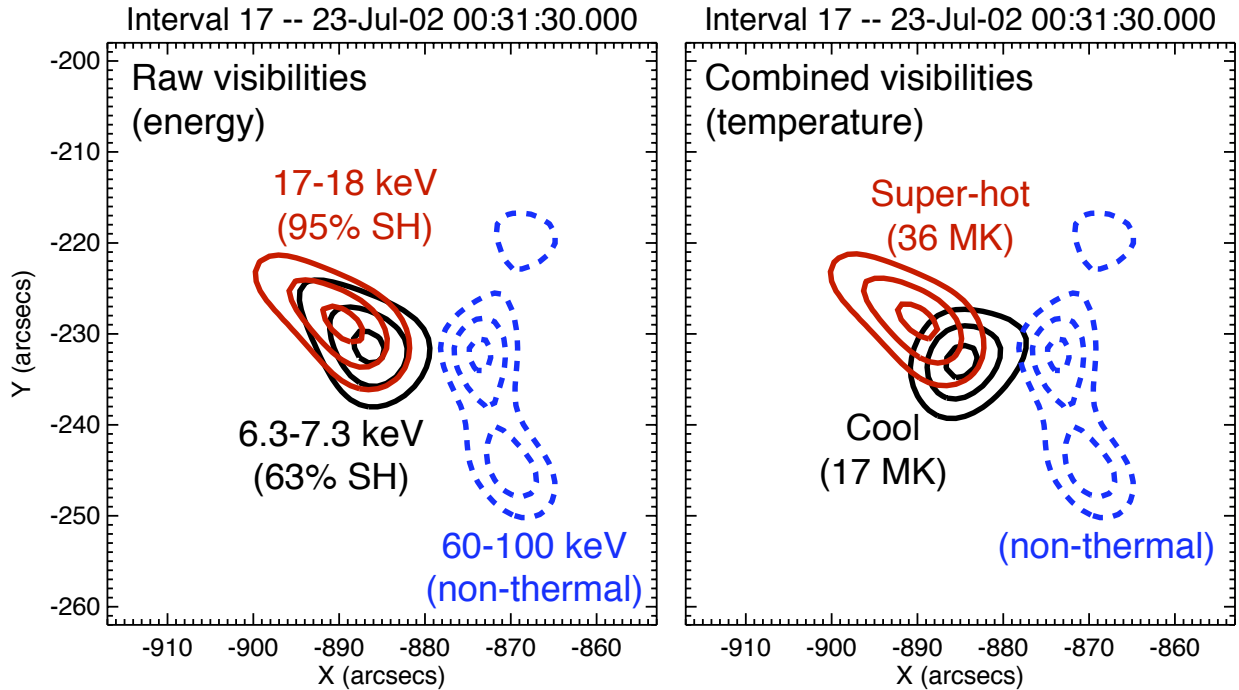

**Figure C.3** – [left] Images at 6.3-7.3, 17-18, and 60-100 keV (50%, 75%, 90% contours) reconstructed with visibilities using UV_SMOOTH; the contribution of the super-hot component to the two thermal bands is determined by spectral modeling. [right] Images of the super-hot and cool thermal sources (UV_SMOOTH, same contours) derived from linear combinations of visibilities in the two thermal energy bands weighted by the fractional contributions of the respective thermal sources; the 60-100 keV non-thermal image is shown for reference.

where we solve for the (unknown) relative visibilities from the super-hot and cool source ($v_{sh}$ and $v_c$, respectively) as the weighted sums of the (observed) relative visibilities in two energy bins $E_1$ and $E_2$; the imaging result of this inversion is shown in Figure C.3. Although only performed here for a single spectrum fit by two temperatures, this method is extensible to any spectrum that is well-fit by an $N$-temperature model and allows for the first time a means of creating precise, high-resolution images of flare emission at specific temperatures (subject to spectral modeling), rather than at specific energies.

# Appendix D: Glossary of acronyms

| | | |
|---|---|---|
| A0 | - | RHESSI attenuator state: no attenuators |
| A1 | - | RHESSI attenuator state: thin attenuator only |
| A3 | - | RHESSI attenuator state: thin+thick attenuators |
| ADC | - | analog-to-digital converter |
| Al | - | aluminum |
| BCS | - | Bragg (or Bent) Crystal Spectrometer |
| Be | - | beryllium |
| Ca | - | calcium |
| Cd | - | cadmium |
| CdTe | - | cadmium telluride |
| CsI(Na) | - | cesium iodide (sodium-doped) |
| CZT | - | cadmium zinc telluride ("cad-zinc-tel") |
| DRM | - | detector response matrix |
| EUV | - | extreme ultraviolet |
| Fe | - | iron |
| FOV | - | field-of-view |
| FWHM | - | full-width at half-max |
| Ge | - | germanium |
| GeD | - | germanium detector |
| GOES | - | *Geostationary Operational Environmental Satellite* |
| HPGe | - | high-purity germanium |
| HXR | - | hard X-ray (defined here as above ~20 keV) |
| HXT | - | Hard X-ray Telescope |
| IDL | - | Interactive Data Language |
| LHS | - | left-hand side |
| Li | - | lithium |
| LLD | - | low-level discriminator |
| MHD | - | magnetohydrodynamics |
| MK | - | mega-Kelvin |
| Mo | - | molybdenum |
| NASA | - | National Aeronautics and Space Administration |
| Ni | - | nickel |
| NoRH | - | *Nobeyama Radio Heliograph* |
| OSO | - | *Orbiting Solar Observatory* |
| OSPEX | - | *Object Spectral Executive* (SSW software) |
| PSF | - | point-spread function |
| RHESSI | - | *Reuven Ramaty High Energy Solar Spectroscopic Imager* |
| RHS | - | right-hand side |
| RMC | - | rotation-modulation collimator |
| S | - | sulfur |
| SAA | - | South Atlantic Anomaly |
| Si | - | silicon |
| SMEX | - | [NASA] Small Explorer |
| SMM | - | *Solar Maximum Mission* |
| SOXS | - | Solar X-ray Spectrometer |
| SSW | - | *SolarSoftWare* (IDL package) |
| SXR | - | soft X-ray (defined here as below ~20 keV) |
| SXS | - | Soft X-ray Spectroscope |
| SXT | - | Soft X-ray Telescope |
| TRACE | - | *Transition Region and Coronal Explorer* |
| UT | - | Universal (Greenwich) Time |
| UV | - | ultraviolet |
| W | - | tungsten |